\documentclass[traditabstract]{aa}
\usepackage{natbib}
\usepackage{graphicx}
\usepackage{lscape}
\usepackage{amsmath}
\usepackage{amssymb}
\usepackage{multirow,bigdelim}
\usepackage{booktabs}

\newcommand{\Spitzer}{\textit{Spitzer}}
\newcommand{\GALEX}{\textit{GALEX}}
\newcommand{\IRAS}{\textit{IRAS}}
\newcommand{\xmm}{\textit{XMM-Newton}}
\def\micron{\hbox{\,$\mu$m}}
\newcommand{\Herschel}{\textit{Herschel}}
\newcommand{\Lsun}{\hbox{$L_{\rm \odot}$}}
\newcommand{\Msun}{\hbox{$M_{\rm \odot}$}}
\newcommand{\LIR}{\hbox{$L_{\rm IR}$}}
\newcommand{\degree}{\ensuremath{^\circ}}
\newcommand{\Halfa}{H$\alpha$}
\newcommand{\HII}{\ion{H}{ii}}

\newcommand\nodata{ ~$\cdots$~ }

\titlerunning{Star-formation histories of local LIRGs}
\authorrunning{Pereira-Santaella et al.}

\begin{document}

\title{Star-formation histories of local luminous infrared galaxies}
\author{Miguel Pereira-Santaella\inst{\ref{inst1}} \and Almudena Alonso-Herrero\inst{\ref{inst2}} \and Luis Colina\inst{\ref{inst1}} \and Daniel Miralles-Caballero\inst{\ref{inst3}} \and  Pablo~G.~P\'erez-Gonz\'alez\inst{\ref{inst4}} \and Santiago Arribas\inst{\ref{inst1}} \and Enrica Bellocchi\inst{\ref{inst1}} \and Sara Cazzoli\inst{\ref{inst5},\ref{inst1}}  \and \\ Tanio D\'iaz-Santos\inst{\ref{inst6},\ref{inst7}} \and Javier Piqueras L\'opez\inst{\ref{inst1}}
}

\institute{
Centro de Astrobiolog\'ia (CSIC/INTA), Ctra de Torrej\'on a Ajalvir, km 4, 28850, Torrej\'on de Ardoz, Madrid, Spain \email{mpereira@cab.inta-csic.es} \label{inst1} \and
Instituto de F\'isica de Cantabria, CSIC-Universidad de Cantabria, 39005 Santander, Spain\label{inst2} \and
Departamento de F\'isica Te\'orica, Universidad Aut\'onoma de Madrid, 28049 Madrid, Spain \label{inst3} \and
Departamento de Astrof\'isica, Facultad de CC. F\'isicas, Universidad Complutense de Madrid, 28040 Madrid, Spain\label{inst4} \and
Cavendish Laboratory, University of Cambridge 19 J. J. Thomson Avenue, Cambridge CB3 0HE, UK \label{inst5} \and
Spitzer Science Center, California Institute of Technology, MS 220-6, Pasadena, CA 91125, USA \label{inst6} \and
N\'ucleo de Astronom\'ia de la Facultad de Ingenier\'ia, Universidad Diego Portales, Av. Ej\'ercito Libertador 441, Santiago, Chile \label{inst7}
}

\abstract{
We present the analysis of the integrated spectral energy distribution (SED) from the ultraviolet (UV) { to the far-infrared and \Halfa\ of a sample of 29 local systems and individual galaxies with infrared (IR) luminosities between 10$^{11}$\Lsun\ and 10$^{11.8}$\Lsun.} We have combined new narrow-band \Halfa+[\ion{N}{ii}] and broad-band $g$, $r$ optical imaging taken with the Nordic Optical Telescope (NOT), with archival \GALEX, 2MASS, \Spitzer, and \Herschel\ data. The SEDs { (photometry and integrated \Halfa\ flux)} have been fitted with a modified version of the \textsc{magphys} code using stellar population synthesis models for the UV--near-IR range and thermal emission models for the IR emission taking into account the energy balance between the absorbed and re-emitted radiation.
{ From the SED fits we derive the star-formation histories (SFH) of these galaxies. For nearly half of them the star-formation rate appears to be approximately constant during the last few Gyrs. In the other half, the current star-formation rate seems to be enhanced by a factor of 3--20 with respect to that occured $\sim$1\,Gyr ago.} Objects with constant SFH tend to be more massive than starbursts and they are compatible with the expected properties of a main-sequence (M-S) galaxy. Likewise, the derived SFHs show that all our objects were { M-S galaxies $\sim$1\,Gyr ago with stellar masses between 10$^{10.1}$ and 10$^{11.5}$\,\Msun.}
{ We also derived from our fits the average extinction ($A_{\rm v}=0.6-3$\,mag)} and the polycyclic aromatic hydrocarbons (PAH) luminosity to \LIR\ ratio (0.03--0.16). We combined the $A_{\rm v}$ with the total IR and H$\alpha$ luminosities into a diagram { which can be used to identify objects with rapidly changing (increasing or decreasing) SFR during the last 100\,Myr.}
}


\maketitle

\section{Introduction}

Galaxies with high infrared (IR) luminosities ($L_{\rm IR}=L(8-1000\,\mu m)>10^{11}$\,\Lsun) are rare in the local Universe (e.g., \citealt{LeFloch2005}), and yet they are a cosmologically important class of objects because they dominate the star-formation rate (SFR) density at high-$z$. They can be classified as luminous ($L_{\rm IR}=10^{11}-10^{12}$\,\Lsun) and ultra-luminous ($L_{\rm IR}>10^{12}$\,\Lsun) IR galaxies (LIRGs and ULIRGs, respectively). LIRGs dominate the SFR density at $z\sim1$, while ULIRGs do so at $z\sim2$ \citep{PerezGonzalez2005, Caputi2007}.

The bulk of the IR luminosity of U\slash LIRGs is produced by strong star-formation (SF) bursts \citep{Sanders96}, although, the number of U\slash LIRGs with an active galactic nuclei (AGN) detected in their optical spectra increases with the IR luminosity, reaching $\sim$50\% when $L_{\rm IR}>10^{12.3}$\,\Lsun\ \citep{Yuan2010}. Similarly, the relative contribution of AGN to the bolometric luminosity increases with increasing IR luminosity providing less than 2--15\% of the total luminosity of local LIRGs \citep{Pereira2011, Petric2011, AAH2012a}, and $\sim$20--25\% of the luminosity of local ULIRGs \citep{Farrah07,Nardini2009}.

These episodes of intense SF and high IR luminosities are mainly triggered by major mergers involving gas-rich progenitors in the case of local ULIRGs \citep{Sanders96}. Actually a majority ($>$80\%) of local ULIRGs are mergers (e.g., \citealt{Veilleux2002}). This is not true for local LIRGs which have more varied morphologies -- isolated disks, disturbed spirals, or mergers (e.g., \citealt{Arribas2004, AAH06s, Hung2014}). \citet{Hammer2005} suggest that episodic SF bursts due to minor mergers (or gas infall) could enhance the IR luminosity above the LIRG threshold. Moreover, according to $N$-body simulations, the most likely remnants of these minor interactions are disturbed spirals \citep{Bournaud2007}. This scenario seems to describe the observed morphologies of local LIRGs well.

High-$z$ ULIRGs differ from local objects with similar luminosities for several reasons. First, the incidence of mergers in high-$z$ ULIRGs is lower than locally, with only 30--40\% of the $z>1$ ULIRGs showing merger morphologies (e.g., \citealt{Elbaz2007, Kartaltepe2010}). In addition, the mid-IR spectra of $z\sim2$ ULIRGs differ from those of local ULIRGs because they are more similar to those local LIRGs \citep{Farrah08, Rigby2008}. Therefore, the triggering mechanisms and the physical conditions of the SF in distant ULIRGs resemble those of local LIRGs. In this context, the detailed study of local LIRGs is needed to better understand their high-$z$ counterparts.

In this work we model the integrated spectral energy distribution (SED) { of a sample of local IR bright galaxies} and derive fundamental physical parameters such as the stellar mass, SFR, star-formation history (SFH), average extinction, etc. The analysis of the integrated emission of local LIRGs is important for providing meaningful comparisons with the integrated SED used in high-$z$ studies. Previous works of the integrated SED of nearby galaxies focus on lower-luminosity galaxies (e.g., \citealt{Noll2009, Skibba2011, Brown2014}) or higher-luminosity U\slash LIRGs (e.g., \citealt{daCunha2010, U2012}), which shows that studying intermediate luminosity objects is needed.

The paper is organized as follows. In Section \ref{s:sample} the sample of local LIRGs is presented. Sections \ref{s:observations} and \ref{s:sed_modeling} describe the data reduction and the models used for the SED fitting, respectively. { In Section \ref{s:discussion} we discuss the SFH of this sample and the age effects on the H$\alpha$ to IR luminosity ratio.}
Finally, the main conclusions are presented in Section \ref{s:conclusions}.  

Throughout this paper we assume the following cosmology: $H_{\rm 0} = 70$\,km\,s$^{-1}$\,Mpc$^{-1}$, $\Omega_{\rm m}=0.3$, and $\Omega_{\rm \Lambda}=0.7$.

\section{Sample}\label{s:sample}

We drew a volume-limited sample of local LIRGs from the \textit{IRAS} Revised Bright Galaxy Sample (RBGS; \citealt{SandersRBGS}). Our selection criteria are similar to those used by \citet{AAH06s}: $v_{\rm hel}=2750-5200$\,km\,s$^{-1}$ and Galactic latitude $|b|>5$, but we slightly decreased the minimum \LIR\ down to $\log$\,\LIR\slash\Lsun$=11.0$. 
There are 59 sources in the RBGS that fulfill these criteria, and 37 of them are observable from the Roque de los Muchachos Obsevatory (Dec.\,$>-16$\degree). 

We obtained good quality $g$, $r$, and narrow-band H$\alpha$ images for 25 of them using the Nordic Optical Telescope (NOT; see Section \ref{ss:optical_data}), and another four had integrated H$\alpha$ flux measurements \citep{Moustakas2006} and Sloan Digital Sky Survey (SDSS; \citealt{Aihara2011}) $g$ and $r$ images available. Therefore, our sample includes 29 (78\%) of the parent sample northern LIRGs. Most of the missing galaxies belong to the lower end of the \LIR\ distribution ($\log$\,\LIR\slash\Lsun$ \sim11.0$), which is already well represented in our sample (see Table \ref{tab:sample}).

In addition, we observed six nearby companions of the RBGS objects (namely, NGC~876, UGC~03405, NGC~2389, MCG~+02-20-002, NGC~6921, and NGC~7769), with $\log$\,\hbox{\LIR\slash\Lsun}$= 10.2-10.7$. They are located between $1-6\arcmin$ ($20-100$\,kpc) away from the main RBGS galaxy and therefore might contribute to the measured \textit{IRAS} fluxes.
We were also able to resolve three of the RBGS targets into two subcomponents (CGCG~468-002 NED01/02, NGC~7752/3, and NGC~7770/1). Our sample contains the 38 sources listed in Table \ref{tab:sample}.

The IR luminosity range is 10$^{10.2}$--10$^{11.8}$\,\Lsun, with a mean and median luminosity of 10$^{11.0}$\,\Lsun. Except for one galaxy, CGCG~468-002 NED01, the IR luminosity is dominated by SF, and the bolometric AGN contribution is small, less than 5\% for most galaxies and up to 12\% in a few of them (see Table \ref{tab:sample} and \citealt{AAH2012a}).

According to their nuclear activity classification, our sample includes 11 \ion{H}{ii} galaxies, 13 composite, one LINER, five Seyfert galaxies, three objects without a clear optical classification\footnote{Two or more lines of the \citet{Baldwin1981} diagrams were not detected in their optical spectra.}, and one galaxy with no available optical spectrum. For nine of them we determined their nuclear classification and [\ion{N}{ii}]6584\,\AA\slash H$\alpha$ ratio using archival optical spectroscopy (see Appendix \ref{apx:optical_class}). 
The fraction of galaxies of each type is similar to what is expected for galaxies with IR luminosities in the range covered by our sample \citep{Yuan2010}.

\begin{table*}
\caption{The sample}
\label{tab:sample}
\setlength{\tabcolsep}{5pt}
\centering
\begin{tabular}{lccccccccc}
\hline \hline
Name &  RA & Dec. & $cz$\tablefootmark{a} & $D_{L}$\tablefootmark{b} & Spectral & [\ion{N}{ii}]\slash  \Halfa\tablefootmark{d} & Ref.\tablefootmark{e} & log\,\LIR\tablefootmark{f} & $L_{\rm AGN}/L_{\rm IR}$\tablefootmark{g}\\
 &  (J2000.0) & (J2000.0) & (km\,s$^{-1}$) & (Mpc) & class\tablefootmark{c} & & &(\Lsun) & \\
\hline
NGC~23 & 00 09 53.4 & +25 55 27 & 4478 & 64.7 & composite & 0.57 & 1 & 11.0 & 0.02 \\
MCG~+12-02-001 & 00 54 04.0 & +73 05 05 & 4722 & 68.3 & \ion{H}{ii} & 0.42 & 1 & 11.4 & $<$0.05 \\
NGC~876 & 02 17 53.2 & +14 31 19 & 3916 & 56.5 & composite & 0.57 & 2 & 10.4 & \nodata \\
NGC~877 & 02 17 59.7 & +14 32 39 & 3963 & 57.2 & composite & 0.52 & 3 & 11.1 & \nodata \\
UGC~01845 & 02 24 08.0 & +47 58 11 & 4600 & 66.5 & composite & 0.72 & 1 & 11.1 & $<$0.05 \\
NGC~992 & 02 37 25.5 & +21 06 02 & 4065 & 58.7 & \ion{H}{ii} & 0.43 & 4 & 11.0 & \nodata \\
UGC~02982 & 04 12 22.6 & +05 32 50 & 5354 & 77.6 & \ion{H}{ii} & 0.43 & 5 & 11.2 & $<$0.05 \\
NGC~1614 & 04 34 00.0 & --08 34 45 & 4778\tablefootmark{\star} & 69.1 & composite & 0.60 & 5 & 11.7 & $<$0.05 \\
CGCG~468-002 NED01 & 05 08 19.7 & +17 21 48 & 5267 & 76.3 & Sy1.9 & 0.98 & 2 & 10.6 & 0.40 \\
CGCG~468-002 NED02 & 05 08 21.2 & +17 22 08 & 4951 & 71.6 & composite & 0.51 & 2 & 11.0 & $<$0.05 \\
NGC~1961 & 05 42 04.7 & +69 22 43 & 3908 & 56.4 & LINER & 1.96 & 6 & 11.1 & \nodata \\
UGC~03351 & 05 45 47.9 & +58 42 03 & 4433 & 64.1 & Sy2 & 1.34 & 7 & 11.2 & 0.02 \\
UGC~03405 & 06 13 57.6 & +80 28 35 & 3799 & 54.8 & composite? & 0.75 & 2 & 10.3 & $<$0.05 \\
UGC~03410 & 06 14 29.6 & +80 27 00 & 3871 & 55.9 & \ion{H}{ii} & 0.45 & 2 & 10.9 & $<$0.05 \\
NGC~2388 & 07 28 53.5 & +33 49 09 & 4078 & 58.9 & \ion{H}{ii} & 0.56 & 1 & 11.2 & $<$0.05 \\
NGC~2389 & 07 29 04.7 & +33 51 40 & 3956\tablefootmark{\star} & 57.1 & composite & 0.53 & 2 & 10.5 & \nodata \\
MCG~+02-20-002 & 07 35 41.5 & +11 36 44 & 5100\tablefootmark{\star} & 73.8 & \nodata & \nodata & \nodata & 10.2 & \nodata \\
MCG~+02-20-003 & 07 35 43.4 & +11 42 35 & 4907 & 71.0 & composite & 0.45 & 1 & 11.0 & $<$0.05\\
NGC~3110 & 10 04 02.1 & --06 28 30 & 5013 & 72.6 & \ion{H}{ii} & 0.42 & 5 & 11.3 & $<$0.05 \\
NGC~3221 & 10 22 20.0 & +21 34 10 & 3959 & 57.1 & \ion{H}{ii} & 0.40 & 8 & 11.0 & \nodata \\
Arp~299 & 11 28 31.0 & +58 33 41 & 3056 & 44.0 & Sy2 & 0.35 & 9 & 11.8 & 0.04 \\
MCG~--02-33-098 & 13 02 19.6 & --15 46 04 & 4713 & 68.2 & composite & 0.33 & 5 & 10.9 & 0.16 \\
IC~860 & 13 15 03.6 & +24 37 08 & 3858 & 55.7 & no & 7.8 & 1 & 11.0 & $<$0.05 \\
NGC~5653 & 14 30 10.5 & +31 12 55 & 3512 & 50.6 & \ion{H}{ii} & 0.38 & 3 & 11.0 & 0.01 \\
Zw~049-057 & 15 13 13.1 & +07 13 32 & 3858 & 55.7 & \ion{H}{ii} & 0.46 & 8 & 11.0 & $<$0.05 \\
NGC~5936 & 15 30 00.9 & +12 59 21 & 3989 & 57.6 & \ion{H}{ii} & 0.48 & 1 & 11.0 & 0.03 \\
NGC~5990 & 15 46 16.4 & +02 24 56 & 3793 & 54.7 & Sy2 & 0.74 & 10 & 11.0 & 0.05 \\
NGC~6052 & 16 05 13.0 & +20 32 33 & 4739 & 68.5 & \ion{H}{ii} & 0.23 & 3 & 10.9 & \nodata \\
NGC~6701 & 18 43 12.5 & +60 39 12 & 3895 & 56.2 & composite & 0.67 & 1 & 11.0 & $<$0.05 \\
NGC~6921 & 20 28 28.9 & +25 43 24 & 4329 & 62.5 & AGN & \nodata & 11,12 & 10.2 & $<$0.05 \\
MCG~+04-48-002 & 20 28 35.1 & +25 44 00 & 4198 & 60.6 & \ion{H}{ii}\slash AGN & 0.42 & 11 & 11.0 & 0.06 \\
NGC~7591 & 23 18 16.3 & +06 35 09 & 4907 & 71.0 & composite & 0.85 & 1 & 11.0 & $<$0.05 \\
NGC~7679 & 23 28 46.7 & +03 30 41 & 5161 & 74.7 & Sy2/Sy1 & 0.59 & 5 & 11.1 & 0.18 \\
NGC~7752 & 23 46 58.5 & +29 27 32 & 4943 & 71.5 & composite & 0.33 & 2 & 10.7 & \nodata \\
NGC~7753 & 23 47 04.8 & +29 29 01 & 5201 & 75.3 & composite? & 1.02 & 2 & 10.9 & \nodata \\
NGC~7769 & 23 51 04.0 & +20 09 02 & 4157 & 60.0 & composite & 0.56 & 2 & 10.9 & 0.10 \\
NGC~7770 & 23 51 22.6 & +20 05 49 & 4127 & 59.6 & \ion{H}{ii} & 0.40 & 13 & 10.4 & 0.17 \\
NGC~7771 & 23 51 24.9 & +20 06 43 & 4276 & 61.8 & \ion{H}{ii} & 0.55 & 1 & 11.3 & 0.02 \\
\hline
\end{tabular}
\tablefoot{\tablefoottext{a}{Heliocentric velocity from the \Spitzer\slash IRS high-resolution spectra \citep{AAH2012a}.}
\tablefoottext{b}{Luminosity distance estimated from the redshift.}
\tablefoottext{c}{Classification of the nuclear activity from optical spectroscopy based on the classification scheme of \citet{Kewley2006}.}
\tablefoottext{d,e}{Nuclear [\ion{N}{ii}]6584\,\AA\slash \Halfa\ ratio and reference.}
\tablefoottext{f}{Logarithm of the total 4--1000\micron\ IR luminosity in solar units calculated in this paper. The AGN torus IR luminosity is included.}
\tablefoottext{g}{Ratio between the bolometric AGN luminosity and the total IR luminosity from \citet{AAH2012a}.}
\tablefoottext{\star}{Heliocentric velocities from NED.}
}
\tablebib{(1) \citet{AAH09PMAS}; (2) This work (Appendix \ref{apx:optical_class}); (3) \citet{Moustakas2006}; (4) \citet{Keel1984}; (5) \citet{Veilleux1995}; (6) \citet{Ho1997}; (7) \citet{Baan1998}; (8) \citealt{Aihara2011}; (9) \citet{GarciaMarin06}; (10) \citet{Kewley2001}; (11) \citet{Masetti2006}; (12) \citet{Tueller2008}; (13) \citet{AAH2012c}.}
\end{table*}

\section{Observations and data reduction}\label{s:observations}

\begin{figure*}[ht]
\centering
\includegraphics[width=0.87\textwidth]{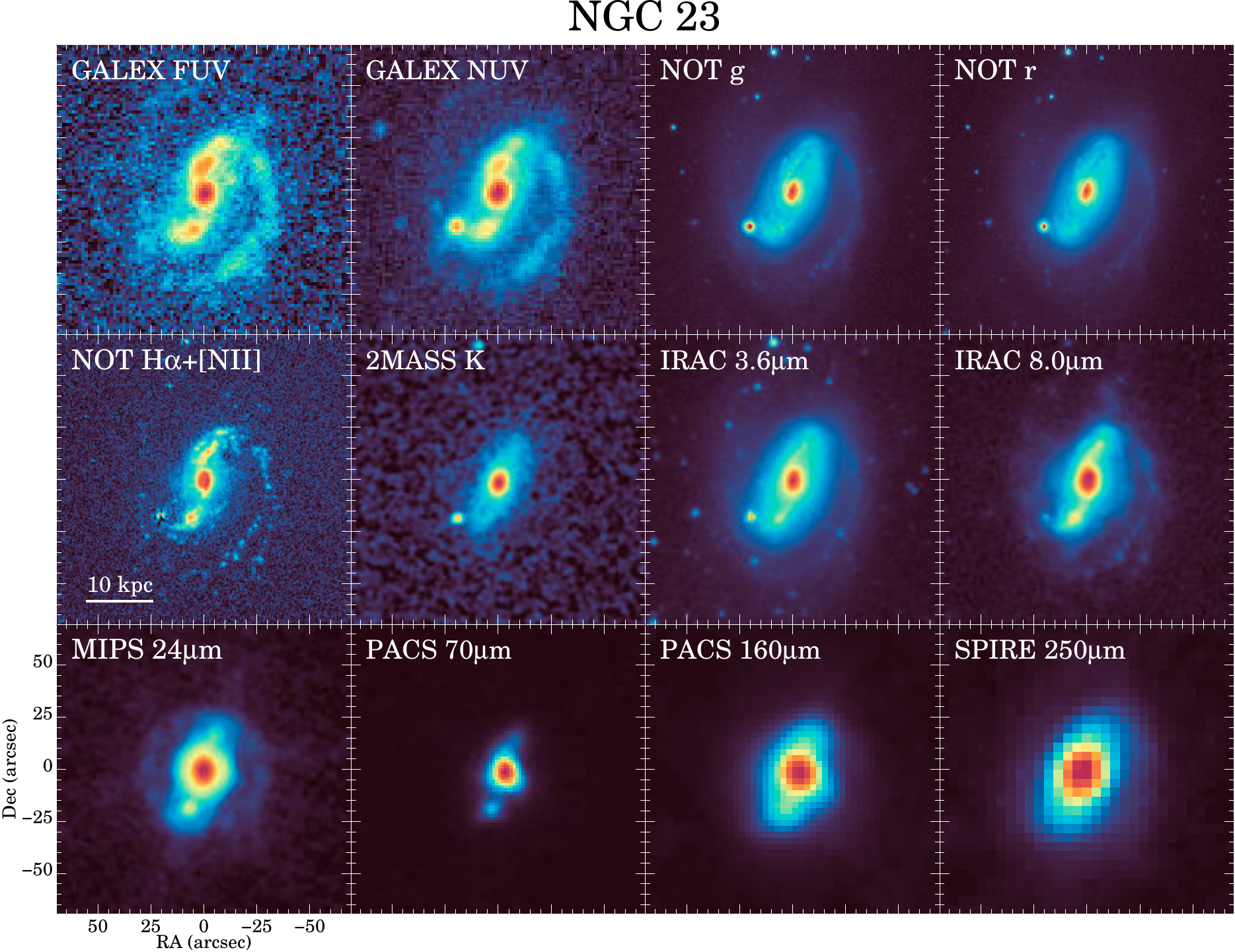}
\caption{Subset of the images used to construct the integrated SED of NGC~23. The new NOT optical observations of the $g$, $r$, and \Halfa +[\ion{N}{ii}] bands are shown. The rest of the images were obtained from the public archives of their respective observatories (see Section \ref{s:observations} for details). Some images used in the SED are not displayed here because they have morphologies similar to those of the presented images (2MASS $J$ and $H$, IRAC 4.5 and 5.8\micron, and PACS 100\micron) or because they have a very low angular resolution (SPIRE 350 and 500\micron). All images are shown in a logarithm scale. North is up and east to
the left. The white bar in the \Halfa +[\ion{N}{ii}] panel represents 10\,kpc at the distance of the object.
\label{fig:images_main_text}}
\end{figure*}

In this section we describe the reduction and analysis of our new optical observations (Section \ref{ss:optical_data}) along with the archival data (Section \ref{ss:archival_data}). The reduced images for each galaxy are shown in Figure \ref{fig:images_main_text} and Appendix \ref{apx:images}.

\subsection{Optical imaging}\label{ss:optical_data}

\begin{table*}
\caption{Log of the NOT\slash ALFOSC observations}
\label{tab:log_obs}
\centering
\begin{tabular}{lcccccccc}
\hline \hline
Date &  Seeing & Targets \\
& (\arcsec) & \\
\hline
11 May 2011 & 0.9--1.4 & Arp~299, MCG~--02-33-098, NGC~5990 \\
12 May 2011 & 0.9 & NGC~5936 \\
18 Sept. 2011 & 0.8 & (NGC~6921, MCG~+04-48-002), NGC~7591\\
19 Sept. 2011 & 0.7--1.1 & MCG~+12-02-001, NGC~6701, NGC~7769, \\
& & (NGC~7770, NGC~7771)  \\
29 Nov. 2011 & 0.7--1.4 & NGC~23, (NGC~876, NGC~877), UGC~02982, \\ 
& & NGC~1614, NGC~2388, NGC~2389 \\
30 Nov. 2011 & 0.6--1.5 & UGC~01845, UGC~03351, MCG~+02-20-002, \\ 
& & MCG~+02-20-003, NGC~3110, NGC~7679 \\
1 Dec. 2011 & 1.2--1.5 & NGC~992, (CGCG~468-002 NED01/02), \\ 
& & NGC~1961, (UGC~03405, UGC~03410), \\
& & NGC~3221, (NGC~7752, NGC~7753)\\
\hline
\end{tabular}
\tablefoot{Galaxies observed with the same telescope pointing are grouped with parentheses.
}
\end{table*}

We obtained broad- and narrow-band imaging of 34 IR bright galaxies using the Andalucia Faint Object Spectrograph and Camera (ALFOSC) on the 2.6\,m NOT at the Roque de los Muchachos Obsevatory during three observing runs between May and December 2011 (see Table \ref{tab:log_obs}) { as part of the programs 115-NOT11/11A and 28-NOT2/11B.}
We used the broad-band SDSS $g$ and $r$ filters (\#120 and \#110), and for the narrow-band images we used the filters \#50 and \#68 from the NOT filters set, and \#65 and \#66 from the Isaac Newton Group of Telescopes (ING). These narrow filters have a full-width half-maximum (FWHM) of about 5\,nm and $\lambda_{\rm c}$ between 663 and 665\,nm (see Table \ref{tab:filtes}). For three objects (Arp~299, NGC~5936, and NGC~5990) not observed by us with the $g$ filter we used their available SDSS $g$ image \citep{Aihara2011}. In addition, the $r$ and $g$ images of SDSS were used for IC~860, NGC~5653, Zw~049-057, and NGC~6052.

We selected the narrow-band filter for each galaxy according to its redshift. The plate scale of ALFOSC is \hbox{0.19\arcsec\,pixel$^{-1}$}, and its field of view (FoV) is 7\arcmin$\times$7\arcmin, large enough to cover the emission of these LIRGs with a single pointing. The atmospheric conditions were photometric, and the seeing varied between 0\farcs6 and 1\farcs5 with a median seeing of 0\farcs9. 

The integration times were 1300 and 800\,s for the $g$ and $r$ filters, respectively, and 3000\,s for the narrow-band filters. Each integration was divided into three to five dithered exposures that were later combined to correct for cosmic ray hits and bad pixels of the detector.

For the data reduction we first subtracted the bias level using the overscan region to scale the master-bias. During the April run, the bias showed noticeable variability between exposures, so we used the overscan region to estimate the bias level of each row in each exposure. Then the resulting images were divided by the sky flat of the corresponding filter. In addition, bad pixels identified in the flat field images were masked. The sky extinction was determined by observing spectrophotometric standard stars from the ING catalog (ING Technical Note 100) and standard SDSS stars \citep{Smith2002} at different air masses between 1 and 4. Besides this, we used these standard stars to calculate the photometric AB zero points for the $g$ and $r$ filters.

We combined individual exposures after subtracting the sky background emission and aligning them to a common reference image using stars in the FoV. The absolute astrometry of each image was determined using the Guide Star Catalog (GSC2.3; \citealt{Lasker2008}). About 10--40 stars were used for each image, and the estimated 1$\sigma$ astrometric uncertainty is 0\farcs1--0\farcs2. 
We checked our absolute flux calibration for those fields that were also observed by the SDSS (15 out of 28). We found good agreement between our measurements and those reported in the Eighth Data Release of the SDSS \citep{Aihara2011} with differences around 0.01--0.05\,mag for objects brighter than 19\,mag.

We used the \citet{Fitzpatrick1999} extinction law and the Galactic color excess $E(B-V)$ (Table \ref{tab:photometry}) from the NASA/IPAC extragalactic database (NED) to correct the observed optical fluxes for Galactic extinction.

\begin{table}
\caption{Characteristics of { the} NOT\slash ALFOSC filters}
\label{tab:filtes}
\centering
\begin{tabular}{lcccccccc}
\hline \hline
Filter \# &  $\lambda_{\rm c}$ & FWHM\\
& (nm) & (nm) \\
\hline
50 & 665.3 & 5.5\\
68 & 664.0 & 4.0 \\
110 $r$ & 618.0 & 148 \\
120 $g$ & 480.0 & 145 \\
65 (ING) & 662.6 & 4.4 \\
66 (ING) & 664.5 & 5.0 \\
\hline
\end{tabular}
\end{table}

\subsubsection{\Halfa\ narrow-band imaging calibration}

To obtain the \Halfa+[\ion{N}{ii}] emission images, we first subtracted the continuum using the $r$ band image. It was scaled using the relative fluxes of 10--30 field stars observed in both images using 4\arcsec\ apertures. These scaling factors are in good agreement with the theoretical factor that can be calculated using the transmission curves of the filters. Some of the nights we obtained observations of standard stars with all the narrow filters, which confirms that the derived scaling factors are accurate within $\sim$15\%.

To calculate the conversion factor from counts~s$^{-1}$ in the narrow filter to the corresponding \Halfa\ flux in erg\,cm$^{-2}$\,s$^{-1}$ units, we first created a synthetic spectrum including only the H$\alpha$ line and the 6548 and 6584\,\AA\ [\ion{N}{ii}] transitions at the redshift of each source. For the nuclear regions ($\sim3$\,kpc) where the [\ion{N}{ii}]\slash\Halfa\ ratio would be more uncertain because of the nuclear activity, we used the ratios derived from the nuclear spectrum of each target (Table \ref{tab:sample}). For the extra-nuclear emission we assumed that [\ion{N}{ii}]6584\AA\slash \Halfa\ $=$ 0.3, typical of \HII\ regions \citep{Kennicutt1983}.
The synthetic spectrum was convolved with the transmission curve of the narrow filter and the result was converted into physical units using the known input H$\alpha$ flux and the relation between the narrow- and broad-band $r$ calibration (see previous section).
The 10$\sigma$ sensitivity of the images is $\sim$10$^{-16}$\,erg\,cm$^{-2}$\,s$^{-1}$, which corresponds to an $H\alpha$ luminosity of 10$^{37}$\,erg\,s$^{-1}$ at the median distance of our galaxies.

\subsubsection{Integrated photometry}\label{ss:integrated_photometry}

We defined the apertures to obtain the integrated emission using the NOT $r$ images. In these images, we considered all the pixels above a surface brightness of 23\,mag\,arcsec$^{-1}$ and fitted an ellipse to them. The resulting elliptical apertures for each galaxy are listed in Table \ref{tab:apertures}.
They encompass most of the $r$ emission for all the sources, although in a few cases some faint \HII\ regions at large galactocentric distances ($r>15$\,kpc) lie outside of them. The diameters of the apertures range between 30 and 200\arcsec\ with a median diameter of 80\arcsec, and are equivalent to $\sim$20-25\,kpc on average.

To perform the photometry in the optical images ($r$, $g$, and \Halfa) we integrated all the emission in the calculated apertures (Table \ref{tab:apertures}) after masking Galactic stars that lie within them. The integrated fluxes are shown in Table \ref{tab:photometry}.

In this paper we use images with a wide range of angular resolutions, from $\sim$1 to 35\arcsec\ (see next section). Therefore, to adapt the calculated elliptical apertures we convolved them using a Gaussian with a FWHM equal to the difference between the desired angular resolution and the $r$ resolution (FWHM$\sim$1\arcsec) subtracted in quadrature.

\subsection{Ancillary archival data}\label{ss:archival_data}

To construct the SEDs of our sources, we looked for observations carried out at different wavelengths from the  UV to the far-IR. In particular, we used images from \GALEX\ (UV; \citealt{Martin2005GALEX}), 2MASS (near-IR; \citealt{Skrutskie2006}), \Spitzer\ (mid-IR; \citealt{Werner2004}), and \Herschel\ (far-IR; \citealt{Pilbratt2010Herschel}), all publicly available in their archives. The reduction and photometry of these archival data is described below.

\subsubsection{\GALEX\ UV data}

In the \GALEX\ archive we found far-UV (1516\AA) and\slash or near-UV (2367\AA) observations for 36 out of 38 galaxies in our sample. Most of them belong to the all-sky imaging survey (AIS). The rest are part of the medium imaging survey (MIS), the nearby galaxies survey (NGS), or guest investigator programs.
We used the images downloaded from the archive to make the photometric measurements using the apertures described in Section \ref{ss:integrated_photometry} (Table \ref{tab:apertures}) taking into account that the angular resolution of the \GALEX\ images is 
4--6\arcsec. As for the optical images, we masked stars inside of the apertures since in some cases they were bright, particularly in the near-UV band. To estimate the background within the apertures, we used the sky background images provided by the \GALEX\ pipeline. Finally to convert from count rates to physical units, we used the conversion factors given in the \GALEX\ Observer's Guide.

For one of the galaxies without any \GALEX\ data (NGC~1614), we took the near-UV flux from imaging obtained by the optical monitor (OM) onboard \xmm\ using the UVW2 (2120\AA) filter \citep{Pereira2011}.

We corrected the UV fluxes (Table \ref{tab:photometry}) for Galactic extinction using the same method as we used for the optical data (see Section \ref{ss:optical_data}).

\subsubsection{2MASS data}

We retrieved the $J$ (1.2\micron), $H$ (1.7\micron), and $K_{\rm s}$ (2.2\micron) near-IR images of our galaxies from the Two Micron All Sky Survey (2MASS). Most of them were from the 2MASS extended source catalog \citep{Jarrett2000}, although a few objects were part of the 2MASS large galaxy atlas \citep{Jarrett2003}. These downloaded images were already flux calibrated with an angular resolution of $\sim$2\arcsec.

The photometry on the images was done considering the same elliptical apertures (Table \ref{tab:apertures}) used for the other bands. In general, there is good agreement between our integrated measurements (Table \ref{tab:photometry}) and those reported in the 2MASS catalogs.

\subsubsection{\Spitzer\ IRAC and MIPS imaging}

In the \Spitzer\ archive we found imaging of our galaxies for the four IRAC bands (at 3.6, 4.5, 5.8, and 8.0\micron; \citealt{Fazio2004IRAC}) and the 24\micron\ MIPS band (\citealt{Rieke2004MIPS}).

We retrieved the basic calibrated data (BCD) for these observations from the \Spitzer\ archive. The BCD processing includes several corrections (e.g., flat field, linearization, and dark subtraction) and flux calibration based on standard stars. We combined these BCD images into mosaics using the version 18 of the MOsaicker and Point source EXtractor (MOPEX) software provided by the SSC using the standard parameters (see the MOPEX User's Guide for details on the data reduction). The FWHM of the point spread function (PSF) of these images vary between 1\farcs7 and 2\farcs0 for the IRAC bands and it is 5\farcs9 for the 24\micron\ MIPS images.

To measure the integrated emission, we used the apertures listed in Table \ref{tab:apertures} and corrected for the lower angular resolution of the \Spitzer\ images. For the IRAC images we applied the extended source aperture correction or the point source aperture correction for very compact objects (see the IRAC Instrument Handbook). These corrections are about 20\% of the measured flux. In the MIPS 24\micron\ images, some galaxies are very compact, so we applied the point source correction described in the MIPS Instrument Handbook.

MCG~+02-20-002\slash 3 were not observed with MIPS, therefore we consider the \IRAS\ 25\micron\ flux from \citet{Surace2004} as an upper limit because of the limited \IRAS\ angular resolution (1\arcmin\ at 25\micron) other sources might contribute to the \IRAS\ flux. The measured integrated fluxes are listed in Tables \ref{tab:photometry} and \ref{tab:photometry_ir}.

\subsubsection{\Herschel\ PACS and SPIRE imaging}

Far-IR imaging of our galaxies taken with \Herschel\slash PACS (70, 100, and 160\micron; \citealt{Poglitsch2010PACS}) and SPIRE (250, 350, and 500\micron; \citealt{Griffin2010SPIRE}) were available in the \Herschel\ archive. Most of them were part of the program ``Herschel-GOALS: PACS and SPIRE Imaging of a Complete Sample of Local LIRGs'' (OT1\_dsanders\_1, PI: D. Sanders).

To produce the images from the downloaded raw data, we first used the \Herschel\ interactive pipeline environment software (HIPE) version 11 to create the flux calibrated timelines for each bolometer of the detectors. HIPE also attaches the pointing information to the timelines. Then, using Scanamorphos version 22 \citep{Roussel2012}, we combined and projected these timelines in a spatial grid and obtained the final images.
For the three PACS bands, the FWHMs of the PSF are $\sim$6\arcsec, 7\arcsec, and 11\arcsec, and for the SPIRE bands they are 18\arcsec, 24\arcsec, and 35\arcsec, respectively.

We performed the photometry on the PACS and SPIRE images using the apertures of Table \ref{tab:apertures} convolved with a Gaussian to account for the lower angular resolution of them. In some cases the galaxies were point-like at the \Herschel\ resolution and we applied the point-source aperture corrections recommended in the PACS and SPIRE observer manuals. It was not possible to resolve the two galaxies of CGCG~468-002 in the SPIRE 500\micron\ image, so we used the measured flux as an upper limit to the emission of each component of the system. For NGC~7769 no PACS images were available, so we used the \IRAS\ 100\micron\ flux from \citet{Surace2004} as an upper limit estimate. In Table \ref{tab:photometry_ir} we list the far-IR fluxes.

\begin{table}
\caption{Photometry apertures}
\label{tab:apertures}
\centering
\begin{tabular}{lccccccccccccc}
\hline \hline
Name & $a$\tablefootmark{a} & $b$\tablefootmark{a} & P.A\tablefootmark{b} & $2a$\tablefootmark{c}\\
&  (\arcsec) & (\arcsec) & (\degree)  & (kpc) \\
\hline
NGC~23 & 40.1 & 32.5 & 150 & 25.2 \\
MCG~+12-02-001 & 16.7 & 8.2 & 117 & 11.0 \\
NGC~876 & 49.4 & 17.0 & 28 & 27.1 \\
NGC~877 & 55.5 & 42.5 & 139 & 30.8 \\
UGC~01845 & 24.6 & 12.2 & 139 & 15.9 \\
NGC~992 & 25.4 & 15.5 & 13 & 14.4 \\
UGC~02982 & 33.9 & 14.7 & 108 & 25.5 \\
NGC~1614 & 29.5 & 24.1 & 163 & 19.8 \\
CGCG~468-002 NED01 & 18.3 & 13.0 & 94 & 13.6 \\
CGCG~468-002 NED02 & 27.2 & 4.6 & 147 & 18.9 \\
NGC~1961 & 106.0 & 68.8 & 91 & 58.0 \\
UGC~03351 & 41.4 & 9.2 & 166 & 25.7 \\
UGC~03405 & 40.6 & 9.4 & 128 & 21.6 \\
UGC~03410 & 48.6 & 13.4 & 120 & 26.3 \\
NGC~2388 & 26.5 & 17.6 & 69 & 15.1 \\
NGC~2389 & 64.4 & 40.1 & 74 & 35.6 \\
MCG~+02-20-002 & 24.9 & 18.0 & 94 & 17.8 \\
MCG~+02-20-003 & 22.9 & 12.8 & 144 & 15.8 \\
NGC~3110 & 38.8 & 18.9 & 171 & 27.3 \\
NGC~3221 & 96.8 & 25.6 & 167 & 53.6 \\
Arp~299 & 43.5 & 41.2 & 7 & 18.6 \\
MCG~--02-33-098 & 46.4 & 11.5 & 62 & 30.7 \\
IC~860 & 29.1 & 16.5 & 18 & 15.7 \\
NGC~5653 & 46.1 & 33.5 & 112 & 22.6 \\
Zw~049-057 & 17.8 & 11.1 & 24 & 9.6 \\
NGC~5936 & 42.0 & 39.3 & 156 & 23.4 \\
NGC~5990 & 37.0 & 23.9 & 116 & 19.6 \\
NGC~6052 & 27.2 & 20.7 & 4 & 18.1 \\
NGC~6701 & 36.3 & 33.6 & 27 & 19.8 \\
NGC~6921 & 34.1 & 12.0 & 140 & 20.7 \\
MCG~+04-48-002 & 24.9 & 10.1 & 67 & 14.6 \\
NGC~7591 & 48.6 & 24.1 & 143 & 33.4 \\
NGC~7679 & 26.5 & 19.4 & 79 & 19.2 \\
NGC~7752 & 20.6 & 9.7 & 98 & 14.3 \\
NGC~7753 & 56.6 & 43.7 & 76 & 41.3 \\
NGC~7769 & 43.2 & 35.5 & 123 & 25.2 \\
NGC~7770 & 20.5 & 15.6 & 12 & 11.9 \\
NGC~7771 & 53.1 & 21.0 & 71 & 31.8 \\
\hline
\end{tabular}
\tablefoot{Elliptical apertures used to measure the integrated emissions based on the 23\,mag\,arcsec$^{-1}$ isophotes. \tablefoottext{a}{Semi-major and -minor axes of the elliptical aperture.}
\tablefoottext{b}{Position angle measured counter-clockwise from the North axis.}
\tablefoottext{c}{Physical size of the major axis of the aperture at the assumed distance (see Table \ref{tab:sample}).}
}
\end{table}

\begin{table*}
\caption{Integrated UV, optical, and near-IR photometry}
\label{tab:photometry}
\centering
\setlength{\tabcolsep}{5pt}
\begin{tiny}
\begin{tabular}{lccccccccccccc}
\hline \hline \\[-2.2ex]
Name & & & \multicolumn{9}{c}{Integrated fluxes (mJy)} \\
\cmidrule{4-12}\\[-2.8ex]
& & & \multicolumn{2}{c}{\GALEX} & \multicolumn{2}{c}{NOT} & \multicolumn{3}{c}{2MASS} & \multicolumn{2}{c}{\Spitzer\slash IRAC}\\
\cmidrule(rl){4-5} \cmidrule(rl){6-7} \cmidrule(rl){8-10} \cmidrule(rl){11-12}\\[-2.8ex]
& $E(B-V)$\tablefootmark{a}  & \Halfa\tablefootmark{b} & FUV & NUV & $g$ & $r$ & $J$ & $H$ & $K$ &  \\
& (mag) & 6563\AA & 1516\AA & 2267\AA & 4770\AA & 6230\AA & 1.23$\mu$m & 1.66$\mu$m & 2.16$\mu$m & 3.6$\mu$m & 4.5$\mu$m \\
\hline
NGC~23 & 0.034 & 13 & 1.2 & 2.8 & 37 & 67 & 170 & 210 & 180 & 100 & 68 \\
MCG~+12-02-001 & 0.550 & 14 & $<$1.80 & $<$2.94 & 7.4 & 14 & 47 & 58 & 56 & 51 & 45 \\
NGC~876 & 0.099 & 1.0 & 0.17 & 0.30 & 5.0 & 9.8 & 34 & 45 & 45 & 26 & 19 \\
NGC~877 & 0.099 & 24 & 4.3 & 7.5 & 61 & 99 & 220 & 260 & 220 & 130 & 93 \\
UGC~01845 & 0.190 & 2.3 & \nodata & $<$0.17 & 6.1 & 14 & 55 & 80 & 78 & 50 & 37 \\
NGC~992 & 0.130 & 14 & 1.5 & 2.7 & 18 & 30 & 76 & 96 & 84 & 63 & 46 \\
UGC~02982 & 0.360 & 7.7 & $<$0.57 & 1.1 $\pm$ 0.2 & 12 & 22 & 68 & 86 & 85 & 62 & 46 \\
NGC~1614 & 0.140 & 18 & \nodata & 2.9\tablefootmark{c} & 26 & 41 & 97 & 120 & 120 & 93 & 76 \\
CGCG~468-002 NED01 & 0.310 & 1.9 & $<$0.35 & 0.47 & 11 & 19 & 50 & 63 & 59 & 42 & 41 \\
CGCG~468-002 NED02 & 0.310 & 1.7 & $<$0.37 & 0.70 & 4.0 & 6.6 & 17 & 22 & 19 & 12 & 11 \\
NGC~1961 & 0.110 & 26 & 4.9 & 10 & 100 & 170 & 470 & 600 & 500 & 260 & 170 \\
UGC~03351 & 0.250 & 3.0 & \nodata & \nodata & 7.2 & 16 & 76 & 110 & 120 & 77 & 58 \\
UGC~03405 & 0.085 & 1.5 & 0.15 $\pm$ 0.02 & 0.31 & 5.7 & 12 & 40 & 53 & 50 & 30 & 20 \\
UGC~03410 & 0.085 & 5.7 & 0.40 & 0.75 & 11 & 23 & 86 & 120 & 110 & 78 & 57 \\
NGC~2388 & 0.051 & 4.0 & 0.049 & 0.21 & 9.7 & 22 & 78 & 100 & 95 & 64 & 48 \\
NGC~2389 & 0.051 & 11 & 3.5 & 5.4 & 23 & 33 & 60 & 75 & 59 $\pm$ 6 & 37 & 26 \\
MCG~+02-20-002 & 0.028 & 2.7 & 0.95 & 1.6 & 8.5 & 12 & 22 & 24 & 21 & 12 & 8.3 \\
MCG~+02-20-003 & 0.028 & 3.0 & 0.42 & 0.79 & 6.5 & 11 & 25 & 32 & 30 & 26 & 33 \\
NGC~3110 & 0.031 & 14 & 1.6 & 3.3 & 25 & 40 & 92 & 110 & 100 & 75 & 54 \\
NGC~3221 & 0.021 & 7.1 & 0.90 & 1.7 & 24 & 45 & 140 & 210 & 180 & 120 & 83 \\
Arp~299 & 0.015 & 80 & 9.2 & 13 & 59\tablefootmark{d} & 95 & 210 & 270 & 270 & 290 & 370 \\
MCG~--02-33-098 & 0.053 & 4.9 & 0.60 & 1.2 & \nodata & 19 & 45 & 55 & 49 & 33 & 26 \\
IC~860 & 0.012 & 0.12\tablefootmark{e} & $<$0.02 & 0.19 & 9.0\tablefootmark{d} & 16\tablefootmark{d} & 37 & 48 & 38 & 19 & 13 \\
NGC~5653 & 0.013 & 14\tablefootmark{e} & 1.6 & 3.4 & 33\tablefootmark{d} & 58\tablefootmark{d} & 130 & 160 & 140 & 92 & 66 \\
Zw~049-057 & 0.035 & 0.51\tablefootmark{e} & $<$0.01 & 0.057 & 3.4\tablefootmark{d} & 7.2\tablefootmark{d} & 21 & 29 & 24 & 14 & 11 \\
NGC~5936 & 0.034 & 12 & 2.3 & 4.2 & 31\tablefootmark{d} & 48 & 110 & 130 & 110 & 72 & 51 \\
NGC~5990 & 0.099 & 14 & 1.5 & 3.4 & 31\tablefootmark{d} & 54 & 150 & 180 & 170 & 130 & 120 \\
NGC~6052 & 0.067 & 26\tablefootmark{e} & 4.5 & 7.0 & 22\tablefootmark{d} & 28\tablefootmark{d} & 39 & 44 & 36 & 28 & 21 \\
NGC~6701 & 0.037 & 11 & \nodata & 3.8 & 32 & 56 & 130 & 150 & 130 & 81 & 57 \\
NGC~6921 & 0.390 & 1.4 & \nodata & $<$0.89 & 40 & 76 & 190 & 240 & 210 & 96 & 61 \\
MCG~+04-48-002 & 0.390 & 6.1 & \nodata & $<$1.04 & 10 & 20 & 64 & 85 & 81 & 60 & 52 \\
NGC~7591 & 0.091 & 5.6 & 0.53 & 1.3 & 17 & 33 & 89 & 110 & 97 & 55 & 39 \\
NGC~7679 & 0.058 & 16 & 2.2 & 4.3 & 25 & 36 & 65 & 73 & 64 & 52 & 39 \\
NGC~7752 & 0.088 & 6.8 & 0.93 & 1.7 & 7.6 & 11 & 23 & 26 & 24 & 19 & 13 \\
NGC~7753 & 0.087 & 10 & 2.5 & 4.6 & 40 & 68 & 170 & 220 & 190 & 84 & 56 \\
NGC~7769 & 0.066 & 11 & 3.9 & 6.1 & 42 & 73 & 170 & 210 & 170 & 95 & 63 \\
NGC~7770 & 0.066 & 6.9 & 1.3 & 2.0 & 9.6 & 16 & 30 & 37 & 31 & 25 & 24 \\
NGC~7771 & 0.066 & 10 & 0.87 & 2.1 & 36 & 73 & 230 & 300 & 270 & 150 & 110 \\
\hline
\end{tabular}
\end{tiny}
\tablefoot{Statistical uncertainties are included only when they are larger than 10\% of the flux. UV, optical, and H$\alpha$ fluxes are corrected for Galactic extinction using the \citet{Fitzpatrick1999} extinction law.
\tablefoottext{a}{Galactic color excess $E(B-V)$ from NED used for the Galactic extinction correction.}
\tablefoottext{b}{The units of the observed \Halfa\ fluxes are 10$^{-13}$\,erg\,cm$^{-2}$\,s$^{-1}$.}
\tablefoottext{c}{\xmm\slash OM UVW2 (2120\,\AA) flux from \citet{Pereira2011}.}
\tablefoottext{d}{Measured from SDSS images.}
\tablefoottext{e}{Integrated H$\alpha$ fluxes from \citet{Moustakas2006}.}
}
\end{table*}

\begin{table*}
\caption{Integrated mid- and far-IR photometry}
\label{tab:photometry_ir}
\centering
\setlength{\tabcolsep}{5pt}
\begin{tabular}{lcccccccccccccccccccccc}
\hline \hline \\[-2.2ex]
Name & \multicolumn{2}{c}{\Spitzer\slash IRAC} & \Spitzer\slash MIPS & \multicolumn{3}{c}{\Herschel\slash PACS} & \multicolumn{3}{c}{\Herschel\slash SPIRE}\\
\cmidrule(rl){2-3} \cmidrule(rl){4-4} \cmidrule(rl){5-7} \cmidrule(rl){8-10}\\[-2.8ex]
& 5.8\micron & 8.0\micron & 24\micron & 70\micron & 100\micron & 160\micron & 250\micron & 350\micron & 500\micron \\
\hline
NGC~23 & 0.17 & 0.43 & 0.86 & 10 & 14 & 13 & 5.2 & 2.1 & 0.63 \\
MCG~+12-02-001 & 0.20 & 0.58 & 2.9 & 25 & 29 & 22 & 7.5 & 2.8 & 0.96 \\
NGC~876 & 0.045 & 0.10 & 0.17 & 3.3 & 5.0 & 5.5 & 2.6 & 1.1 & 0.39 \\
NGC~877 & 0.29 & 0.77 & 0.88 & 12 & 23 & 28 & 14 & 5.8 & 1.9 \\
UGC~01845 & 0.12 & 0.34 & 0.84 & 13 & 17 & 15 & 4.8 & 1.8 & 0.56 \\
NGC~992 & 0.19 & 0.55 & 1.0 & 12 & 16 & 15 & 5.4 & 2.1 & 0.66 \\
UGC~02982 & 0.19 & 0.56 & 0.60 & 9.9 & 16 & 16 & 7.2 & 3.1 & 1.2 \\
NGC~1614 & 0.30 & 0.88 & 5.7 & 36 & 36 & 23 & 6.5 & 2.3 & 0.65 \\
CGCG~468-002 NED01 & 0.056 & 0.088 & 0.33 & 2.4 & 2.9 & 2.5 & 0.82 & 0.30 & $<$0.19\tablefootmark{a} \\
CGCG~468-002 NED02 & 0.033 & 0.086 & 0.60 & 9.2 & 9.6 & 6.8 & 2.0 & 0.73 & $<$0.19\tablefootmark{a} \\
NGC~1961 & 0.32 & 0.73 & 0.71 & 11 & 24 & 36 & 18 & 8.1 & 3.0 \\
UGC~03351 & 0.19 & 0.52 & 0.66 & 18 & 31 & 32 & 14 & 5.5 & 1.6 \\
UGC~03405 & 0.054 & 0.14 & 0.12 & 2.1 & 4.3 & 5.6 & 2.9 & 1.3 & 0.45 \\
UGC~03410 & 0.19 & 0.51 & 0.62 & 9.9 & 17 & 20 & 8.8 & 3.6 & 1.2 \\
NGC~2388 & 0.15 & 0.42 & 1.6 & 20 & 25 & 21 & 7.4 & 3.0 & 0.97 \\
NGC~2389 & 0.079 & 0.20 & 0.27 & 3.2 & 6.3 & 5.2 & 2.3 & 1.1 & 0.50 \\
MCG~+02-20-002 & 0.021 & 0.057 & $<$0.50\tablefootmark{b} & 0.85 & 1.6 & 2.1 & 0.96 & 0.43 & 0.16 \\
MCG~+02-20-003 & 0.11 & 0.22 & $<$0.73\tablefootmark{b} & 10 & 13 & 10 & 3.5 & 1.4 & 0.45 \\
NGC~3110 & 0.21 & 0.60 & 0.91 & 13 & 20 & 20 & 7.9 & 3.1 & 0.94 \\
NGC~3221 & 0.24 & 0.64 & 0.72 & 10 & 19 & 25 & 12 & 5.1 & 1.7 \\
Arp~299 & 1.1 & 2.4 & 23\tablefootmark{c} & 130 & 120 & 72 & 22 & 7.6 & 2.3 \\
MCG~--02-33-098 & 0.076 & 0.21 & 1.1 & 7.2 & 8.4 & 6.5 & 2.6 & 1.1 & 0.34 \\
IC~860 & 0.018 & 0.038 & 0.87 & 20 & 18 & 11 & 3.6 & 1.5 & 0.47 \\
NGC~5653 & 0.22 & 0.62 & 1.1 & 14 & 21 & 21 & 8.1 & 3.1 & 1.0 \\
Zw~049-057 & 0.028 & 0.071 & 0.54 & 28 & 32 & 24 & 7.9 & 3.0 & 0.98 \\
NGC~5936 & 0.17 & 0.48 & 1.0 & 11 & 16 & 16 & 6.7 & 2.6 & 0.87 \\
NGC~5990 & 0.23 & 0.48 & 1.2 & 12 & 16 & 15 & 6.3 & 2.4 & 0.84 \\
NGC~6052 & 0.075 & 0.20 & 0.66 & 7.4 & 9.9 & 8.9 & 3.3 & 1.3 & 0.43 \\
NGC~6701 & 0.17 & 0.45 & 0.98 & 12 & 19 & 18 & 7.4 & 2.9 & 0.97 \\
NGC~6921 & 0.060 & 0.077 & 0.086 & 2.2 & 3.1 & 2.3 & 0.79 & 0.30 & 0.10 $\pm$ 0.03 \\
MCG~+04-48-002 & 0.17 & 0.46 & 0.68 & 10 & 14 & 12 & 4.1 & 1.5 & 0.49 \\
NGC~7591 & 0.10 & 0.28 & 0.65 & 8.9 & 13 & 13 & 6.4 & 2.7 & 0.91 \\
NGC~7679 & 0.14 & 0.40 & 0.81 & 8.0 & 10 & 8.4 & 3.1 & 1.2 & 0.42 \\
NGC~7752 & 0.052 & 0.14 & 0.28 & 3.4 & 4.9 & 4.5 & 1.7 & 0.70 & 0.21 \\
NGC~7753 & 0.11 & 0.26 & 0.34 & 4.1 & 8.6 & 12 & 6.5 & 3.0 & 1.1 \\
NGC~7769 & 0.14 & 0.35 & 0.49 & \nodata & $<$13.58\tablefootmark{b} & \nodata & 4.9 & 1.9 & 0.61 \\
NGC~7770 & 0.060 & 0.14 & 0.38 & 2.9 & 3.8 & 3.2 & 1.2 & 0.44 $\pm$ 0.07 & 0.14 $\pm$ 0.03 \\
NGC~7771 & 0.26 & 0.63 & 1.2 & 23 & 38 & 39 & 17 & 6.6 & 2.1 \\
\hline
\end{tabular}
\tablefoot{The units of the fluxes are Jy. Uncertainties are included only when they are larger than 10\% of the flux.
\tablefoottext{a}{The two components of CGCG~468-002 are not resolved in the 500\micron\ \Herschel\slash SPIRE image, therefore we assume the integrated flux of the two components as an upper limit.}
\tablefoottext{b}{These upper limits are the \IRAS\ 25\micron\ and 100\micron\ fluxes measured by \citet{Surace2004}.}
\tablefoottext{c}{The MIPS 24\micron\ image of Arp~299 is saturated so we took the \textit{IRAS} 25\micron\ flux from \citet{SandersRBGS} corrected by the conversion factor between \textit{IRAS} 25\micron\ and MIPS 24\micron\ fluxes
given \citet{Calzetti2010}. }
}
\end{table*}

\section{SED modeling}\label{s:sed_modeling}

To investigate the properties of these galaxies from their SED, we used a method based on the models and fitting procedures presented by \citet{daCunha2008}. They compute the model SEDs from the UV to the far-IR wavelengths taking the energy balance between the absorbed UV-optical radiation and that emitted in the IR by dust into account. First they calculate the emission from stars that is attenuated according to an extinction law, and this absorbed energy is re-emitted in the IR distributed into several components (PAH bands, hot grains, and warm and cold dust). These models are used together with the \textsc{magphys} code \citep{daCunha2008} to calculate the likelihood distributions of the physical parameters adopting a Bayesian approach.

{This method produces good results when applied to nearby star-forming galaxies \citep{daCunha2008}}. However, we tried to use it directly with the SED of our LIRGs and the results were not always satisfactory as already noted by \citet{daCunha2010}.
{This is because models with physical conditions typical of LIRGs (higher extinction and dust temperatures than in normal star-forming galaxies) are not numerous in their set, so the obtained likelihood distributions are not reliable.}
Therefore, to analyze our data better and to include the \Halfa\ emission in the fit, we decided to generate a new set of models for the stellar and IR emissions and to modify the original \textsc{magphys} code. In the following we describe our models and modifications.

\begin{figure*}[ht!]
\centering
\includegraphics[width=.9\textwidth]{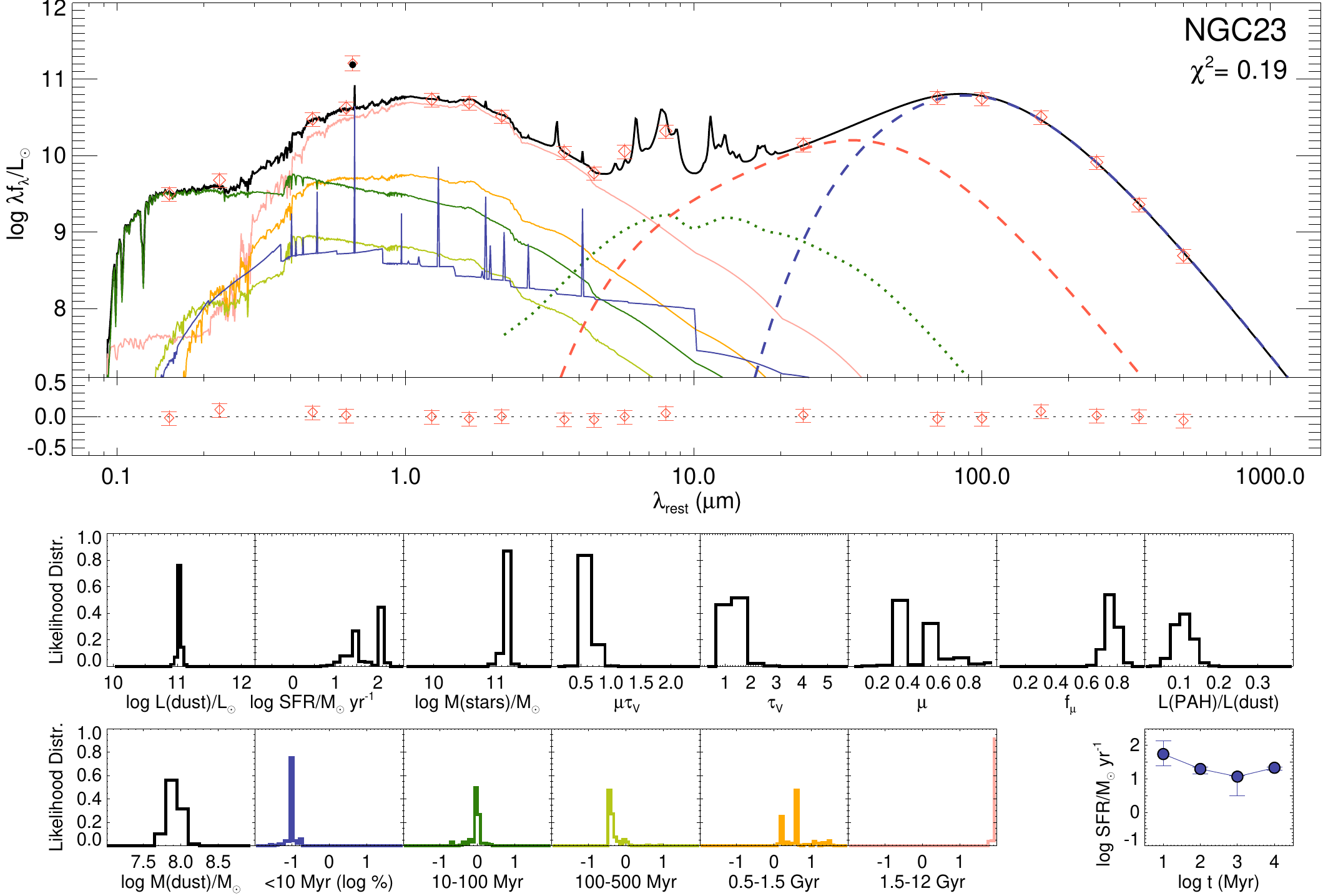}
\caption{{\it Top panel}: Best-fitting model for the SED of NGC~23 (constant SFH, see Section \ref{s:sfrh}) in a solid black line, together with the observed data (red diamonds). The \Halfa\ luminosity and model prediction in \Lsun\ units are plotted at $\lambda =$ 6563\AA\ (red diamond and solid black circle, respectively) multiplied by a factor of 10$^3$. The solid color lines indicate the contribution of the different stellar populations with ages $<$10\,Myr (blue), 10--100\,Myr (dark green), 100--500\,Myr (light green), 0.5--1.5\,Gyr (orange), and $>1.5$\,Gyr (pink). The blue and red dashed lines are the dust emission for dust temperatures lower and higher than 50\,K, respectively. The dotted green line is the AGN torus model derived by \citet{AAH2012a}. The residuals are shown in the lower part of the panel. {\it Bottom panels:} Likelihood distribution for several of the physical parameters (IR dust luminosity, SFR, stellar mass, ISM extinction $\mu\tau_{\rm v}$, young stars extinction $\tau_{\rm v}$, ratio between the ISM and young stars extinctions $\mu$, fraction of IR luminosity produced by cold dust with $T<50$\,K removing the AGN contribution $f_\mu$, fraction of IR luminosity due to PAH emission, and dust mass). The next five panels show the logarithm of the percentage of the stellar mass for different stellar age ranges using the same color coding as in the top panel. The last panel represents the SFH.\label{fig:sedfitting}}
\end{figure*}

\begin{figure*}[ht!]
\centering
\includegraphics[width=0.9\textwidth]{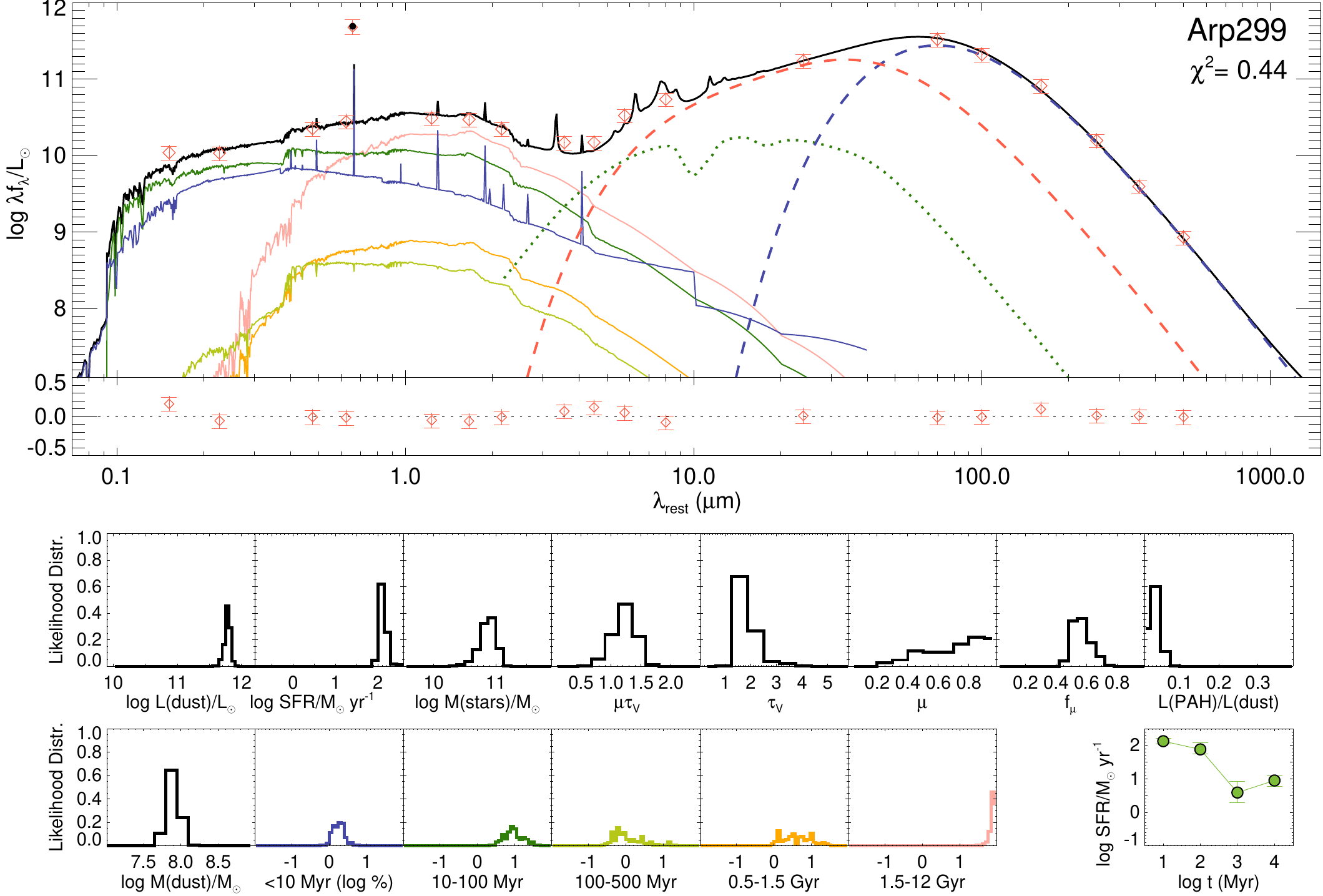}
\caption{Same as Figure \ref{fig:sedfitting} but for Arp~299 (recent SF burst, see Section \ref{s:sfrh}).\label{fig:sedfitting_arp299}}
\end{figure*}

\subsection{Stellar emission}\label{ss:stellar_emission}

We modeled the stellar emission by combining the emission of single stellar population bursts of different ages weighted by the stellar mass of each age. The input stellar spectra are from the \textsc{POPSTAR}\footnote{http://www.fractal-es.com/PopStar} library \citep{Molla2009,MartinManjon2010}. 
Specifically, we used only those models created with the \citet{Kroupa2001} initial mass function (IMF) and solar metallicity.

In this library there is a large number of models (106) for different stellar ages but many of them have very similar photometric colors.
Because of this, we selected a set of four representative spectra of several age ranges (10--100\,Myr, 100--500\,Myr, 0.5--1.5\,Gyr, and $>$1.5\,Gyr; stars younger than 10\,Myr are considered below through a recent star-formation burst, see below). We chose these age ranges because the variation in the mass-to-luminosity ratio is lower than a factor of two, and they have similar photometric colors, so age variations within these ranges would be almost indistinguishable using photometric information alone. 
We compared the \textsc{POPSTAR} models with the \citet{Maraston2005} models, which include a detailed treatment of the thermally pulsing asymptotic giant branch (TP-AGB). The main differences appear in the near-IR range for populations with 0.5--1.5\,Gyr, although the differences in luminosity are small, a factor of 2--3.

To combine the single stellar population models, we considered that a random fraction of the stellar mass was formed at a constant rate during the five age intervals mentioned above. We also added to the SFH a recent burst of star-formation that began between 1 and 300\,Myr ago ($t_{\rm SB}$) and continues until today with a constant intensity $I_{\rm SB}$ between 0.03 and 10\,000 times the average previous SFR. 

Then, we extinguished the combined stellar spectrum following the prescription given by \citet{Charlot2000} and used by \citet{daCunha2008} in the original \textsc{magphys} code. That is, we assumed that stars younger than 10\,Myr are still embedded in their birth clouds (BC) and have higher extinctions than older stars, which are only affected by the interstellar medium (ISM) extinction. 
Thus, for a given wavelength, the total effective absorption, $\tau_{\lambda}$, is $\tau_{\lambda}^{\rm BC} + \tau_{\lambda}^{\rm ISM}$ for stars younger than 10\,Myr and $\tau_{\lambda}^{ISM}$ for older stars. 
For the wavelength dependence of $\tau_{\lambda}$, we used {the power-law dependence assumed by \citet{Charlot2000}}.
Similar to \citet{daCunha2008}, the input parameters regarding the extinction are $\tau_{\rm v}$ and $\mu$, where $\tau_{\rm v}$\footnote{$A_{\rm v}\slash \tau_{\rm v} = 2.5\log e \simeq 1.086$. Both $A_{\rm v}$ and $\tau_{\rm v}$ are used interchangeably along this paper taking into account this factor.} is the total extinction affecting young stars 
, and $\mu = \tau_{\rm v}^{\rm ISM}\slash \tau_{\rm v}$. 
We also computed the amount of energy that is absorbed by dust ($L_{\rm dust}$) and the fraction of this absorbed energy that is produced by stars older than 10\,Myr ($f_{\mu}$). These two parameters are used later in Section \ref{ss:bayes} to combine the stellar models with the IR emission models.

The luminosities of the hydrogen recombination lines were calculated from the number of ionizing photons in the unextinguished stellar spectrum using the case B \citet{Storey1995} recombination coefficients. 
These emission lines are affected by the BC and\slash or the ISM extinctions as well.

In total we generated 50\,000 different models, which is enough to produce smooth likelihood distributions in the fits.

\subsection{IR emission}\label{ss:ir_model_emission}

For the IR emission we used a two-component model: dust thermal emission and PAH emission. For the former component we assumed that the temperature distribution of the dust mass follows a power-law, $dM_{\rm d}\slash dT \propto T^{-\gamma}$ with a low temperature cut-off, $T_{\rm min}$ (see, e.g., \citealt{Dale2001, Kovacs2010}). 
We assumed that the dust emission for a given temperature is a graybody with fixed $\beta=2$ and an absorption coefficient $\kappa=0.517$\,m$^2$\,kg$^{-1}$ at 240\micron\ \citep{Li2001}. We estimated the fraction of the IR luminosity produced by dust with $T< 50$\,K ($f_{\mu}^{\rm IR}$) to separate the fraction of the IR luminosity that is produced in photodissociation regions (PDR, $U\sim200$\footnote{{Where $U=1$ is the interstellar radiation field in the solar neighborhood.}}) and that produced in more diffuse regions \citep{Draine07}.

The PAH emission consists of several emission bands in the mid-IR range between 3 and 20\micron. To obtain a ``pure'' PAH template we used the \citet{Smith07} average {5--35\micron\ } mid-IR spectra of local star-forming galaxies after subtracting the underlying hot dust continuum using the \textsc{pahfit} code \citep{Smith07}.
We added the PAH feature at 3.3\micron\ 
using a Drude profile with an intensity equal to one third of the 6.2\micron\ PAH feature \citep{Draine07}. 

The dust emission model and the PAH template were combined into the final model assuming that the PAH luminosity can represent between 1 and 40\% of the total IR luminosity. We produced 20\,000 IR emission models with random values for $\gamma$, $T_{\rm min}$, and PAH luminosity fractions ($q_{\rm PAH}$).

\subsection{AGN contribution}

Most of the galaxies in our sample are part of the larger sample of LIRGs studied by \citet{AAH2012a}. In that work the AGN contribution to the mid-IR emission was estimated decomposing their 5--38\micron\ \Spitzer\ low-resolution spectra using star-formation templates \citep{Brandl06, Rieke2009} and clumpy torus models \citep{Nenkova2008}.
Although, in general, the AGN energy output { in the IR} is small compared to that of star-formation in these galaxies (see Table \ref{tab:sample}), at certain wavelengths the AGN contribution can be noticeable. For this reason we used the torus model fitted by \citet{AAH2012a}, when an AGN was detected, to subtract the AGN IR emission from our integrated measurements.

Only for CGCG~468-002 NED01 and NGC~7770 does the AGN dominate the mid-IR emission.

\subsection{Bayesian parameter inference}\label{ss:bayes}

To determine the likelihood distributions of the physical parameters we modified the \textsc{magphys} code so we could make use of the models we constructed. {First, stellar and IR models are combined to obtain the complete SED requiring that $f_{\mu}^{\rm IR}=f_{\mu}\pm 30 \%$. In total, we find about 470 million combinations that fulfill this requirement. Then, the SED models are scaled to match the observed photometric fluxes and \Halfa\ emission. A probability ($e^{-\chi^2\slash 2}$) is assigned to each model by considering upper limits when present in the SED (see Appendix \ref{apx:likelihood}). Finally, the likelihood distributions of the parameters are derived from these probabilities (see \citealt{daCunha2008} for details).}

\begin{figure*}[ht!]
\centering
\includegraphics[width=0.87\textwidth]{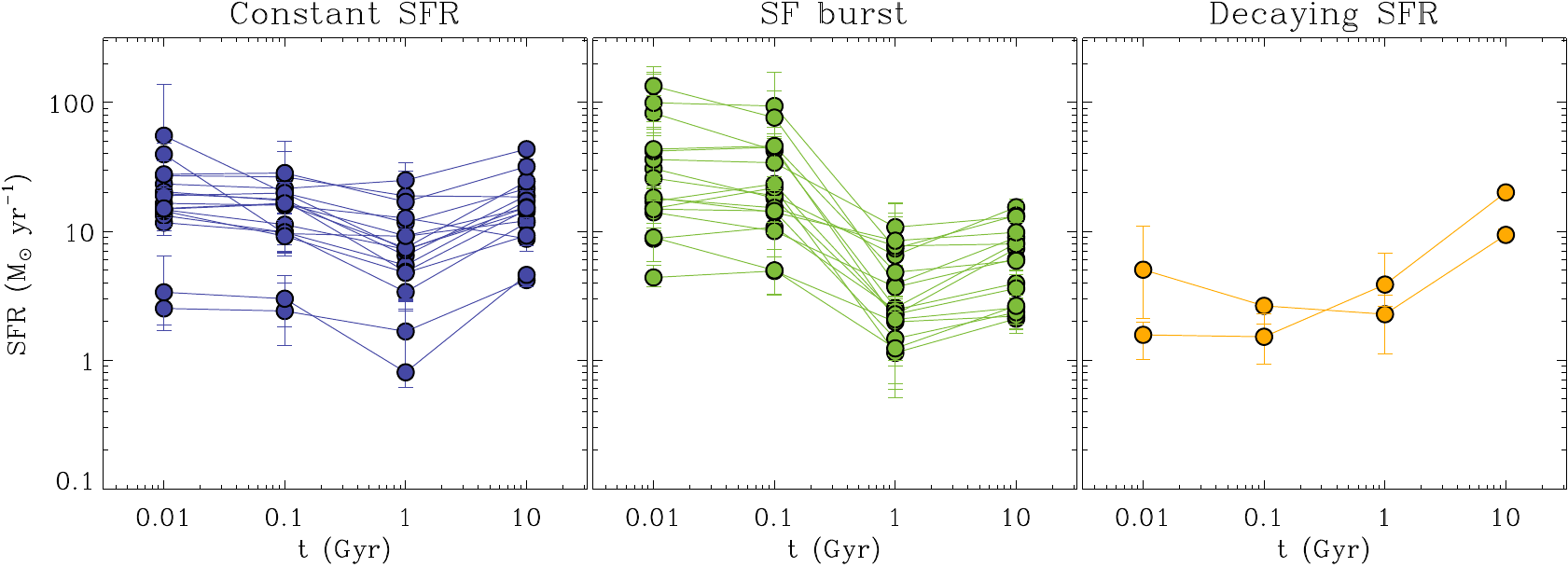}
\caption{Star-formation histories derived from the SED fitting. In the left panel we show the SFH of those galaxies with a constant SFH in blue. In the central panel, we plot those SFH with a recent burst of SF (green). In the right panel we show those SFH with a decaying SFR (orange). The numbers represent the galaxy ID (see Table \ref{tab:sfh}).\label{fig:sfr_history_1}}
\end{figure*}

\begin{figure*}[ht!]
\centering
\includegraphics[width=0.87\textwidth]{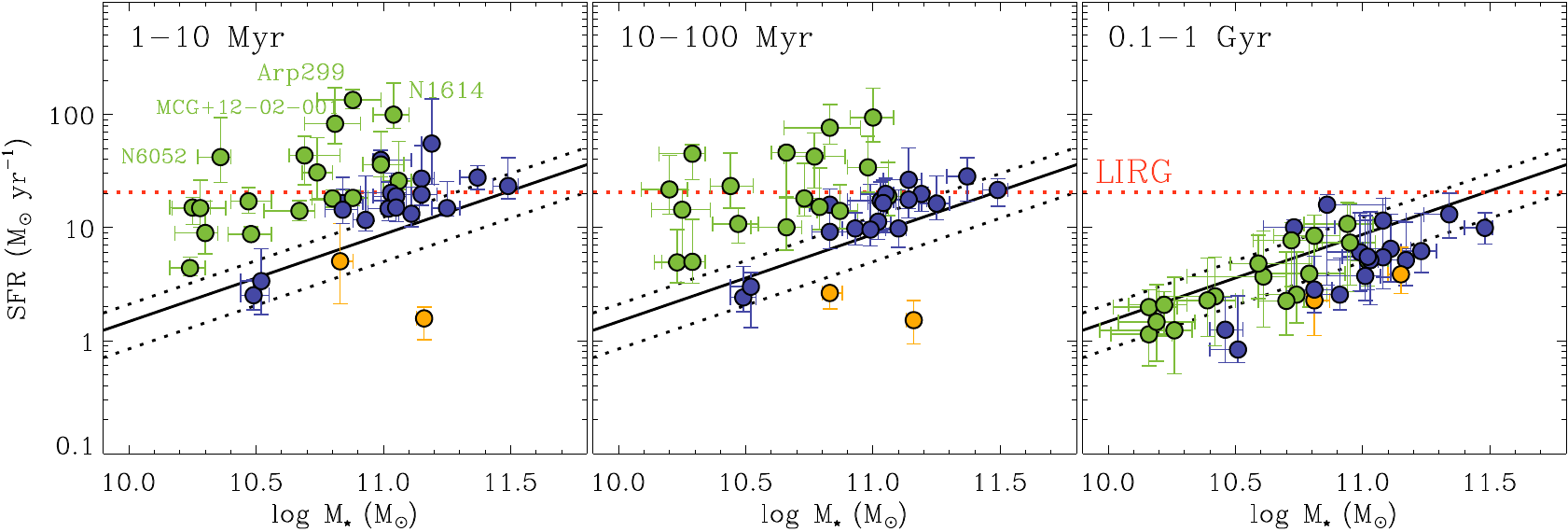}
\caption{SFR vs. stellar masses at three different age intervals calculated from the SFH. The color of the symbols is as in Figure \ref{fig:sfr_history_1}. The solid and dashed black lines are the M-S relation and uncertainty, respectively, derived by \citet{Elbaz2007} for SDSS galaxies at z$\sim$0. The horizontal dashed red line marks the approximate threshold SFR needed to reach a $\log L_{\rm IR}/L_{\rm \odot}>11$ (excluding the AGN luminosity) based on our calibration (Section \ref{ss:halfa_lir}). Galaxies below the LIRG threshold are close companions of the sources identified as LIRGs using low angular resolution (2\arcmin) \IRAS\ data (see Section \ref{s:sample}).
\label{fig:sfr_history_2}}
\end{figure*}

\subsection{Fitting results}

To fit the SED of our sample of LIRGs (Tables \ref{tab:photometry} and \ref{tab:photometry_ir}) we used the procedure described in the previous section. For most of these integrated measurements, the statistical error is low ($<5$\%) so most of the uncertainty comes from systematic errors, {such as the absolute flux calibration, the uncertain aperture correction for semi-extended sources, and possible aperture mismatches}. 
Therefore, we assumed a conservative 20\% systematic error for all the photometric points in our SED added in quadrature to the, typically very low, statistical error.

In Figures \ref{fig:sedfitting} and \ref{fig:sedfitting_arp299} we show the results of SED fitting for two of our galaxies: one with a constant SFH and the other with a strong burst of recent SF (see Section \ref{s:sfrh}). For the rest of the sample, they are shown in Appendix \ref{apx:bestfit}. The fitted parameters are listed in Table \ref{tab:results}.

The best-fitting model for each galaxy is able to closely reproduce the observed SED for most of the galaxies except for IC~860 and Zw~049-057. Their optical $g$ and $r$ emissions are clearly underestimated by the best-fitting models (see Appendix \ref{apx:bestfit}). This can be caused by the presence of a deeply obscured energy source (AGN or SF), which is only detected through its far-IR emission or an extinction behavior that is more complex than the one assumed in the models. In either case, the derived parameters may be uncertain, so we decided to exclude them from the discussion of Sections \ref{s:sfrh} and \ref{s:sfr_tracers}.

\section{Results and discussion}\label{s:discussion}

\subsection{Star-formation histories}\label{s:sfrh}

From the Bayesian fitting of the SEDs we derived likelihood distributions for the SFH of these galaxies based on the different SFH of the models.
As explained in Section \ref{ss:stellar_emission}, with the photometric information alone, we are able to distinguish stellar populations in limited age intervals. Because of this, we calculated the SFH in four intervals with the same duration in log scale (i.e., 0--10\,Myr, 10--100\,Myr, 0.1--1\,Gyr, 1--10\,Gyr). These intervals are almost exactly the same as we used to construct the SED models.

To classify these SFH we first tested whether they are compatible with a constant SFR (see Figure \ref{fig:sfr_history_1}). { For 17 out of 36 galaxies} we found that their SFH deviate less than 3$\sigma$ from a constant SFR during their histories (averaged over the time intervals specified above). The mean uncertainty of the SFR in each interval is $\sim$0.5\,dex, so we are not able to detect variations lower than a factor of $\sim$3 in the SFR. { These 17 galaxies} are mostly spirals, some of them with clearly disturbed morphologies (e.g., NGC~1961 or NGC~5653).
{ For two cases we found a decaying SFR, and the remaining 17 galaxies (50\% of the sample) show a SFH with a} SF burst in the 0--100\,Myr intervals. We did not find significant differences between the SFR in the 0--10\,Myr and 10--100\,Myr ranges for any of the galaxies, including those with a burst of SF (starbursts in the following) and those with nearly constant SFR.

The SFH of the identified starbursts indicates that the current burst began on average between { 30--300\,Myr} ago. The upper limit of this range is calculated by estimating how long it would be possible to sustain the burst intensity during the 0.1--1\,Gyr interval and still detect a relative SF burst in the 10--100\,Myr interval. Similarly, the lower limit is calculated by doing the same for the 1--10\,Myr and 10--100\,Myr intervals.
This burst duration is similar to the one derived by \citet{Marcillac2006}, 40--260\,Myr from the modeling of Balmer absorption lines and the 4000\AA\ break in $z=0.7$ LIRGs.
The intensity of the current burst is between { 2 and 20 times with a median of 7.5 times} the previous averaged SFR (Table \ref{tab:sfh}). 

Galaxies with the higher burst intensities { ($>10$)} are those whose morphologies are highly disturbed indicating recent or ongoing interactions (MCG~+12-02-001, NGC~1614, CGCG~468-002 NED02, Arp~299, and NGC~6052). However, in our sample there is one object (MCG~--02-33-098) that is also an interacting system with two nuclei separated by $\sim$4\,kpc, but its burst intensity is only { (2.5$^{+5.4}_{-0.8}$)}. Combining the age of the burst with its intensity, we calculate the stellar mass produced by the current burst of SF with respect to the current stellar mass (Table \ref{tab:sfh}), which can be $>$10--30\% in these mergers. For the rest of the galaxies the mass formed tends to be $<$10\%, although there is a continuous distribution of formed mass fractions between the extreme bursts of mergers and the weaker SF bursts of other galaxies in our sample.

In Figure \ref{fig:sfr_history_2} we plot the SFR averaged over the 0--10\,Myr, 10--100\,Myr, and 0.1-1\,Gyr intervals vs. the stellar mass today, 100\,Myr ago, and 1\,Gyr ago. From the lefthand panel of this figure, it is clear that those galaxies classified as starbursts lie above the SFR-$M_{\star}$ main-sequence (M-S), as expected, while those with a relatively constant SFH are consistent with the M-S relation within 2$\sigma$.

Figure \ref{fig:sfr_history_2} shows the luminosity threshold for a galaxy to be classified as LIRG ($\log  L_{\rm IR}\slash \Lsun>11$). According to this luminosity criterion, { there are 8 starbursts and 5 M-S galaxies with LIRG luminosities} due to star-formation in our sample. The latter set of galaxies { (NGC~23, NGC~877, NGC~1961, NGC~2388, and NGC~7771)} are spirals classified as LIRGs, although they do not have a particularly high specific SFR { (sSFR = SFR\slash $M_{\star}<0.4$\,Gyr$^{-1}$). They lie within 1$\sigma$ in the M-S, and since they are relatively massice ($\log M_\star\slash M_\odot>11.0$), their expected SFR imply IR luminosities above or close to the LIRG luminosity threshold.}

The starburst LIRGs have enhanced SFR with respect to the M-S, with the most extreme cases { (Arp~299, NGC~6052, MCG~+12-02-001, and NGC~1614)} being mergers with high specific sSFR $>$ 1.0\,Gyr$^{-1}$ (see Table \ref{tab:results}). The rest of starbursts include { spiral} galaxies with different degrees of disturbed morphologies (e.g., NGC~6701 and NGC~5936).
The AGN contribution to the IR luminosities is small in our sample (Table \ref{tab:sample} { and \citealt{AAH2012a}}), therefore the presence of an AGN { does} not affect their classification as LIRGs. 

Both the lefthand and middle panels of Figure \ref{fig:sfr_history_2} show {similar distributions of the galaxies in the $M_{\star}$-SFR plane}. 
As indicated before, this is probably because the star-formation bursts have been active during the past $\sim$100\,Myr.
However, in the righthand panel (0.1--1\,Gyr range; $z\sim0.075$), the situation is completely different. Almost all the galaxies are in agreement with the M-S relation, and only one or two objects, with $\log M_{\star}/M_{\odot}>11.1$, would have been classified as LIRGs if they had been observed $\sim$1\,Gyr ago.

\subsubsection{Comparison with previous results for local U\slash LIRGs}

\citet{RodriguezZaurin2009, RodriguezZaurin2010} used optical spectroscopy to study the SFH of local ULIRGs and find that young stellar populations (age $<$100\,Myr) usually dominate the stellar mass of these systems. Most of their ULIRGs are mergers in different evolutionary stages, so we compare their results with { the five mergers in our sample with burst intensities $>10$. In our merger LIRGs, the stellar mass formed during the ongoing burst is up to $\sim$30\%, slightly lower than in ULIRGs, although they are also compatible with burst stellar mass fractions as low as 5--10\% (see Table \ref{tab:sfh}).}

\citet{RodriguezZaurin2009} argue that an old stellar population ($>2$\,Gyr) is not present in most of the optical spectra of ULIRGs, which is consistent with the best-fitting models for our merger LIRGs (Figure \ref{fig:sedfitting_arp299} and Appendix \ref{apx:bestfit}) where the optical light is dominated by young stars (10--100\,Myr).

Similar results regarding the stellar masses of ULIRGs were obtained by \citet{daCunha2010} using \textsc{magphys}. In that study they also found that the median dust mass in their ULIRGs is 10$^{8.6}$\Msun, which is almost a factor of 10 higher than the dust masses derived for these LIRGs. However, the dust mass determination in ULIRGs is uncertain because it is not easy to constrain the cold dust temperature, which contributes significantly to the dust mass but not to the IR luminosity \citep{daCunha2010}. Actually, they obtain a median $f_\mu$ of 0.1, while $f_\mu$ ranges between 0.6 and 0.9 in our sample, indicating that, in contrast to ULIRGs, dust with $T<50$\,K dominates the IR SEDs of LIRGs.

The resolved stellar populations of local LIRGs were examined by \citet{AAH2010} for nine objects in our sample using optical integral field spectroscopy. These nine objects are mostly spirals or weakly interacting galaxies. They found a higher contribution of old stellar populations to the total stellar mass in these LIRGs than in ULIRGs. This agrees with our result of lower mass fractions formed during the current burst of SF in spiral-like LIRGs than in merging LIRGs.

\setcounter{table}{8}
\begin{table}
\caption{Star-formation burst properties}
\label{tab:sfh}
\setlength{\tabcolsep}{5pt}
\centering
\begin{tabular}{llcccccccccc}
\hline \hline
Name & Intensity\tablefootmark{a} & Age\tablefootmark{b} & Mass\tablefootmark{c} \\
&   & (Myr) & (\%) \\
\hline
MCG~+12-02-001 & 13$^{+38}_{-7}$ & 40--200 & 3--16 \\
NGC~992 & 6$^{+15}_{-2}$ & 20--440 & 1--15 \\
UGC~02982 & 2.2$^{+4.9}_{-0.1}$ & 40--210 & 1--3 \\
NGC~1614 & 10$^{+13}_{-0.7}$ & 50--370 & 4--27 \\
CGCG~468-002 NED02 & 12$^{+11}_{-5}$ & 40--290 & 4--25 \\
NGC~2389 & 3.1$^{+4.0}_{-1.6}$ & 50--310 & 2--8 \\
MCG~+02-20-002 & 2.23$^{+0.47}_{-0.83}$ & 30--190 & 1--4 \\
MCG~+02-20-003 & 7.0$^{+8.5}_{-2.2}$ & 40--170 & 3--10 \\
NGC~3110 & 3.0$^{+1.1}_{-1.5}$ & 30--200 & 1--6 \\
Arp~299 & 17$^{+44}_{-8}$ & 70--260 & 9--29 \\
MCG~--02-33-098 & 2.5$^{+5.4}_{-0.8}$ & 40--140 & 1--3 \\
NGC~5936 & 2.1$^{+0.7}_{-1.0}$ & 30--110 & 1--3 \\
NGC~6052 & 18.8$^{+2.0}_{-5.0}$ & 60--420 & 12--60 \\
NGC~6701 & 1.8$^{+1.1}_{-0.8}$ & 30--40 & 1--1 \\
NGC~7679 & 8.4$^{+0.4}_{-3.1}$ & 60--290 & 6--22 \\
NGC~7752 & 7.9$^{+6.8}_{-1.3}$ & 30--210 & 2--13 \\
NGC~7770 & 4$^{+10}_{-1}$ & 20--260 & 1--7 \\
\hline
\end{tabular}
\tablefoot{{ Galaxies with constant or decaying SFR are not included in this table.}
\tablefoottext{\star}{Mergers with burst intensities\,$>20$.}
\tablefoottext{a}{Intensity of the current SF burst with respect to the previous average SFR.}
\tablefoottext{b}{Age of the current burst.}
\tablefoottext{c}{Fraction of the stellar mass formed in the current SF burst.}
}
\end{table}

\begin{figure}[!ht]
\centering
\includegraphics[width=0.47\textwidth]{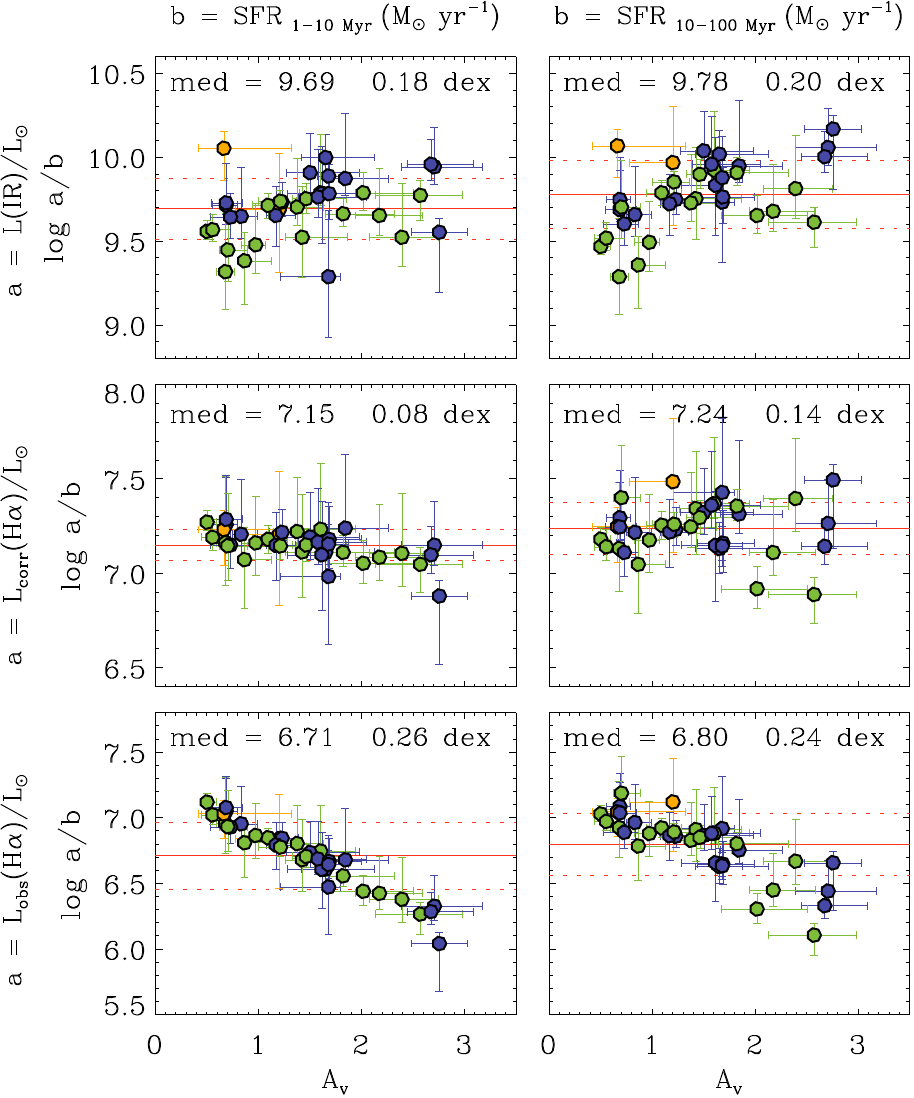}
\caption{IR and H$\alpha$ luminosity\slash SFR ratio as a function of the fitted total extinction, $A_{\rm v}$ 
considering the 1--10 and 10--100\,Myr to average the SFR. The median ratio (solid red line) and the scatter (dashed red line) are indicated in each plot. The contribution of the AGN torus has been subtracted when present. The color of the symbols is as in Figure \ref{fig:sfr_history_1}.
\label{fig:sfr_ratios}}
\end{figure}

\subsection{H$\alpha$ vs. IR luminosity. Tracing the last 100\,Myr SFH}\label{ss:halfa_lir}\label{s:sfr_tracers}

H$\alpha$ and IR luminosities are two fundamental SFR tracers. The former traces the most recent SFR since its luminosity is proportional to the number of ionizing photons, although it is necessary to apply an extinction correction that is not always accurately known. The latter, however, traces the obscured SFR, and it is virtually unaffected by foreground extinction, although it depends on the extinction level in the star-forming regions (e.g., objects with no extinction do not emit in the IR; \citealt{Kennicutt2012}).

Figure \ref{fig:sfr_ratios} shows that our calibrations between the IR luminosity and the SFR at 1--10\,Myr and 10--100\,Myr, both $\sim$(9.64 $\pm$ 0.18)\,\Lsun\slash (\Msun\,yr$^{-1}$), are in very good agreement with the \citet{Kennicutt1998} calibration of 9.57\,\Lsun\slash (\Msun\,yr$^{-1}$) (after correcting for the different IMF assumed). We also show in this figure the relation between the extinction corrected H$\alpha$ luminosities derived from the SED analysis with the SFR. The $\log L_{\rm corr}({\rm H}\alpha)$\slash SFR ratio is (7.14 $\pm$ 0.09)\,\Lsun\slash (\Msun\,yr$^{-1}$), so similar, although slightly lower than the calibration given by \citet{Kennicutt1998} for the H$\alpha$ luminosity, 7.33\,\Lsun\slash (\Msun\,yr$^{-1}$).

In the same way, we obtained the SFR calibration for the uncorrected H$\alpha$ luminosity. The correlation between the  $L_{\rm obs}({\rm H}\alpha)$\slash SFR ratio and the extinction ($A_{\rm v}$) is clear. However, since the average integrated extinction of these LIRGs is not extremely high, $\tau_{\rm v}=$0.5--3 or $A_{\rm v}=$0.6--3.3\,mag (see also \citep{AAH06s}), the observed H$\alpha$ luminosity could be used to estimate the integrated SFR in similar galaxies within a factor of $\sim$3.

\begin{figure}[h!]
\centering
\includegraphics[width=0.4\textwidth]{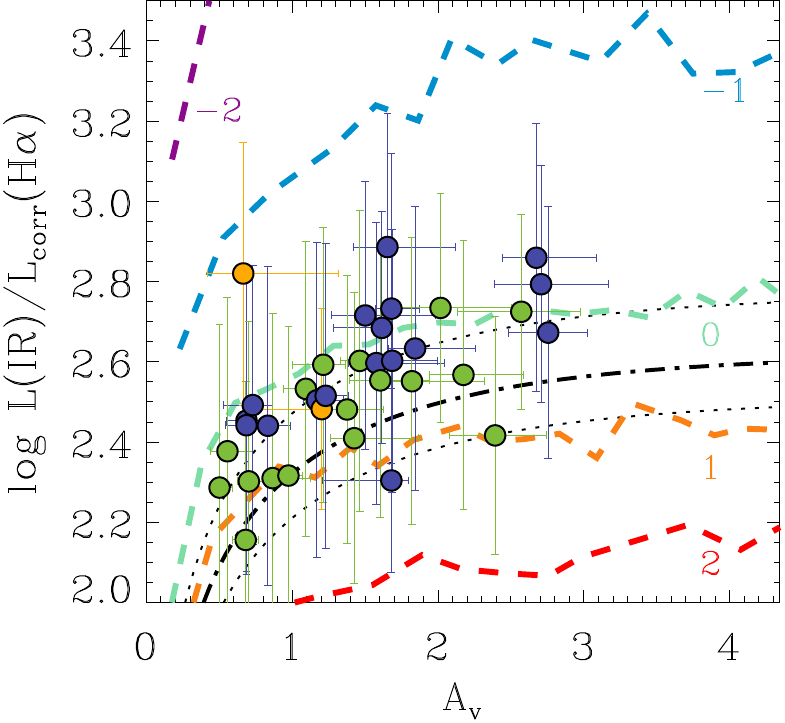}
\caption{
\LIR\slash corrected H$\alpha$  luminosity ratio as a function of the total extinction, $A_{\rm v}$. The dashed lines represent the expected ratio derived from our models for various $\log$\,SFR$_{\rm 1-10\,Myr}$\slash SFR$_{\rm 10-100\,Myr}$ values (red$=$2, orange$=$1, turquoise$=$0, blue$=$--1, and purple$=$--2). The dot-dashed black line is the ratio calculated from empirical relation $L_{\rm corr}({\rm H}\alpha)$ = $L_{\rm obs}({\rm H}\alpha)$ + 0.0024 \LIR\ derived by \citet{Kennicutt2009} for a sample of nearby galaxies and the dotted black lines are the uncertainty of that calibration. The contribution of the AGN torus to the IR luminosity has been subtracted when present. The color of the symbols is as in Figure \ref{fig:sfr_history_1}.
\label{fig:sfr_lir_lha_ratio}}
\end{figure}

We find that both the \LIR\slash SFR and the $L_{\rm corr}({\rm H}\alpha)$\slash SFR ratios are approximately constant and do not clearly depend on the age interval used to average the SFR.  However, part of the scatter of these relations can be explained by the differences in the SFH and the extinction level.
This is clearer in Figure \ref{fig:sfr_lir_lha_ratio} where we plot the \LIR\slash $L_{\rm corr}({\rm H}\alpha)$ ratio as a function of the extinction comparing the observed galaxies and our models predictions.
We expect this ratio to depend mainly on the relative intensities of the current SF and the average SFR during the last 100\,Myr because H$\alpha$ emission is only produced by young stars, whereas older stars contribute to the IR luminosity too. But this ratio also depends on the extinction because the fraction of absorbed photons later re-emitted in the IR increases with the extinction. The latter effect is apparent in this figure since the \LIR\slash $L_{\rm corr}({\rm H}\alpha)$ ratio increases, with $A_{\rm v}$ tending asymptotically to the ratio expected when 100\% of the stellar light is absorbed and re-emitted by dust.

In our sample, the age effect is more subtle because, according to the derived SFH, the SFR during the last 100\,Myr has been approximately constant (see Section \ref{s:sfrh}). In any case, to test this dependence we computed the expected \LIR\slash $L_{\rm corr}({\rm H}\alpha)$ ratio as a function of the extinction for several $\log$\,SFR$_{\rm 1-10\,Myr}$\slash SFR$_{\rm 10-100\,Myr}$ ratios. The model ratios are plotted in Figure \ref{fig:sfr_lir_lha_ratio} and we can see that our sample of LIRGs is, as expected, in agreement with the ratios for nearly constant SFH in the last 100\,Myr. However, it should be noted that this ratio shows a strong dependence on the SFH.

For comparison, the empirical relation between the corrected H$\alpha$ luminosity and a linear combination of the observed H$\alpha$ luminosity and the IR luminosity derived by \citet{Kennicutt2009} for a sample of nearby galaxies is plotted in Figure \ref{fig:sfr_lir_lha_ratio}. The agreement between the data, our models with $\log$\,SFR$_{\rm 1-10\,Myr}$\slash SFR$_{\rm 10-100\,Myr}=0-1$, and the \citet{Kennicutt2009} relation is good within the uncertainties. This comparison suggests that the empirical relation is only valid for galaxies with constant or nearly constant SFH during the last 100\,Myr. Likewise, this diagram can be readily used to distinguish between galaxies with approximately constant SFH, such as those in the M-S, and galaxies with a recent (less than 100\,Myr old) SF burst that would lie below the \citet{Kennicutt2009} relation. 

\section{Summary and conclusions}\label{s:conclusions}

We modeled the integrated SED of a sample of 38 local IR bright sources with IR luminosities ranging from $\log \LIR\slash \Lsun = 10.2$ to 11.8 and a median of $\log \LIR\slash \Lsun = 11.0$ { belonging to 29 systems with $\log \LIR\slash \Lsun = 11.0-11.8$}. The SEDs include new optical $g$, $r$, and H$\alpha$ narrow-band imaging obtained with the NOT telescope combined with archival observations from UV to far-IR. To fit the SED, we modified the Bayesian code \textsc{magphys} \citep{daCunha2008} and created a set of stellar population synthesis models and dust models optimized for these objects. This SED fitting approach takes into account the balance between the energy absorbed in the UV/optical spectral ranges and that re-emitted in the IR. Except for three LIRGs that might host a deeply obscured energy source (AGN or SF), the SED models are able to reproduce the observed data well.
The main results are summarized in the following:

\begin{enumerate}

\item We classified the galaxies of our sample in three groups according to their SFH: { objects with a recent burst of SF (47\%), objects with a constant SFH (47\%), and objects with decaying SFR (6\%)}. In all cases the averaged SFR during the last 100\,Myr seems to be relatively constant within a factor of three.

\item The intensity of the recent SF burst with respect to the previous averaged SFR varies between a factor { of 2 and 20} and it began { $\sim$30--300\,Myr} ago. The most extreme bursts { (intensities $>$ 10) are associated with mergers and high sSFR ($>$ 0.8\,Gyr$^{-1}$)}. Similar to local ULIRGs, these objects would be compatible with a large part { (up to 30\%)} of their current stellar mass being formed during the ongoing SF burst event. The rest of the stabursts include galaxies with varied morphologies: mergers, disturbed spirals, interacting pairs, and isolated disks.

\item Galaxies with constant SFH in our sample tend to be more massive than those with a burst of SF (median $\log M_{\rm \star}\slash \Msun = 11.0$ vs. 10.6). Most of them lie just slightly above ($<$2$\sigma$) the M-S of galaxies, but they reach the LIRG luminosity threshold due to their high stellar masses.

\item The calculated SFHs of the galaxies in our sample reveal that all of them had SFR and stellar masses in very good agreement with local M-S $\sim$1\,Gyr ago ($z\sim0.075$). Their stellar masses were between 10$^{9.7}$ and 10$^{11.4}$\,\Msun\ and their SFR between 0.5 and 20\,\Msun\,yr$^{-1}$. It is likely that only one or two of them would have been classified as LIRGs if they would had observed 1\,Gyr ago.

\item We find that the $L({\rm IR})$\slash $L_{\rm corr}$(H$\alpha$) vs. integrated optical extinction ($A_{\rm v}$) relation derived for our galaxies is in good agreement with the empirical correlation of \citet{Kennicutt2009}. Our models show that this relation holds only if the SFR has been approximately constant during the last 100\,Myr. Deviations from this relation can be used to identify galaxies with rapidly changing (increasing or decreasing) SFR during the last 100\,Myr period.

\end{enumerate}

Similar studies covering wider IR luminosity ranges and including integrated spectroscopic information will be crucial for obtaining a more detailed evolutionary history of local U\slash LIRGs, as well as for studying the differences and similarities with their high-$z$ counterparts.

\begin{acknowledgements}
{ We thank the anonymous referee for comments and suggestions.}
{ We also thank the Nordic Optical Telescope and Isaac Newton Group staff for their support during the observations.}
We acknowledge support from the Spanish Plan Nacional de Astronom\'ia y Astrof\'isica through grants AYA2010-21161-C02-1, AYA2010-21887-C04-03, AYA2012-32295, AYA2012-31447, AYA2012-31277, and AYA2012-39408-C02-01.
Based on observations made with the Nordic Optical Telescope, operated by the Nordic Optical Telescope Scientific Association at the Observatorio del Roque de los Muchachos, La Palma, Spain, of the Instituto de Astrofisica de Canarias.
The data presented here were obtained in part with ALFOSC, which is provided by the Instituto de Astrofisica de Andalucia (IAA) under a joint agreement with the University of Copenhagen and NOTSA.
This research has made use of the NASA/IPAC Extragalactic Database (NED) which is operated by the Jet Propulsion Laboratory, California Institute of Technology, under contract with the National Aeronautics and Space Administration.
\end{acknowledgements}

\setcounter{table}{7}
\begin{landscape}
\begin{table}
\caption{Results}
\label{tab:results}
\setlength{\tabcolsep}{4pt}
\centering
\begin{tabular}{lccccccccccccccccc}
\hline \hline
Name & $\chi^2_{\rm red}$ & $\log L_{dust}\slash L_\odot$ & SFR$_{\rm 1-10\,Myr}$  & $\log M_{\star}\slash M_\odot$ & log\,sSFR\slash Gyr$^{-1}$ & $\tau_{\textnormal{v}}$(total) & $\tau_{\rm v}$(ISM) &  log\,$M_{\rm dust}\slash M_\odot$ & $f_\mu$ & $T_{\rm dust}$ & $L_{\rm PAH}\slash L_{\rm dust}$  \\
 &  & & ($M_\odot$\,yr$^{-1}$) & & & & & & & (K) & (\%) \\
\hline
NGC~23 & 0.19 & 11.03 $\pm$ 0.02 & 55$^{+82}_{-31}$ & 11.19 $\pm$ 0.04 & --0.4 $\pm$ 0.4 & 1.55$^{+0.11}_{-0.47}$ & 0.56$^{+0.13}_{-0.02}$ & 7.93 $\pm$ 0.09 & 0.77 $\pm$ 0.04 & 21.1 $\pm$ 1.2 & 11.0 $\pm$ 2.2 \\
MCG~+12-02-001 & 0.15 & 11.44 $\pm$ 0.04 & 83$^{+90}_{-27}$ & 10.8 $\pm$ 0.1 & 0.2 $\pm$ 0.3 & 2.20$^{+0.36}_{-0.31}$ & 1.75$^{+0.27}_{-0.21}$ & 8.03 $\pm$ 0.09 & 0.65 $\pm$ 0.07 & 23.2 $\pm$ 1.6 & 5.7 $\pm$ 1.6 \\
NGC~876 & 0.18 & 10.40 $\pm$ 0.03 & 2.52$^{+0.96}_{-0.65}$ & 10.49 $\pm$ 0.06 & --1.1 $\pm$ 0.2 & 1.52$^{+0.47}_{-0.24}$ & 1.06 $\pm$ 0.08 & 7.7 $\pm$ 0.1 & 0.84 $\pm$ 0.03 & 18.3 $\pm$ 1.2 & 10.8 $\pm$ 2.4 \\
NGC~877 & 0.30 & 11.08 $\pm$ 0.04 & 27$^{+26}_{-7}$ & 11.15 $\pm$ 0.10 & --0.7 $\pm$ 0.4 & 0.77$^{+0.15}_{-0.09}$ & 0.56$^{+0.07}_{-0.08}$ & 8.50 $\pm$ 0.08 & 0.82 $\pm$ 0.04 & 17.3 $\pm$ 1.0 & 15.8 $\pm$ 2.8 \\
UGC~01845 & 0.29 & 11.11 $\pm$ 0.04 & 15$^{+11}_{-3}$ & 11.02 $\pm$ 0.08 & --0.8 $\pm$ 0.2 & 2.49$^{+0.46}_{-0.32}$ & 1.89$^{+0.21}_{-0.13}$ & 7.82 $\pm$ 0.09 & 0.79 $\pm$ 0.04 & 23.1 $\pm$ 1.6 & 8.2 $\pm$ 1.9 \\
NGC~992 & 0.19 & 11.01 $\pm$ 0.05 & 31$^{+31}_{-13}$ & 10.74 $\pm$ 0.06 & --0.3 $\pm$ 0.3 & 1.31$^{+0.44}_{-0.17}$ & 0.83$^{+0.11}_{-0.18}$ & 7.88 $\pm$ 0.09 & 0.74 $\pm$ 0.05 & 20.9 $\pm$ 1.4 & 14.0 $\pm$ 2.7 \\
UGC~02982 & 0.24 & 11.20 $\pm$ 0.04 & 26$^{+32}_{-4}$ & 11.06 $\pm$ 0.06 & --0.7 $\pm$ 0.4 & 1.48$^{+0.34}_{-0.21}$ & 1.19$^{+0.14}_{-0.09}$ & 8.5 $\pm$ 0.1 & 0.83 $\pm$ 0.03 & 18.4 $\pm$ 1.3 & 16.1 $\pm$ 2.7 \\
NGC~1614 & 0.18 & 11.65 $\pm$ 0.06 & 99$^{+90}_{-24}$ & 11.04 $\pm$ 0.08 & 0.0 $\pm$ 0.3 & 2.00$^{+0.41}_{-0.23}$ & 1.37$^{+0.15}_{-0.22}$ & 7.77 $\pm$ 0.09 & 0.58 $\pm$ 0.06 & 27.8 $\pm$ 2.2 & 5.5 $\pm$ 1.7 \\
CGCG~468-002 NED01 & 0.50 & 10.39 $\pm$ 0.06 & 5.0$^{+5.9}_{-2.9}$ & 10.83 $\pm$ 0.05 & --1.2 $\pm$ 0.4 & 1.11$^{+0.02}_{-0.43}$ & 0.49$^{+0.09}_{-0.07}$ & 7.2 $\pm$ 0.1 & 0.87 $\pm$ 0.02 & 22.7 $\pm$ 1.4 & 10.7 $\pm$ 2.4 \\
CGCG~468-002 NED02 & 0.19 & 10.95 $\pm$ 0.05 & 15.0$^{+3.5}_{-4.4}$ & 10.25 $\pm$ 0.09 & --0.1 $\pm$ 0.2 & 2.37$^{+0.40}_{-0.44}$ & 1.33$^{+0.22}_{-0.16}$ & 7.4 $\pm$ 0.1 & 0.75 $\pm$ 0.05 & 26.2 $\pm$ 2.1 & 3.40 $\pm$ 0.70 \\
NGC~1961 & 0.30 & 11.08 $\pm$ 0.03 & 23$^{+18}_{-5}$ & 11.49 $\pm$ 0.06 & --1.1 $\pm$ 0.2 & 0.63$^{+0.10}_{-0.14}$ & 0.41$^{+0.04}_{-0.11}$ & 8.7 $\pm$ 0.1 & 0.84 $\pm$ 0.04 & 15.6 $\pm$ 0.9 & 15.2 $\pm$ 3.8 \\
UGC~03351 & 0.20 & 11.25 $\pm$ 0.03 & 19.6$^{+8.1}_{-3.8}$ & 11.15 $\pm$ 0.09 & --0.8 $\pm$ 0.2 & 2.46$^{+0.41}_{-0.23}$ & 2.03$^{+0.19}_{-0.21}$ & 8.36 $\pm$ 0.09 & 0.89 $\pm$ 0.02 & 20.4 $\pm$ 1.1 & 8.5 $\pm$ 1.7 \\
UGC~03405 & 0.14 & 10.31 $\pm$ 0.03 & 3.4$^{+3.1}_{-1.7}$ & 10.52 $\pm$ 0.08 & --1.0 $\pm$ 0.3 & 1.49$^{+0.07}_{-0.34}$ & 0.95$^{+0.10}_{-0.05}$ & 7.85 $\pm$ 0.09 & 0.86 $\pm$ 0.03 & 16.7 $\pm$ 0.9 & 16.3 $\pm$ 2.3 \\
UGC~03410 & 0.36 & 10.94 $\pm$ 0.02 & 12$^{+17}_{-2}$ & 10.93 $\pm$ 0.05 & --0.9 $\pm$ 0.4 & 1.70$^{+0.41}_{-0.19}$ & 1.25$^{+0.12}_{-0.02}$ & 8.2 $\pm$ 0.1 & 0.83 $\pm$ 0.03 & 18.4 $\pm$ 1.0 & 15.1 $\pm$ 2.7 \\
NGC~2388 & 0.36 & 11.15 $\pm$ 0.06 & 40$^{+9}_{-22}$ & 11.0 $\pm$ 0.1 & --0.4 $\pm$ 0.4 & 2.54 $\pm$ 0.27 & 2.04$^{+0.10}_{-0.22}$ & 7.96 $\pm$ 0.09 & 0.74 $\pm$ 0.05 & 21.9 $\pm$ 1.4 & 6.3 $\pm$ 1.6 \\
NGC~2389 & 0.31 & 10.50 $\pm$ 0.04 & 8.7$^{+1.4}_{-0.9}$ & 10.48 $\pm$ 0.09 & --0.5 $\pm$ 0.1 & 0.46$^{+0.09}_{-0.06}$ & 0.28$^{+0.08}_{-0.05}$ & 7.8 $\pm$ 0.1 & 0.78 $\pm$ 0.05 & 18.1 $\pm$ 1.2 & 15.9 $\pm$ 3.6 \\
MCG~+02-20-002 & 0.31 & 10.21 $\pm$ 0.08 & 4.4$^{+1.1}_{-0.7}$ & 10.24 $\pm$ 0.08 & --0.6 $\pm$ 0.2 & 0.51$^{+0.17}_{-0.12}$ & 0.27$^{+0.09}_{-0.03}$ & 7.65 $\pm$ 0.10 & 0.7 $\pm$ 0.2 & 16.1 $\pm$ 1.4 & 12.6 $\pm$ 4.9 \\
MCG~+02-20-003 & 0.54 & 11.02 $\pm$ 0.06 & 17.1$^{+5.4}_{-3.7}$ & 10.47 $\pm$ 0.09 & --0.2 $\pm$ 0.2 & 1.86$^{+0.49}_{-0.35}$ & 1.29$^{+0.14}_{-0.12}$ & 7.78 $\pm$ 0.09 & 0.82 $\pm$ 0.07 & 23.0 $\pm$ 1.5 & 9.7 $\pm$ 2.3 \\
NGC~3110 & 0.28 & 11.26 $\pm$ 0.06 & 36$^{+34}_{-8}$ & 10.99 $\pm$ 0.09 & --0.4 $\pm$ 0.4 & 1.27$^{+0.25}_{-0.23}$ & 0.87$^{+0.07}_{-0.15}$ & 8.24 $\pm$ 0.09 & 0.80 $\pm$ 0.04 & 20.7 $\pm$ 1.4 & 13.3 $\pm$ 2.5 \\
NGC~3221 & 0.29 & 11.03 $\pm$ 0.03 & 13.2$^{+9.5}_{-3.1}$ & 11.11 $\pm$ 0.05 & --1.0 $\pm$ 0.3 & 1.38$^{+0.48}_{-0.23}$ & 0.96$^{+0.09}_{-0.07}$ & 8.44 $\pm$ 0.09 & 0.84 $\pm$ 0.03 & 17.3 $\pm$ 1.0 & 15.8 $\pm$ 2.8 \\
Arp~299 & 0.44 & 11.79 $\pm$ 0.05 & 134$^{+31}_{-21}$ & 10.9 $\pm$ 0.1 & 0.3 $\pm$ 0.2 & 1.68$^{+0.50}_{-0.18}$ & 1.22$^{+0.21}_{-0.20}$ & 7.91 $\pm$ 0.09 & 0.56 $\pm$ 0.06 & 27.6 $\pm$ 1.9 & 3.4 $\pm$ 1.4 \\
MCG~--02-33-098 & 0.07 & 10.90 $\pm$ 0.04 & 14.0$^{+3.2}_{-2.5}$ & 10.7 $\pm$ 0.1 & --0.5 $\pm$ 0.3 & 1.35$^{+0.39}_{-0.13}$ & 0.96$^{+0.11}_{-0.13}$ & 7.64 $\pm$ 0.10 & 0.66 $\pm$ 0.06 & 21.5 $\pm$ 1.6 & 7.1 $\pm$ 1.9 \\
IC~860 & 1.73 & 10.94 $\pm$ 0.03 & 1.7$^{+1.3}_{-1.1}$ & 10.5 $\pm$ 0.1 & --1.3 $\pm$ 0.5 & 3.65$^{+0.73}_{-0.99}$ & 2.39$^{+0.05}_{-0.10}$ & 7.42 $\pm$ 0.06 & 0.83 $\pm$ 0.03 & 26.0 $\pm$ 1.0 & 1.10 $\pm$ 0.10 \\
NGC~5653 & 0.28 & 10.95 $\pm$ 0.03 & 16.5$^{+5.2}_{-2.9}$ & 10.84 $\pm$ 0.08 & --0.6 $\pm$ 0.2 & 1.13$^{+0.16}_{-0.13}$ & 0.76$^{+0.17}_{-0.06}$ & 7.95 $\pm$ 0.09 & 0.79 $\pm$ 0.04 & 20.4 $\pm$ 1.2 & 12.4 $\pm$ 2.4 \\
Zw~049-057 & 1.25 & 11.08 $\pm$ 0.04 & 8.6$^{+8.7}_{-1.9}$ & 10.40 $\pm$ 0.09 & --0.5 $\pm$ 0.3 & 3.74$^{+0.95}_{-0.32}$ & 3.13$^{+0.08}_{-0.15}$ & 7.89 $\pm$ 0.07 & 0.892 $\pm$ 0.005 & 23.5 $\pm$ 1.3 & 1.10 $\pm$ 0.40 \\
NGC~5936 & 0.21 & 10.97 $\pm$ 0.04 & 18.1$^{+3.4}_{-2.9}$ & 10.80 $\pm$ 0.08 & --0.6 $\pm$ 0.2 & 1.01$^{+0.19}_{-0.15}$ & 0.66$^{+0.12}_{-0.08}$ & 8.02 $\pm$ 0.10 & 0.76 $\pm$ 0.06 & 19.6 $\pm$ 1.3 & 10.7 $\pm$ 2.0 \\
NGC~5990 & 0.41 & 10.96 $\pm$ 0.05 & 20$^{+10}_{-7}$ & 11.0 $\pm$ 0.1 & --0.7 $\pm$ 0.2 & 1.08$^{+0.18}_{-0.10}$ & 0.73 $\pm$ 0.11 & 7.9 $\pm$ 0.1 & 0.76 $\pm$ 0.06 & 20.6 $\pm$ 1.5 & 11.2 $\pm$ 2.2 \\
NGC~6052 & 0.25 & 10.94 $\pm$ 0.04 & 42$^{+52}_{-17}$ & 10.36 $\pm$ 0.09 & 0.3 $\pm$ 0.3 & 0.63$^{+0.09}_{-0.09}$ & 0.45$^{+0.03}_{-0.10}$ & 7.75 $\pm$ 0.08 & 0.73 $\pm$ 0.05 & 21.3 $\pm$ 1.3 & 8.1 $\pm$ 1.5 \\
NGC~6701 & 0.16 & 11.00 $\pm$ 0.03 & 18.3$^{+3.0}_{-2.7}$ & 10.88 $\pm$ 0.09 & --0.6 $\pm$ 0.2 & 1.12$^{+0.15}_{-0.21}$ & 0.69 $\pm$ 0.10 & 8.02 $\pm$ 0.08 & 0.77 $\pm$ 0.04 & 20.1 $\pm$ 1.3 & 10.1 $\pm$ 2.2 \\
NGC~6921 & 0.10 & 10.25 $\pm$ 0.04 & 1.57$^{+0.41}_{-0.56}$ & 11.16 $\pm$ 0.03 & --1.9 $\pm$ 0.3 & 0.61$^{+0.66}_{-0.24}$ & 0.18$^{+0.03}_{-0.03}$ & 7.02 $\pm$ 0.09 & 0.85 $\pm$ 0.02 & 23.3 $\pm$ 1.3 & 10.0 $\pm$ 2.8 \\
MCG~+04-48-002 & 0.19 & 10.92 $\pm$ 0.04 & 14$^{+13}_{-3}$ & 10.84 $\pm$ 0.07 & --0.7 $\pm$ 0.3 & 1.45$^{+0.47}_{-0.19}$ & 1.13$^{+0.14}_{-0.12}$ & 7.72 $\pm$ 0.09 & 0.85 $\pm$ 0.03 & 23.0 $\pm$ 1.4 & 15.2 $\pm$ 2.7 \\
NGC~7591 & 0.16 & 11.06 $\pm$ 0.04 & 18.9$^{+9.4}_{-5.7}$ & 11.05 $\pm$ 0.09 & --0.8 $\pm$ 0.2 & 1.55$^{+0.31}_{-0.25}$ & 0.93$^{+0.19}_{-0.05}$ & 8.2 $\pm$ 0.1 & 0.79 $\pm$ 0.04 & 18.9 $\pm$ 1.1 & 8.3 $\pm$ 1.8 \\
NGC~7679 & 0.42 & 11.02 $\pm$ 0.04 & 43$^{+21}_{-20}$ & 10.7 $\pm$ 0.1 & --0.1 $\pm$ 0.3 & 0.80$^{+0.26}_{-0.06}$ & 0.53$^{+0.12}_{-0.03}$ & 7.81 $\pm$ 0.09 & 0.76 $\pm$ 0.05 & 22.1 $\pm$ 1.5 & 13.3 $\pm$ 2.1 \\
NGC~7752 & 0.36 & 10.65 $\pm$ 0.04 & 15$^{+11}_{-5}$ & 10.3 $\pm$ 0.1 & --0.1 $\pm$ 0.3 & 0.90$^{+0.09}_{-0.15}$ & 0.64$^{+0.14}_{-0.11}$ & 7.56 $\pm$ 0.09 & 0.77 $\pm$ 0.05 & 21.0 $\pm$ 1.4 & 13.0 $\pm$ 2.4 \\
NGC~7753 & 0.40 & 10.90 $\pm$ 0.04 & 15$^{+11}_{-2}$ & 11.25 $\pm$ 0.06 & --1.1 $\pm$ 0.2 & 0.63$^{+0.16}_{-0.15}$ & 0.31$^{+0.09}_{-0.04}$ & 8.6 $\pm$ 0.1 & 0.83 $\pm$ 0.04 & 15.5 $\pm$ 1.2 & 12.5 $\pm$ 2.9 \\
NGC~7769 & 0.10 & 10.8 $\pm$ 0.1 & 15.0$^{+5.7}_{-3.8}$ & 11.05 $\pm$ 0.06 & --0.9 $\pm$ 0.1 & 0.67$^{+0.13}_{-0.20}$ & 0.37$^{+0.12}_{-0.11}$ & 7.9 $\pm$ 0.2 & 0.82 $\pm$ 0.09 & 20.1 $\pm$ 3.4 & 13.1 $\pm$ 4.8 \\
NGC~7770 & 0.36 & 10.40 $\pm$ 0.04 & 9.0$^{+8.1}_{-3.1}$ & 10.3 $\pm$ 0.1 & --0.3 $\pm$ 0.3 & 0.65$^{+0.19}_{-0.08}$ & 0.51$^{+0.05}_{-0.15}$ & 7.17 $\pm$ 0.09 & 0.73 $\pm$ 0.06 & 22.1 $\pm$ 1.8 & 10.2 $\pm$ 2.2 \\
NGC~7771 & 0.15 & 11.33 $\pm$ 0.03 & 27.7$^{+7.6}_{-5.9}$ & 11.37 $\pm$ 0.07 & --0.9 $\pm$ 0.2 & 1.55$^{+0.19}_{-0.11}$ & 1.11$^{+0.03}_{-0.12}$ & 8.45 $\pm$ 0.09 & 0.85 $\pm$ 0.03 & 19.8 $\pm$ 1.2 & 7.8 $\pm$ 1.2 \\
\hline
\end{tabular}

\end{table}
\end{landscape}

\clearpage

\Online
\appendix

\section{Optical spectra of nine galaxies}\label{apx:optical_class}

We present the optical spectra of nine galaxies in our sample without a previously published [\ion{N}{ii}]\slash H$\alpha$ ratio, to best of our knowledge.
Their reduced spectra were available through the Smithsonian Astrophysical Observatory Telescope Data Center.
They were obtained between 1998 and 2006 with the FAST Spectrograph \citep{Tokarz1997} on the Mount Hopkins Tillinghast 1.5\,m reflector. The slit width was 3\arcsec\ and the spectra cover the spectral range between 3700 and 7500\,\AA\ with a dispersion of 1.5\,\AA\ per pixel. The integration times were between 600 and 1500\,s. The spectra are not flux calibrated but they can be used to measure line ratios between transitions close in wavelength.

\begin{figure*}
\centering
\includegraphics[width=0.42\textwidth]{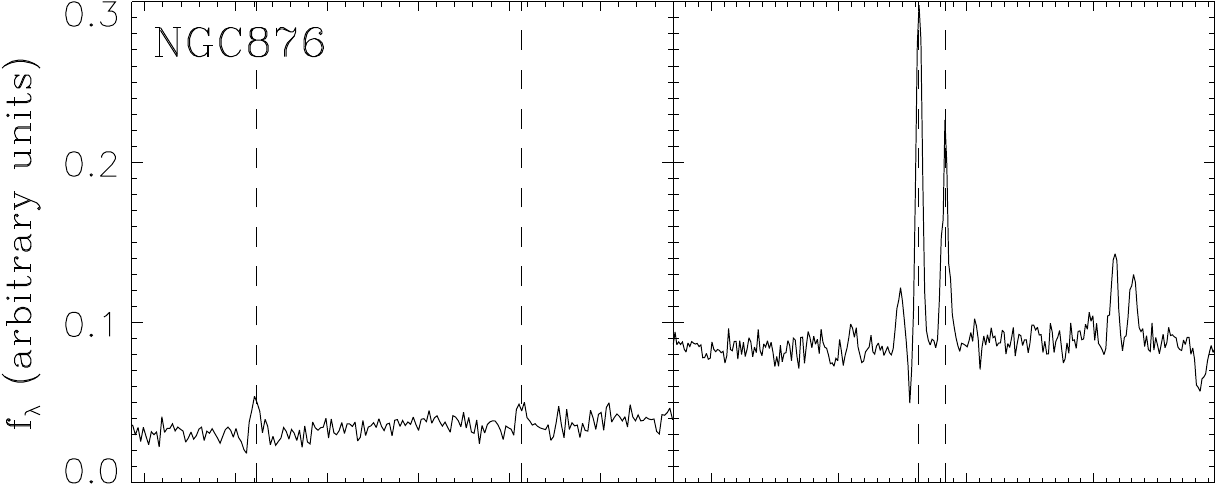}
\includegraphics[width=0.42\textwidth]{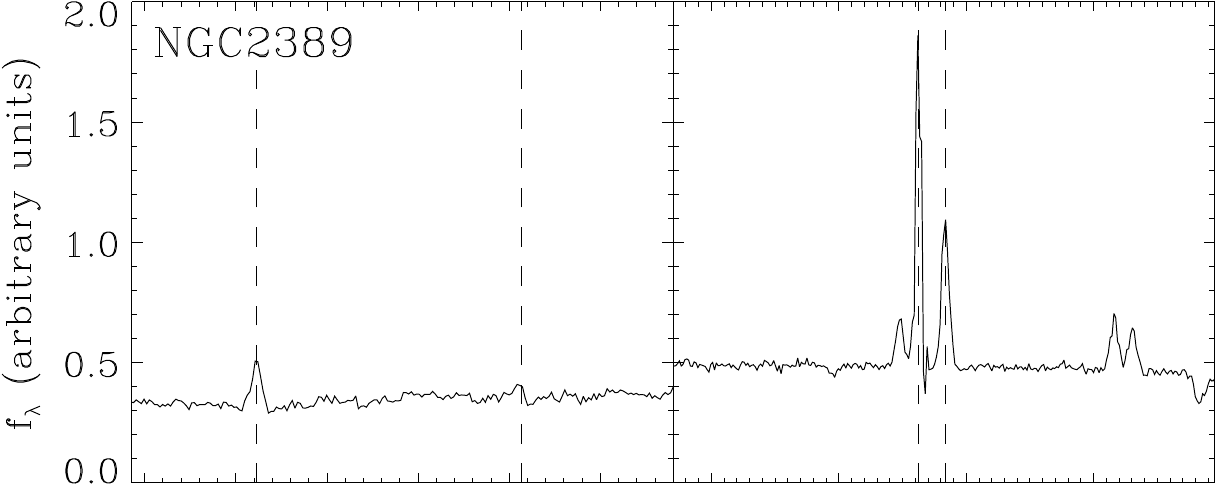}
\includegraphics[width=0.42\textwidth]{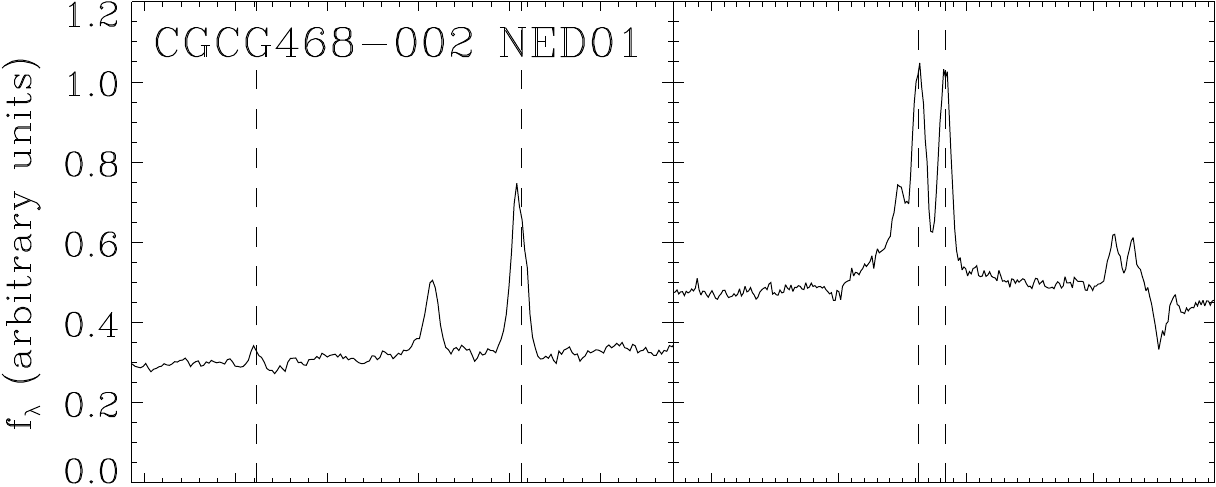}
\includegraphics[width=0.42\textwidth]{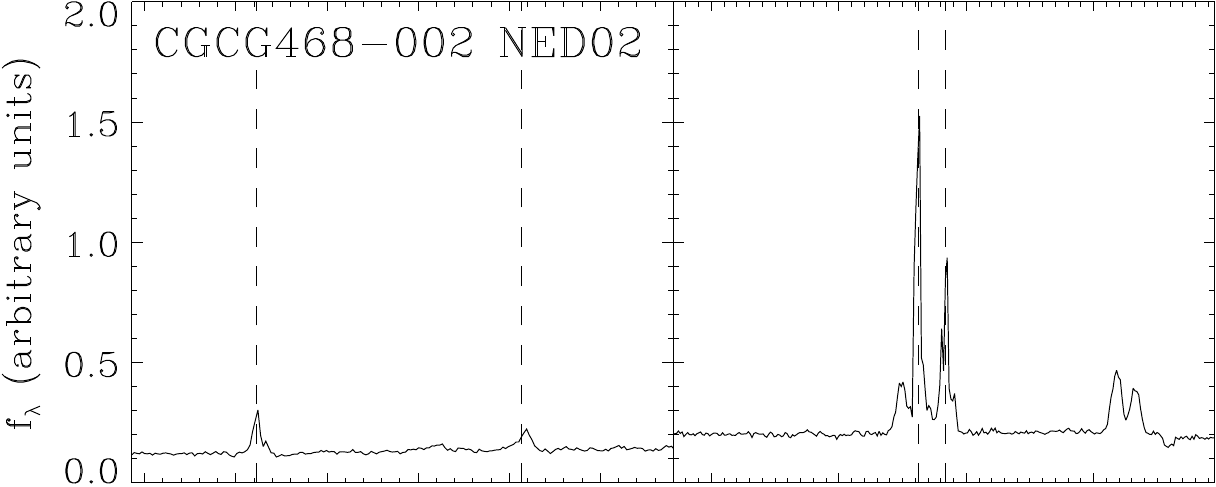}
\includegraphics[width=0.42\textwidth]{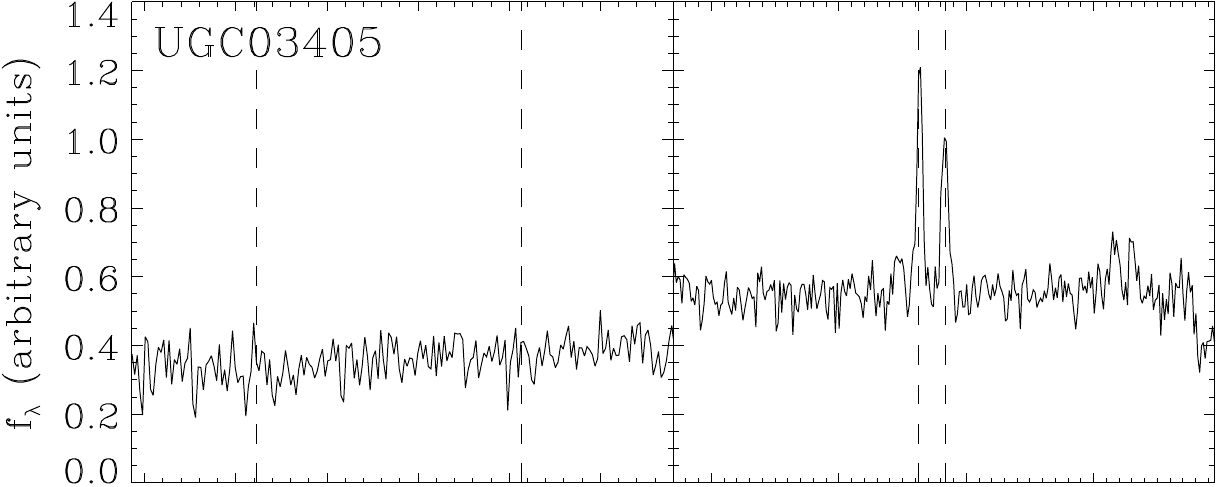}
\includegraphics[width=0.42\textwidth]{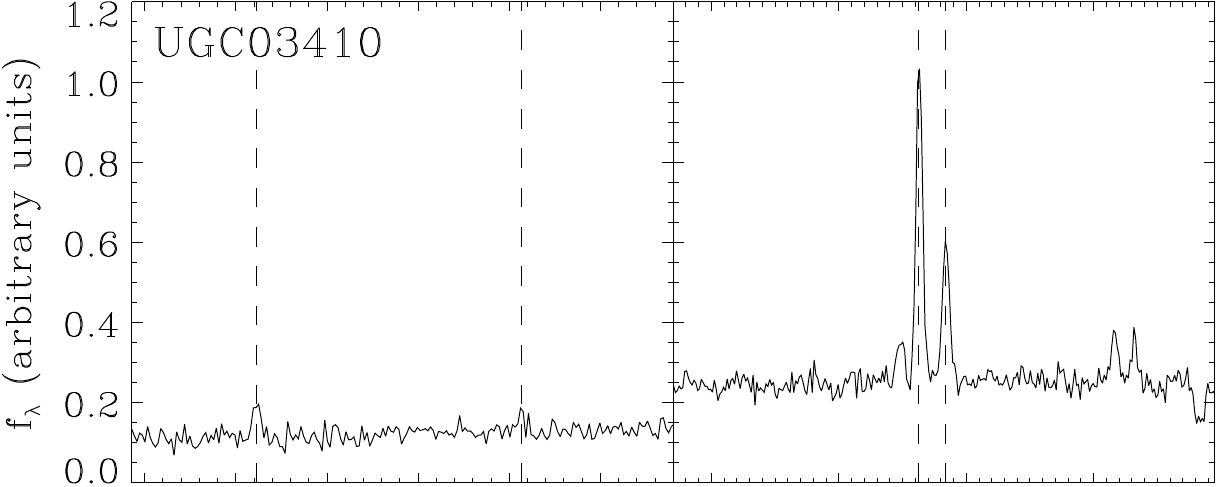}
\raisebox{.7cm}{\includegraphics[width=0.42\textwidth]{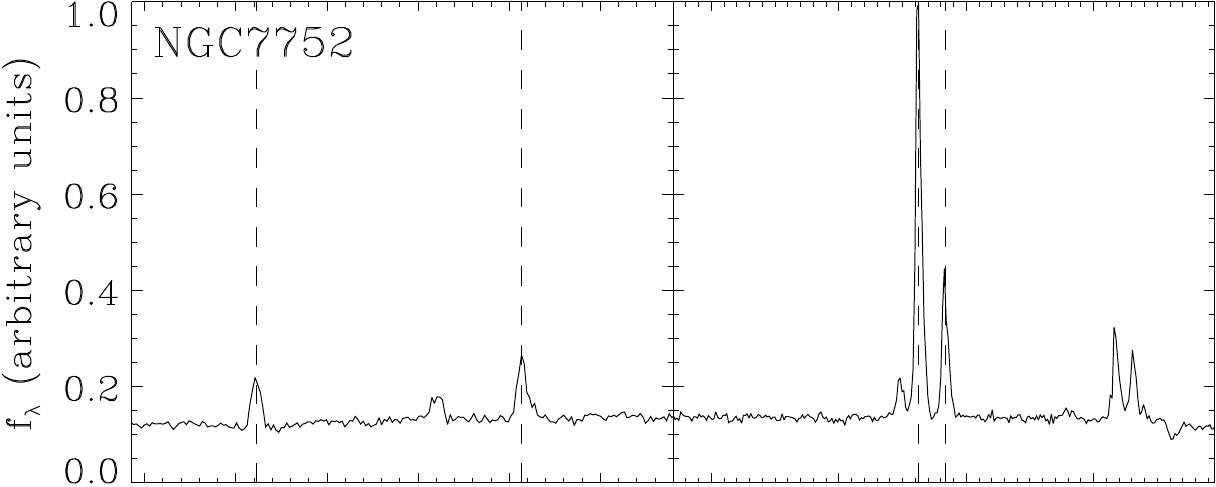}}
\vspace*{-.65cm}
\includegraphics[width=0.42\textwidth]{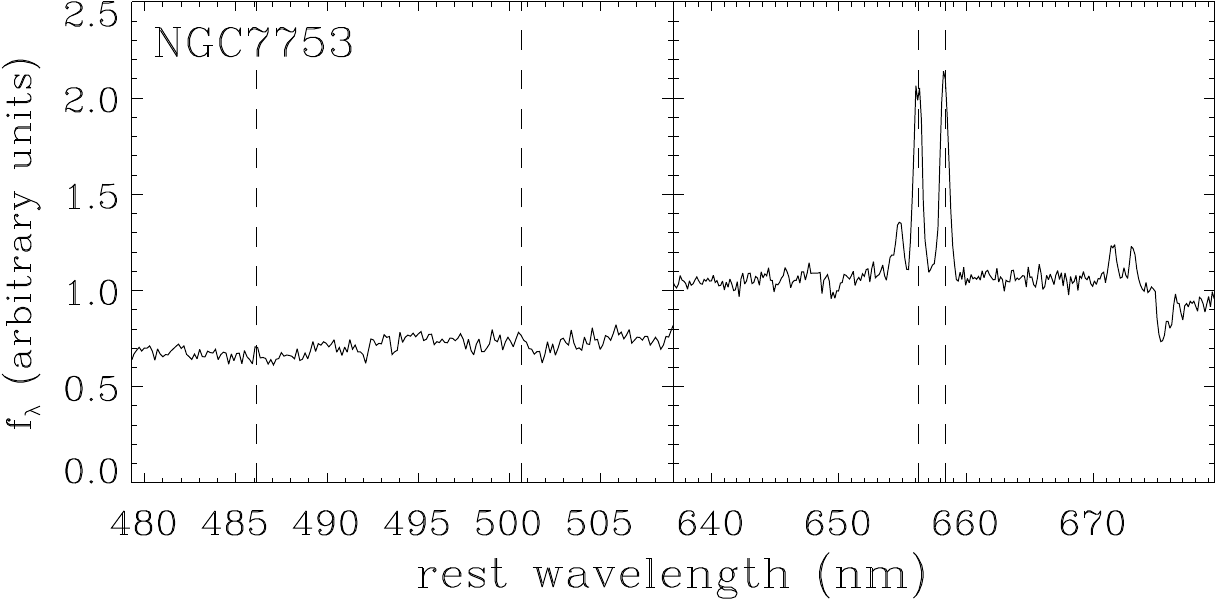}
\includegraphics[width=0.42\textwidth]{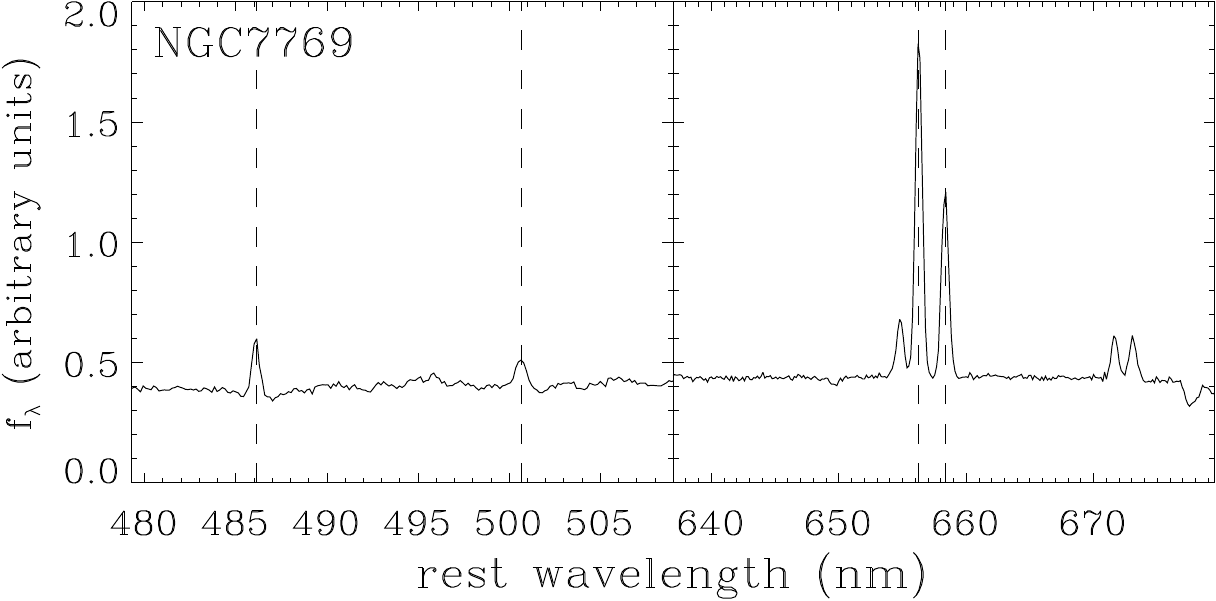}
\hspace{0.42\textwidth}
\caption{Nuclear optical spectra of nine galaxies of our sample obtained with the FAST Spectrograph in the ranges 480--510\AA\ and 637--680\AA. The dashed lines mark the position of the H$\beta$, [\ion{O}{iii}]$\lambda$5007\AA, H$\alpha$, and [\ion{N}{ii}]$\lambda$6584\AA\ transitions.\label{fig:optical_spectra}}
\end{figure*}

In the spectra shown in Figure\ref{fig:optical_spectra}, we measured the fluxes of the H$\beta$, [\ion{O}{iii}]$\lambda$5007\AA, H$\alpha$, and [\ion{N}{ii}]$\lambda$6584\AA\ emission lines using a single component Gaussian fit, except for CGCG~468-002 NED01, which shows a blue-shifted broad H$\alpha$ component. The FWHM of the narrow component in this galaxy is $\sim$500\,km\,s$^{-1}$ \citep{AAH2013}, whereas the broad H$\alpha$ component has a FWHM of 3100$\pm$190\,km\,s$^{-1}$ and is blue-shifted by 180$\pm$60\,km\,s$^{-1}$.
The observed line ratios are listed in Table \ref{tab:opt_ratio}. $H\beta$ is not corrected for stellar absorption.

\begin{figure}[h]
\includegraphics[width=0.4\textwidth]{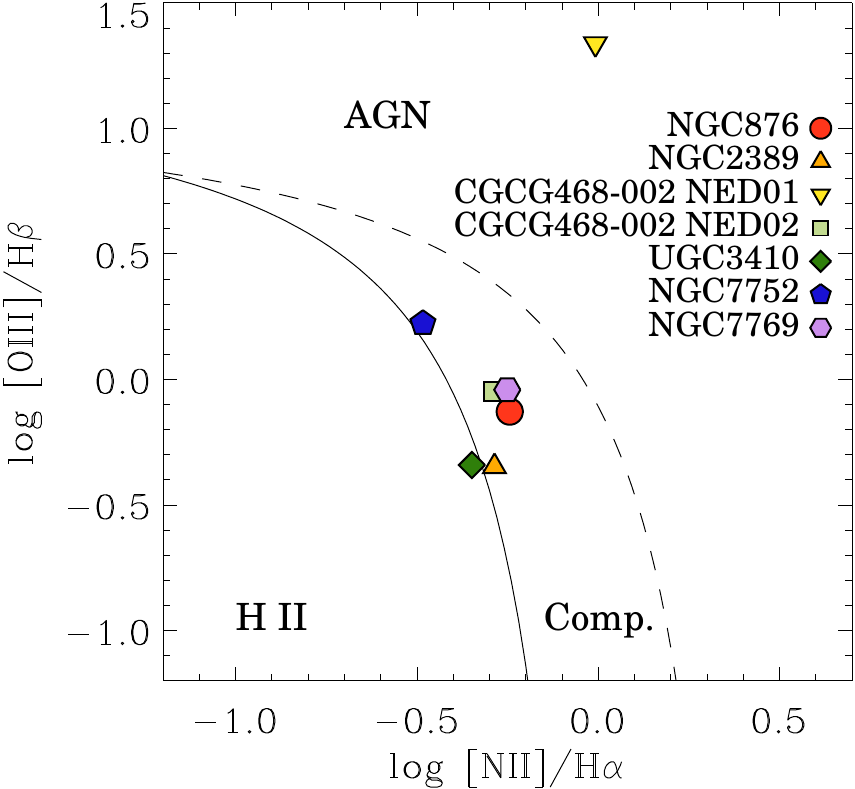}
\caption{[\ion{N}{ii}]$\lambda$6584\AA\slash H$\alpha$ vs. [\ion{O}{iii}]$\lambda$5007\AA\slash H$\beta$ diagnostic diagram for the nuclear spectra of six LIRGs. The solid and dashed black lines mark the empirical separation between \ion{H}{ii}, composite, and AGN galaxies from \citet{Kewley2006}.\label{fig:optical_bpt}}
\end{figure}

We used the standard optical diagnostic diagram [\ion{N}{ii}]\slash H$\alpha$ vs. [\ion{O}{iii}]\slash H$\beta$
\citep{Baldwin1981} to determine the nuclear activity classification. We used the boundary limits between \ion{H}{ii}, composite and AGN galaxies proposed by \citet{Kewley2006}. Figure \ref{fig:optical_bpt} shows that four of the galaxies lie in the composite region of the diagram, one in the \ion{H}{ii} region, although close to the \ion{H}{ii}--composite border, and one galaxy, CGCG~468-002 NED01, is classified as AGN. Since in this object we only detect a broad component in H$\alpha$ we classify it as Sy1.9. For UGC~03405 and NGC~7753, the H$\beta$ and [\ion{O}{iii}]$\lambda$5007\AA\ transitions were not detected so we excluded these objects from the diagram. However, the high  [\ion{N}{ii}]\slash H$\alpha$ ratio in these two sources, together with the absence of [\ion{O}{iii}] detections, which is bright in AGNs, suggests that these are composite galaxies.

\begin{table}[h!]
\caption{Optical line ratios and nuclear activity classification}
\label{tab:opt_ratio}
\centering
\setlength{\tabcolsep}{4pt}
\begin{tabular}{lcccccccccccccc}
\hline \hline
Name & [\ion{O}{iii}]\slash H$\beta$ & [\ion{N}{ii}]\slash H$\alpha$ & Class.\\
\hline
NGC~876 & 0.75 & 0.57 & composite\\
NGC~2389 & 0.45 & 0.53 & composite\\
CGCG~468-002 NED01\tablefootmark{\star} & 21.7 & 0.98 & Sy1.9 \\
CGCG~468-002 NED02 & 0.89 & 0.51 & composite\\
UGC~03405 & \nodata & 0.75& composite? \\
UGC~03410 & 0.46 & 0.45 & \ion{H}{ii} \\
NGC~7753 & \nodata & 1.02 & composite? \\
NGC~7752 & 0.60 & 0.33 & composite\\
NGC~7769 & 0.91 & 0.56 & composite\\
\hline
\end{tabular}
\tablefoot{\tablefoottext{\star}{Line ratios corresponding to the narrow component.}}
\end{table}

\section{Likelihood with detections and upper limits}\label{apx:likelihood}

In this appendix we briefly describe how the upper limits are included in our Bayesian analysis (see, e.g, \citealt{Gregory2005, Bohm2010} for a general description of the Bayesian approach).
We let $F_i$ and $\sigma_i$ be the flux and 1$\sigma$ uncertainty measured for a galaxy in the band $i$. If the galaxy is not detected, we measure the $n\sigma_i$ upper limit. Likewise, the prediction of the model $k$ for the flux of band $i$ is $M_i(k)$.
The likelihood is defined as

\begin{equation}\label{eqn:likedef}
 L_k = \prod_i f_i(F_i, \sigma_i, M_i(k))
\end{equation}

For the detections we assume that $f_i$ follows a normal distribution

\begin{equation}
 f^d_i(F_i, \sigma_i, M_i) = \frac{1}{\sqrt{2 \pi \sigma_i}} \exp \left[ -\frac{\left( F_i - M_i\right)^2}{2\sigma_i^2}  \right]
\end{equation}

On the other hand, to obtain the likelihood for the upper limits, $f^u_i$, we first calculate the probability for the flux to have an arbitrary value $R_i$. The unknown background in the aperture used to measure the flux is $B_i$. The standard deviation of the background is $\sigma_i$, and for simplicity we assume that the mean background of the image is zero. If the galaxy flux is $R_i$ but we do not detect it on the image at a $n\sigma_i$ level is because $R_i + B_i < n\sigma_i$. If the background follows a normal distribution the probability of this is
\begin{equation}\label{eqn:gauss}
f^u_i(F_i, \sigma_i, R_i) = \Phi\left(\frac{n\sigma_i - R_i}{\sigma_i}\right)
\end{equation}

\noindent where $\Phi$ is the cumulative distribution function of the standard normal distribution. Therefore the likelihood value for the upper limits is

\begin{equation}\label{eqn:upper}
f_i^u(F_i, \sigma_i, R_i) = \frac{1}{2}\left[ 1 + {\rm erf}\left(\frac{n\sigma_i - R_i}{\sigma_i\sqrt{2}}\right) \right]
\end{equation}

Then when substituting Equations \ref{eqn:gauss} and \ref{eqn:upper} into Equation \ref{eqn:likedef}

\begin{align}\label{eqn:like1}
 L_k =& \prod_{i_1} \frac{1}{\sqrt{2 \pi \sigma_{i_1}}} \exp \left[ -\frac{\left( F_{i_1} - M_{i_1}(k)\right)^2}{2\sigma_{i_1}^2}  \right] \times \\
\nonumber & \prod_{i_2} \frac{1}{2}\left[ 1 + {\rm erf}\left(\frac{n\sigma_{i_2} - M_{i_2}(k)}{\sigma_{i_2}\sqrt{2}}\right) \right]
\end{align}

where $i_1$ and $i_2$ are the subindices for the detections and non-detections respectively. The logarithm of Equation \ref{eqn:like1} is

\begin{align}
 \ln L_k = & \sum_{i_1}  -\frac{\left( F_{i_1} - M_{i_1}(k)\right)^2}{2\sigma_{i_1}^2} + \\
\nonumber & \sum_{i_2} \ln \left[ 1 + {\rm erf}\left(\frac{n\sigma_{i_2} - M_{i_2}(k)}{\sigma_{i_2}\sqrt{2}}\right) \right] + C\\
\nonumber = & -\frac{1}{2} \chi^2_k + \sum_{i_2} \ln \left[ 1 + {\rm erf}\left(\frac{n\sigma_{i_2} - M_{i_2}(k)}{\sigma_{i_2}\sqrt{2}}\right) \right] + C
\end{align}

Thus, we assign the following probability to model $k$ in the parameter inference process

\begin{equation}
 L_k = \exp \left(-\frac{1}{2} \chi^2_k \right) \prod_{i_2} \left[ 1 + {\rm erf}\left(\frac{n\sigma_{i_2} - M_{i_2}(k)}{\sigma_{i_2}\sqrt{2}}\right) \right]
\end{equation}
\clearpage
\onecolumn
\section{Multi-wavelength imaging of the LIRGs}\label{apx:images}

\begin{figure*}[!h]
\centering
\includegraphics[width=0.8\textwidth]{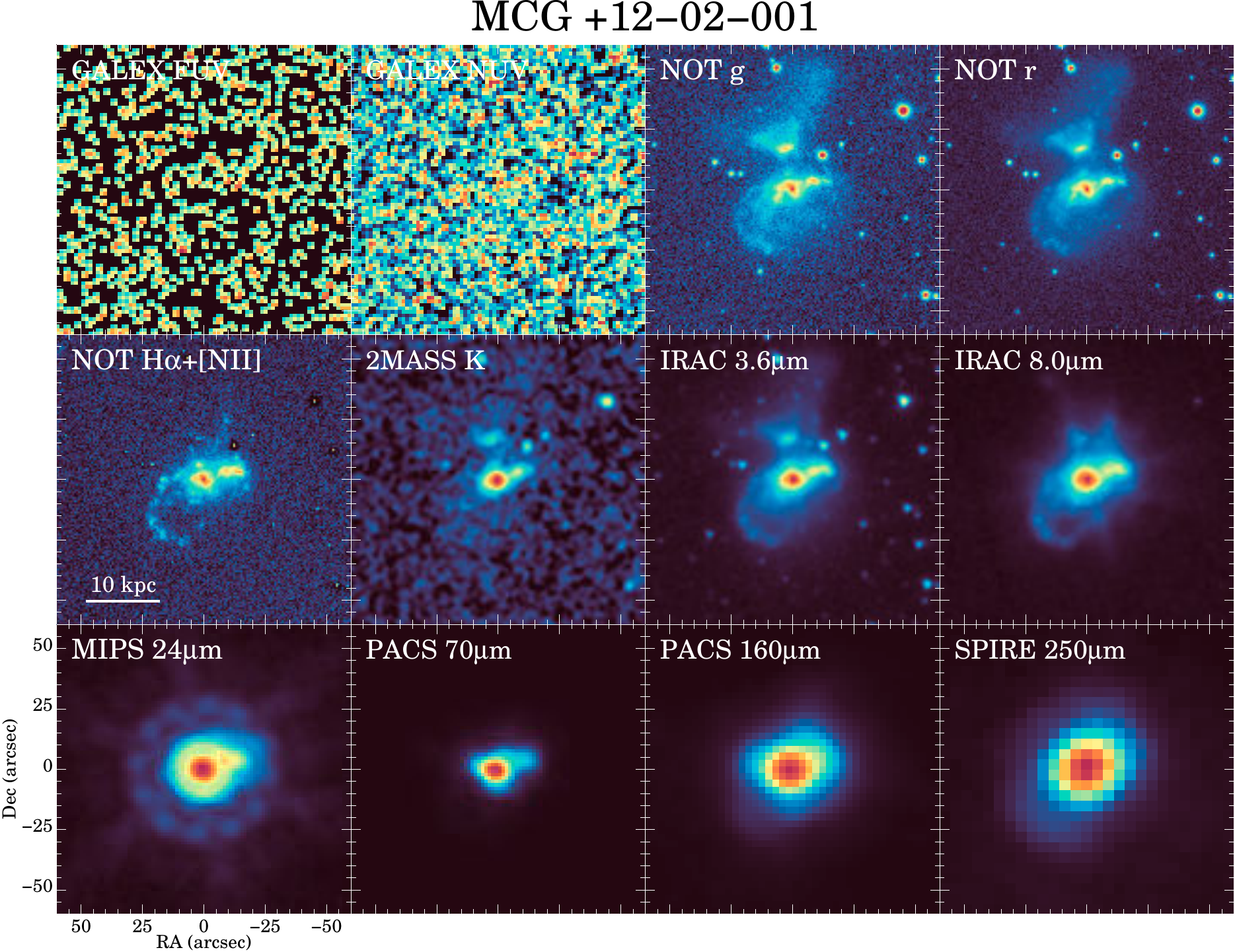}
\includegraphics[width=0.8\textwidth]{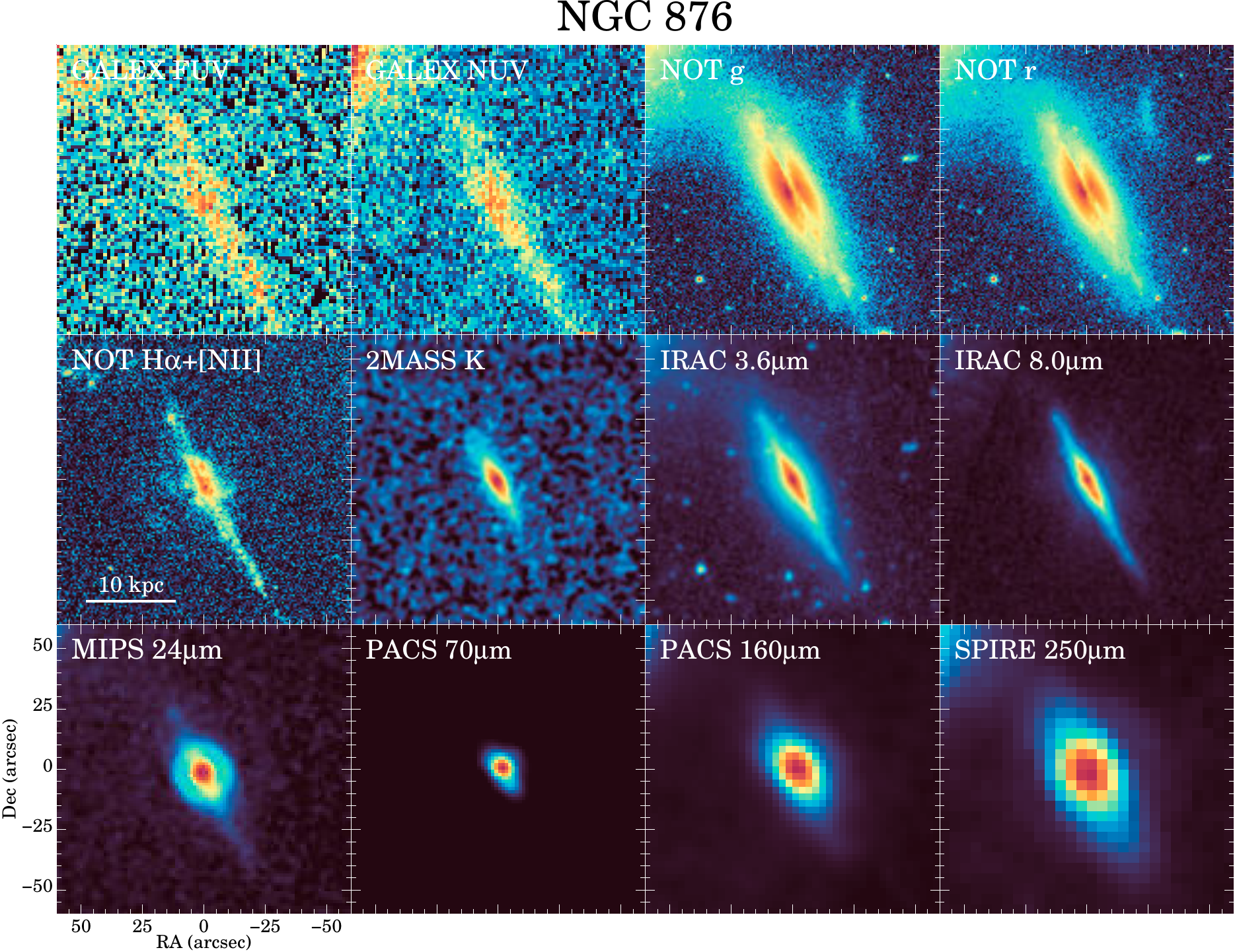}
\caption{Same as Figure \ref{fig:images_main_text}.}
\end{figure*}
\clearpage

\begin{figure*}[!h]
\centering
\addtocounter{figure}{-1}
\includegraphics[width=.8\textwidth]{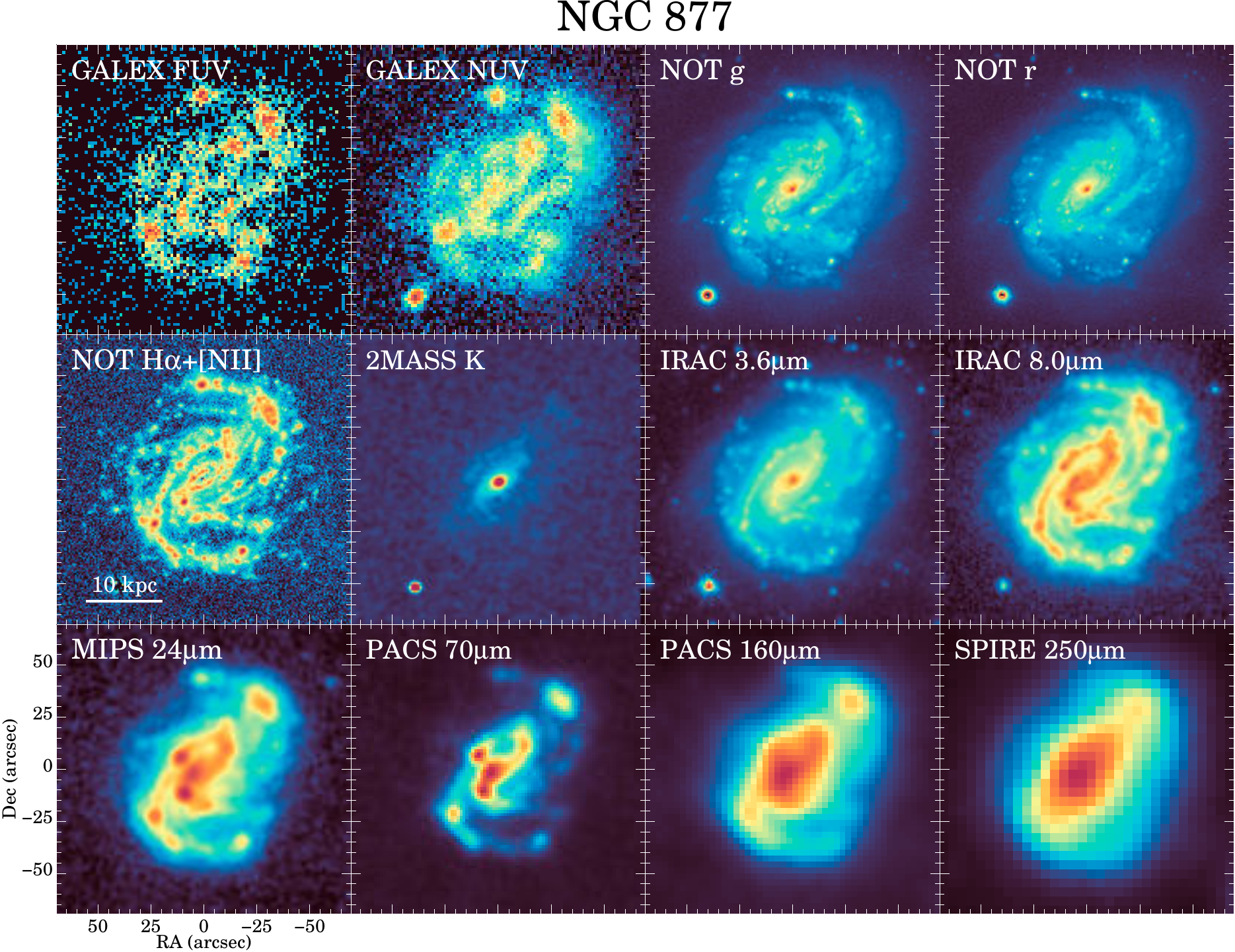}
\includegraphics[width=.8\textwidth]{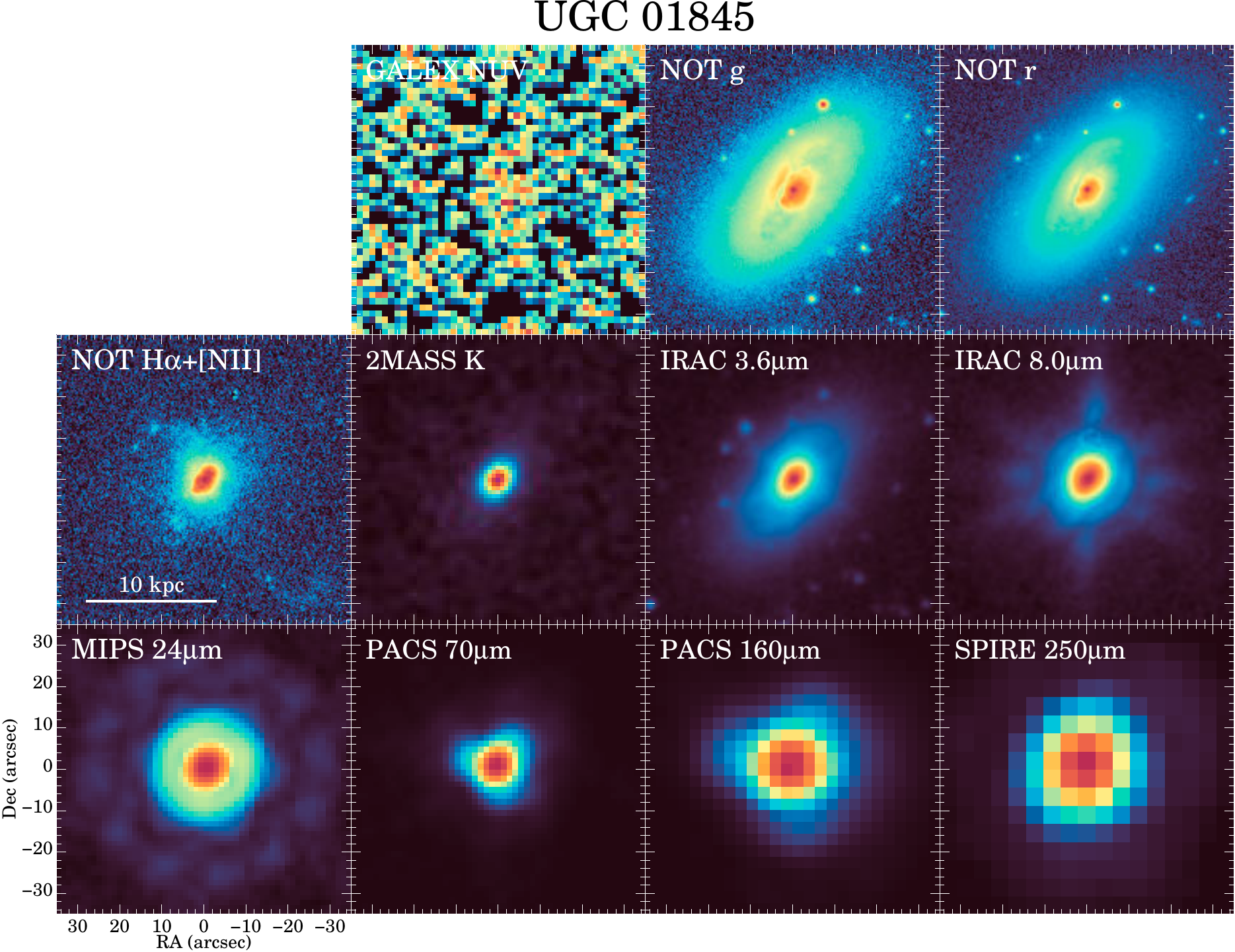}
\caption{Continued.}
\end{figure*}
\clearpage

\begin{figure*}[!h]
\centering
\addtocounter{figure}{-1}
\includegraphics[width=.8\textwidth]{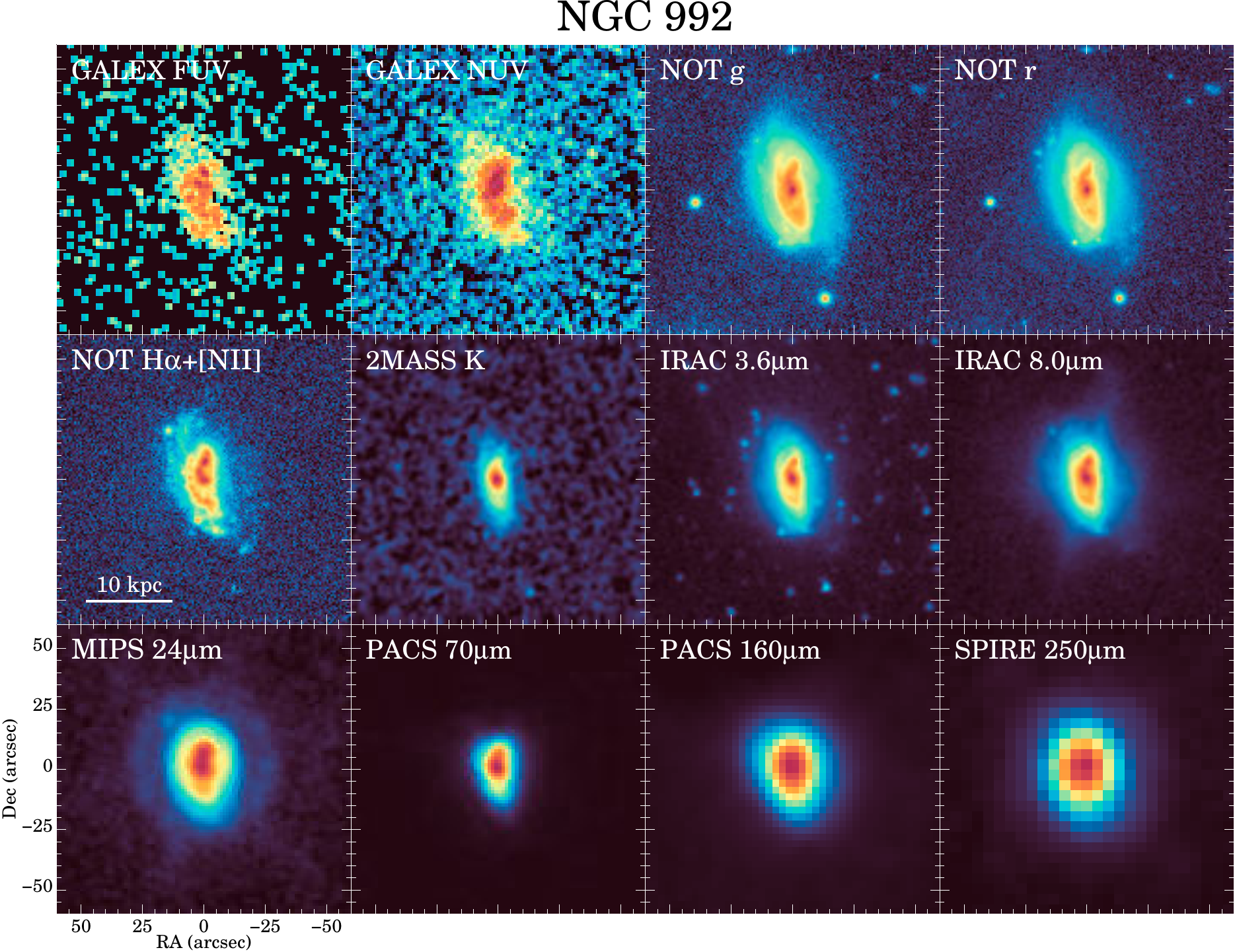}
\includegraphics[width=.8\textwidth]{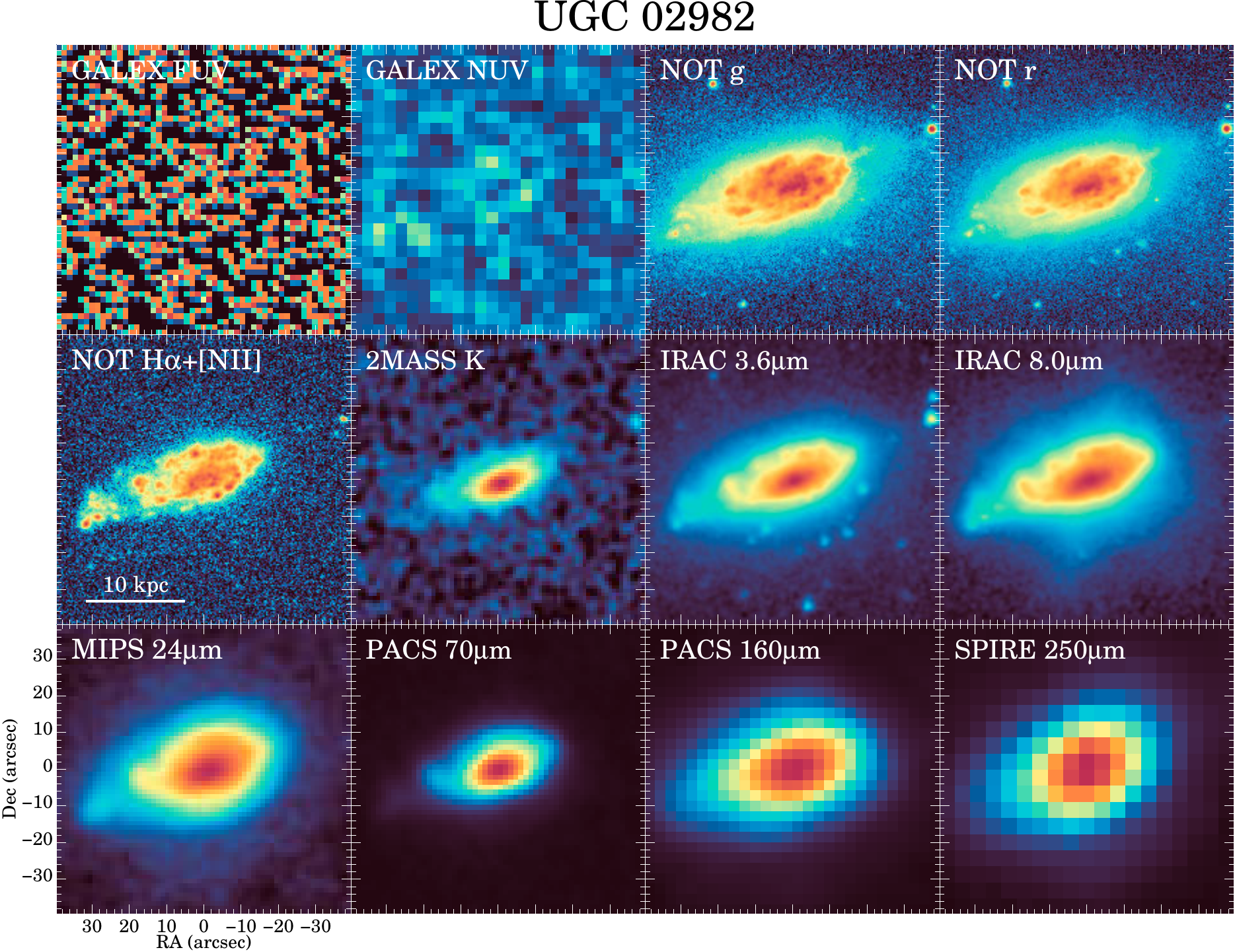}
\caption{Continued.}
\end{figure*}
\clearpage

\begin{figure*}[!h]
\centering
\addtocounter{figure}{-1}
\includegraphics[width=.8\textwidth]{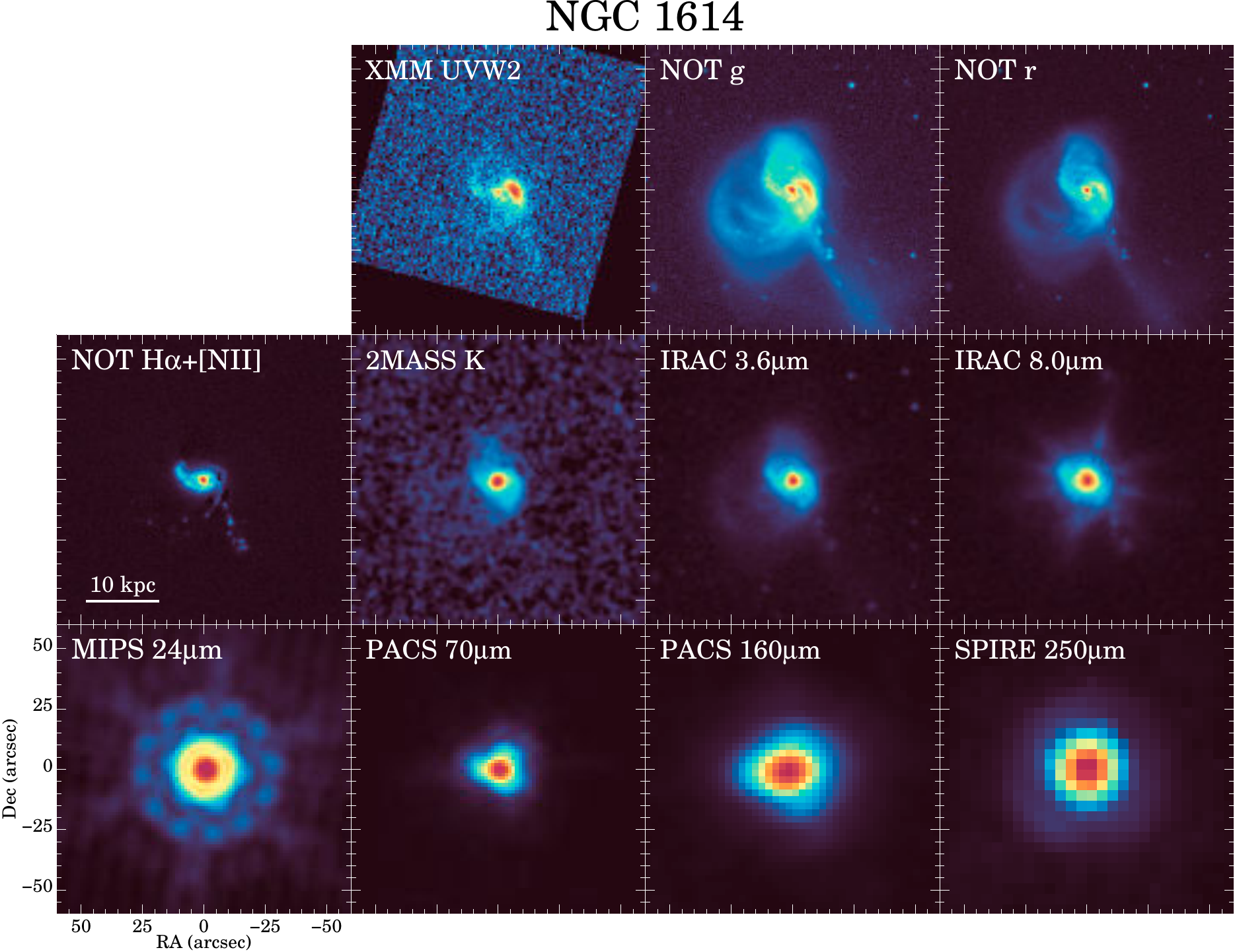}
\includegraphics[width=.8\textwidth]{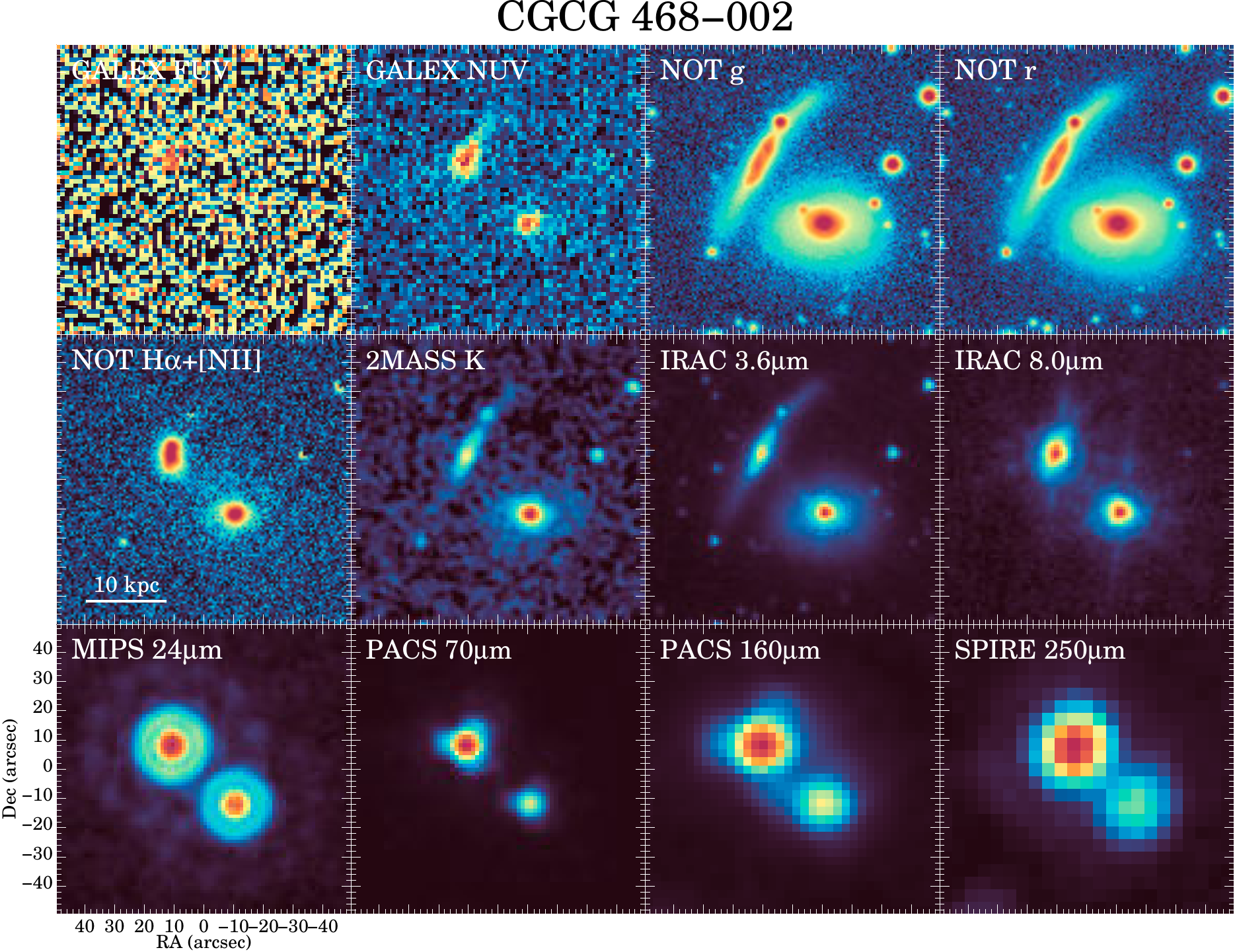}
\caption{Continued.}
\end{figure*}
\clearpage

\begin{figure*}[!h]
\centering
\addtocounter{figure}{-1}
\includegraphics[width=.8\textwidth]{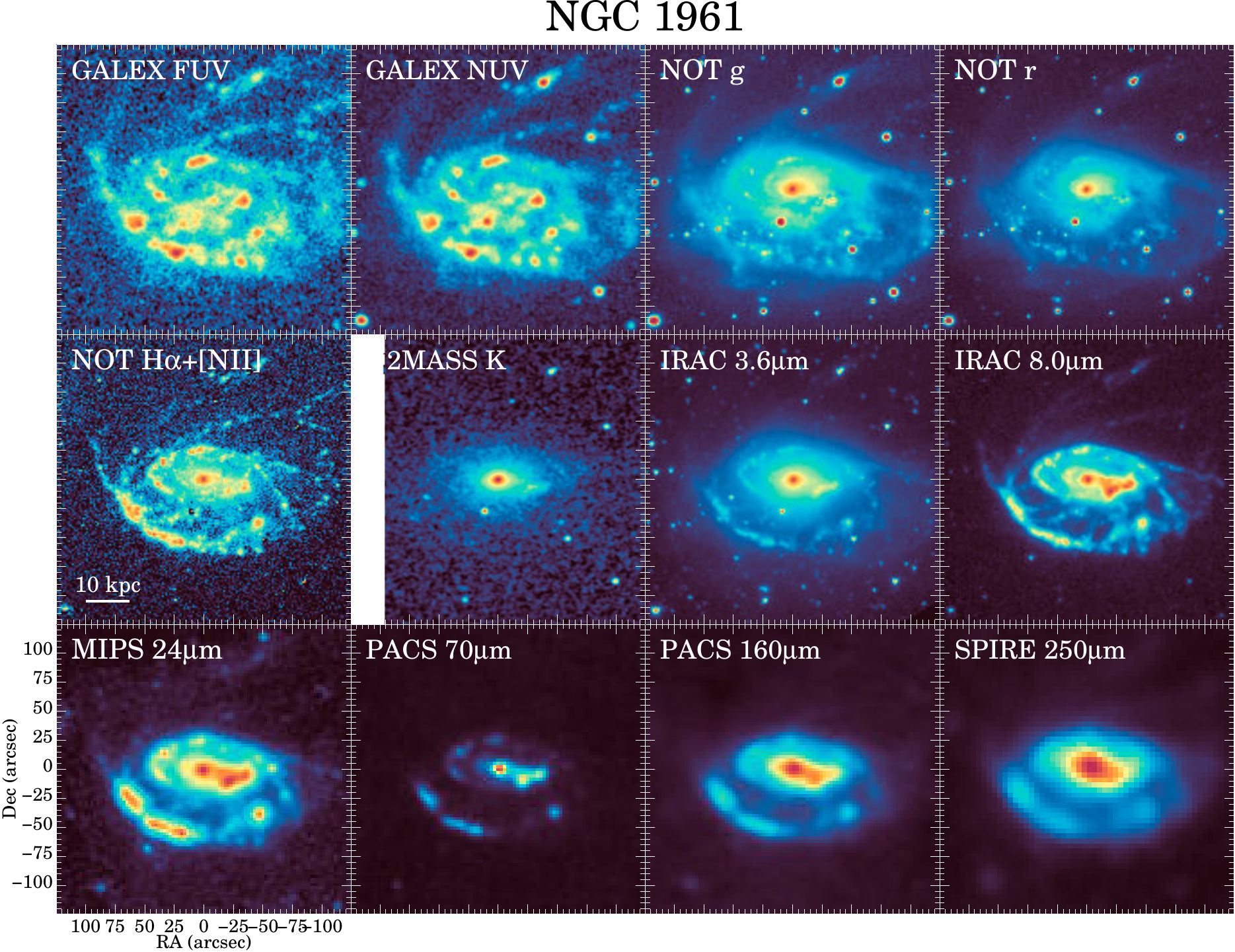}
\includegraphics[width=.8\textwidth]{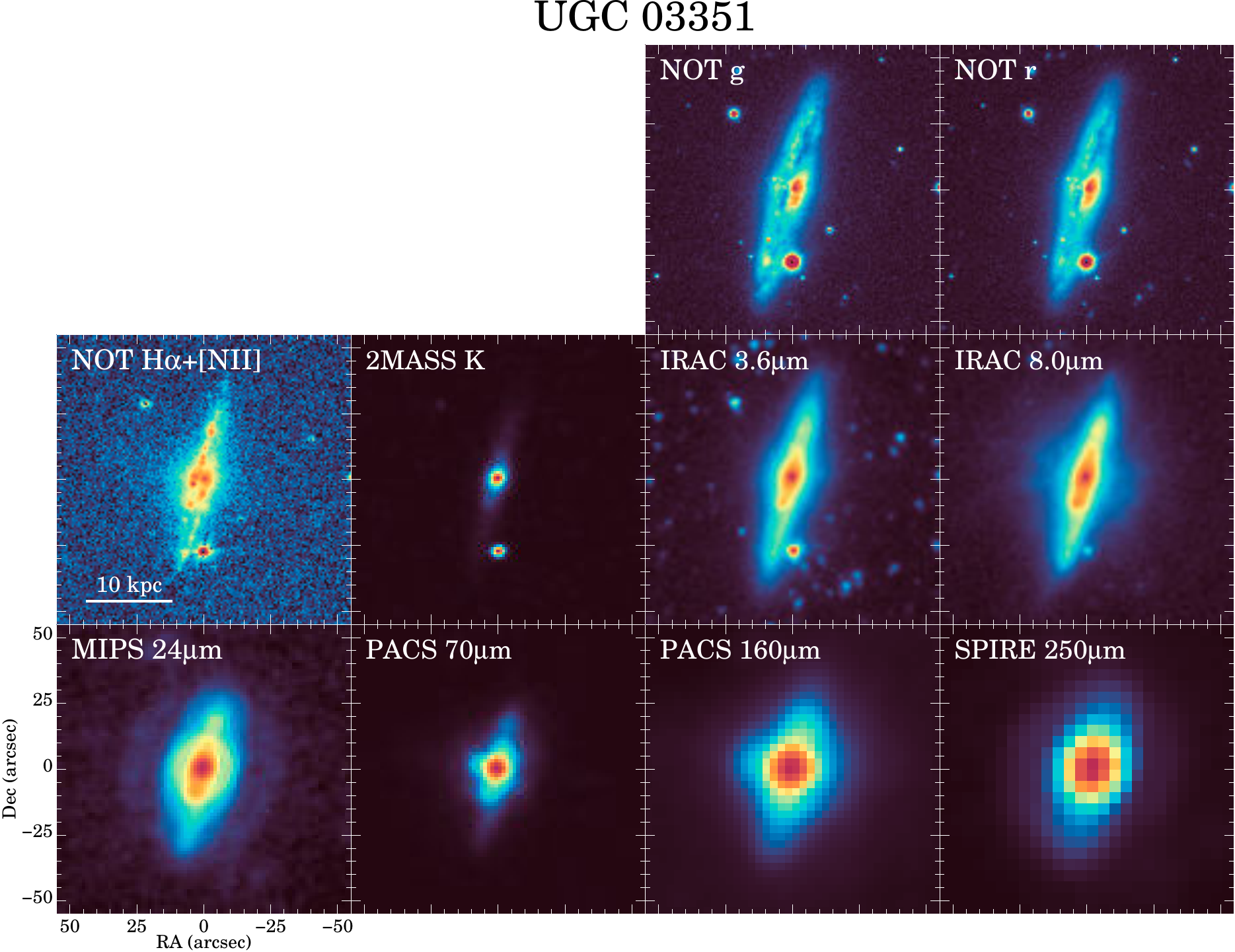}
\caption{Continued.}
\end{figure*}
\clearpage

\begin{figure*}[!h]
\centering
\addtocounter{figure}{-1}
\includegraphics[width=.8\textwidth]{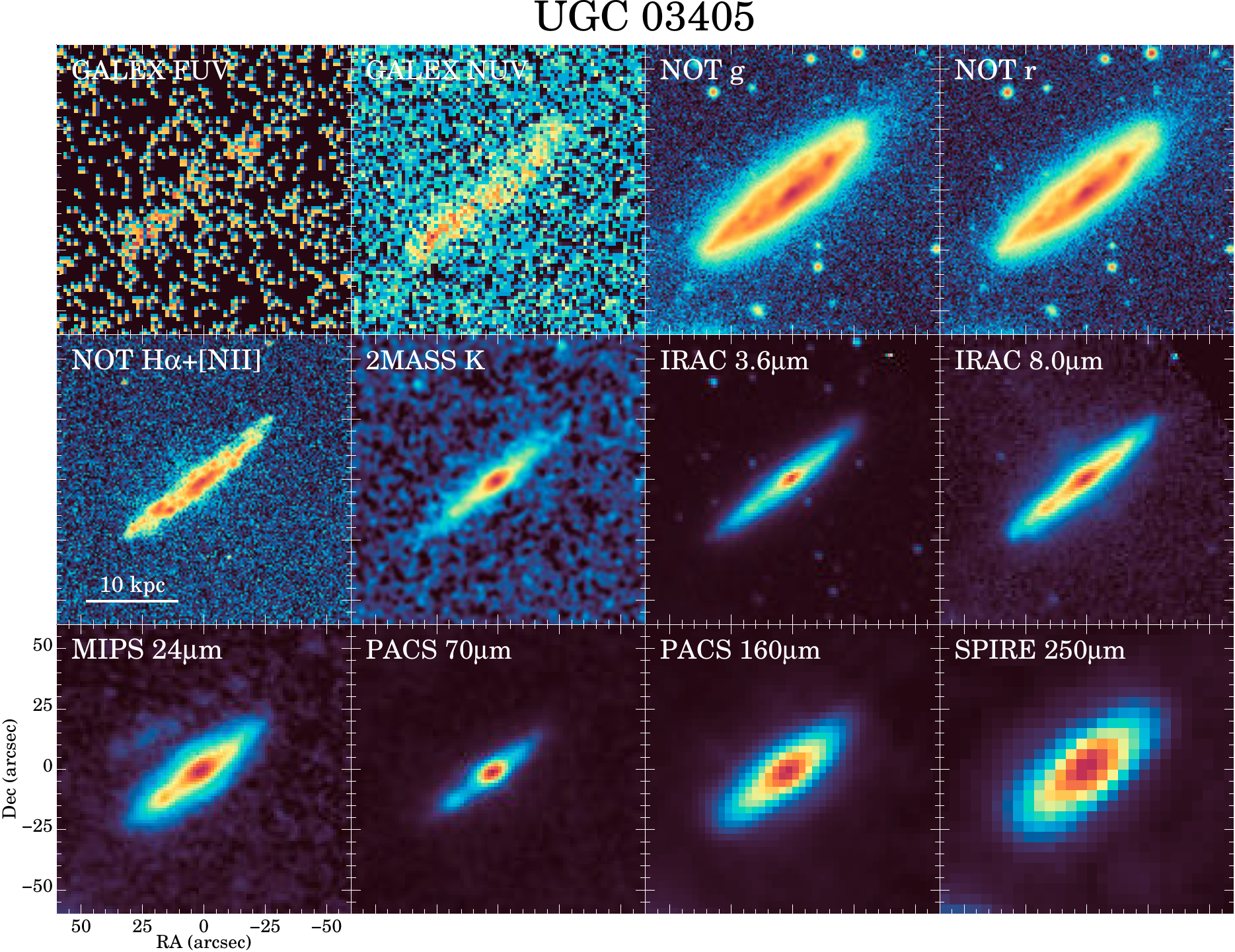}
\includegraphics[width=.8\textwidth]{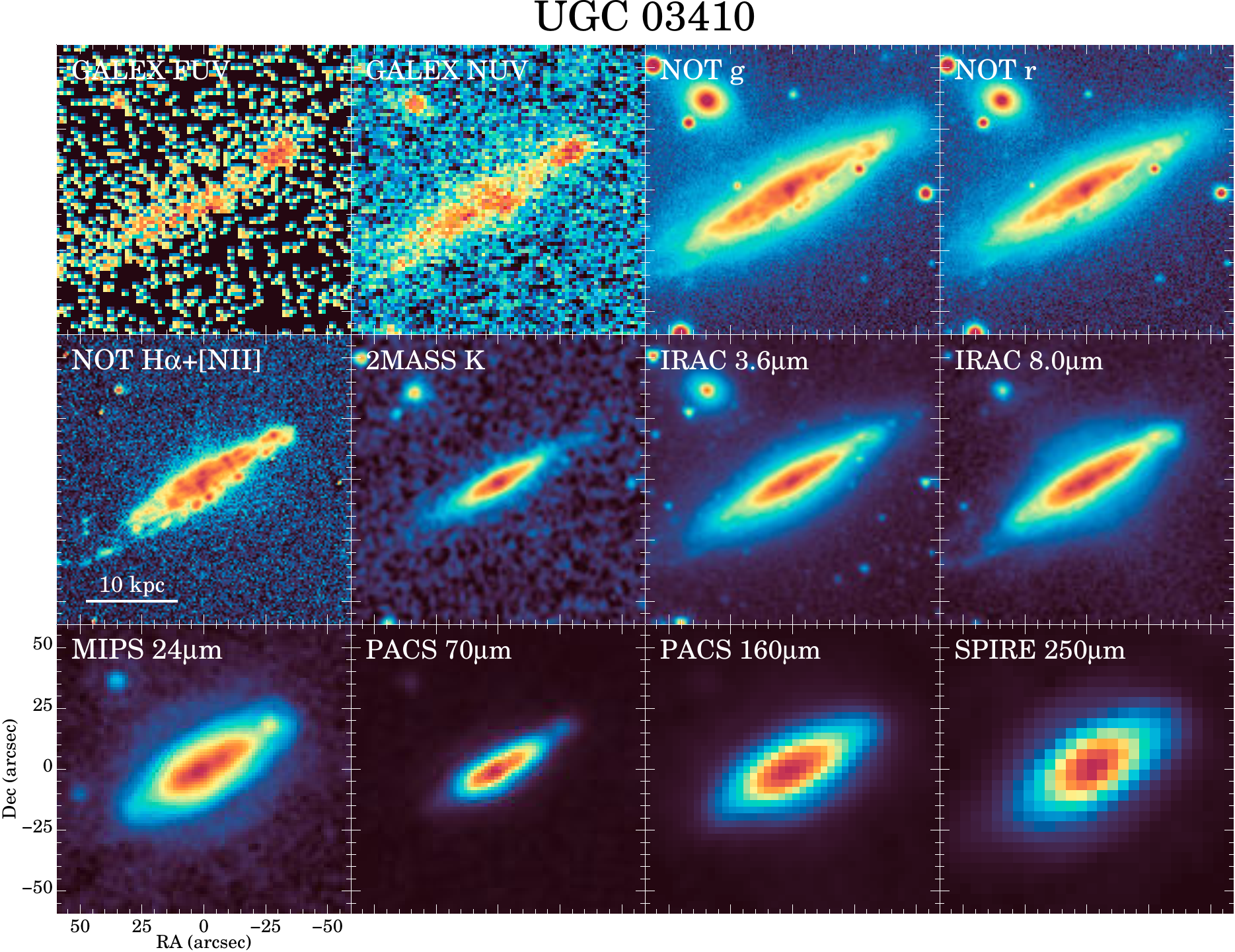}
\caption{Continued.}
\end{figure*}
\clearpage

\begin{figure*}[!h]
\centering
\addtocounter{figure}{-1}
\includegraphics[width=.8\textwidth]{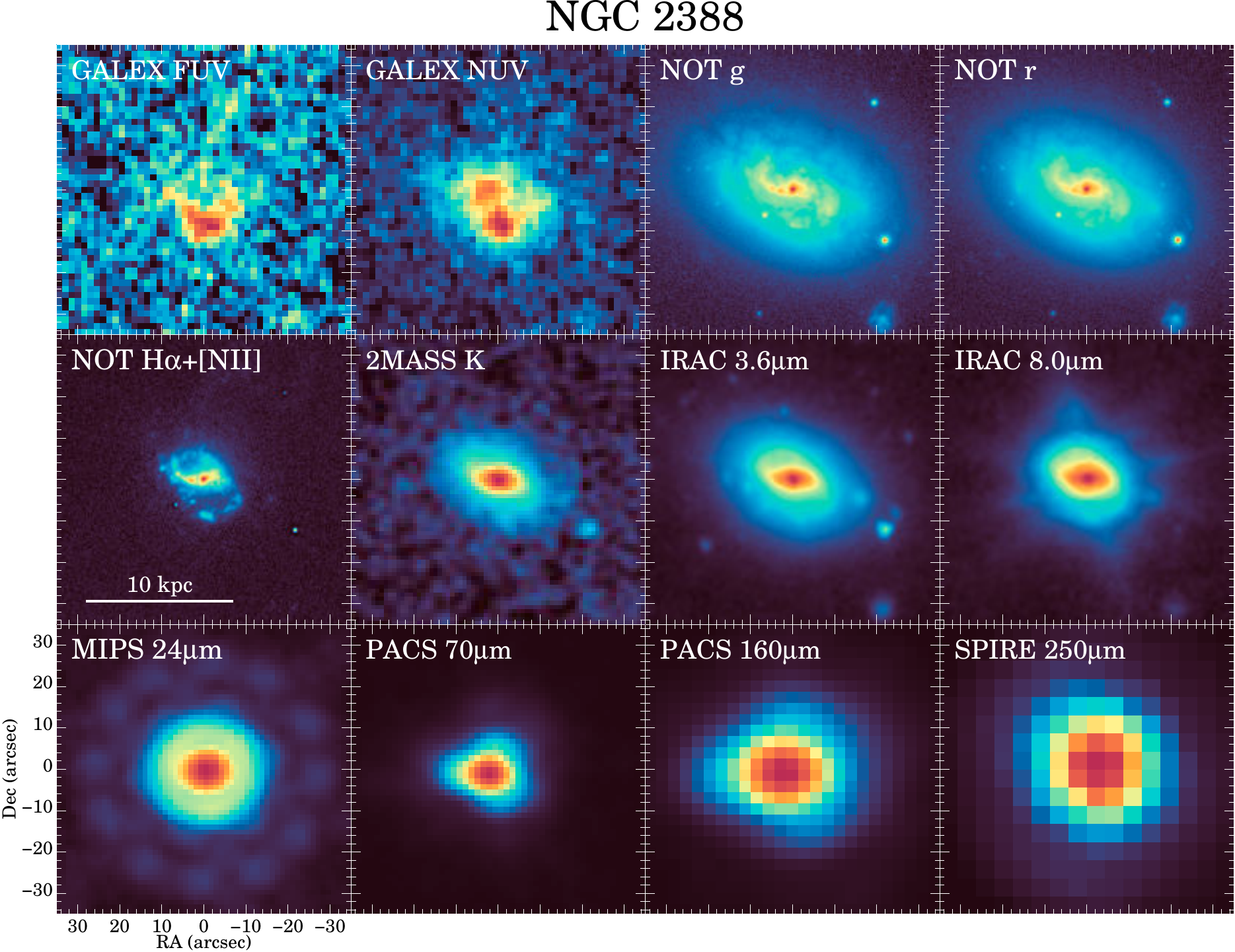}
\includegraphics[width=.8\textwidth]{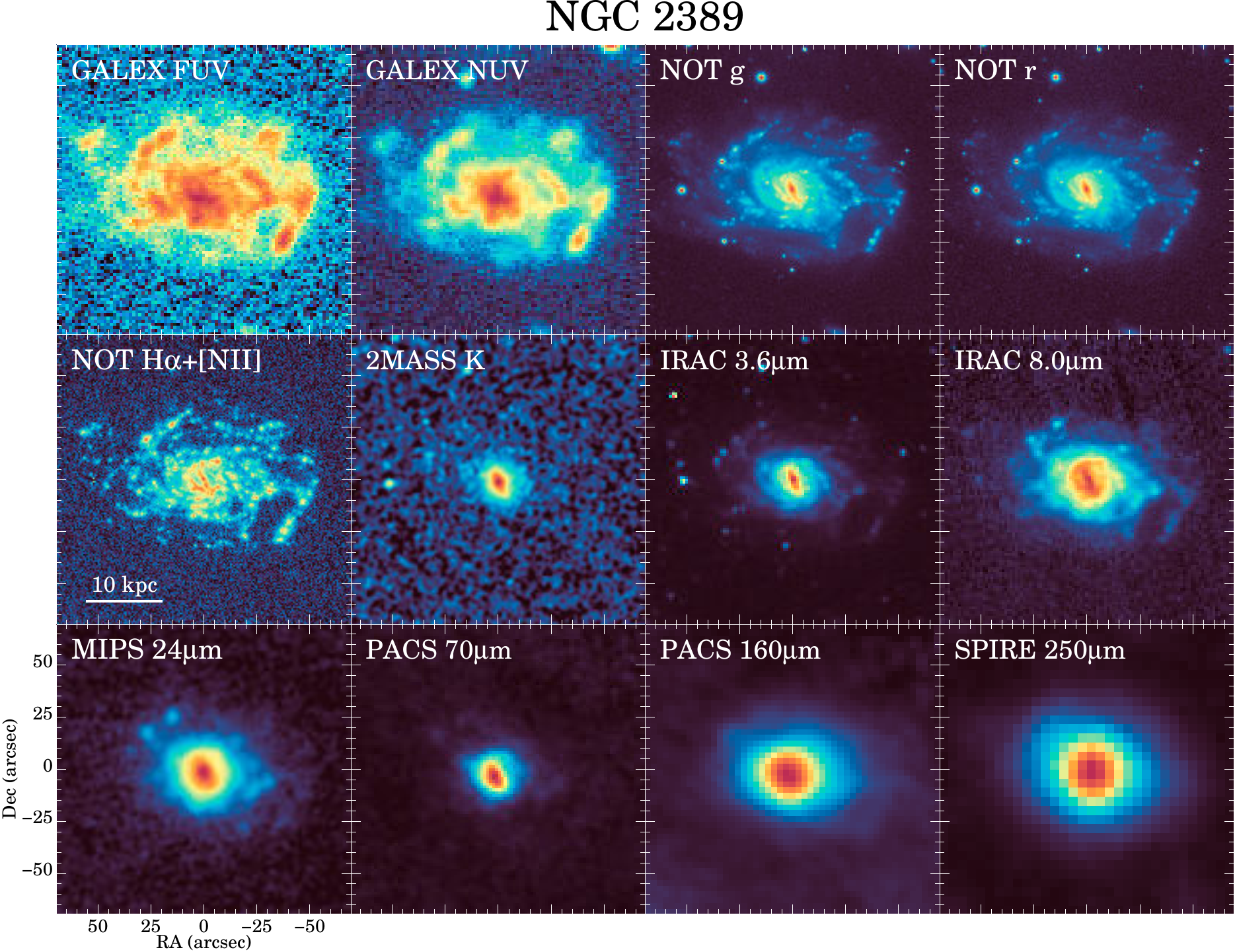}
\caption{Continued.}
\end{figure*}
\clearpage

\begin{figure*}[!h]
\centering
\addtocounter{figure}{-1}
\includegraphics[width=.8\textwidth]{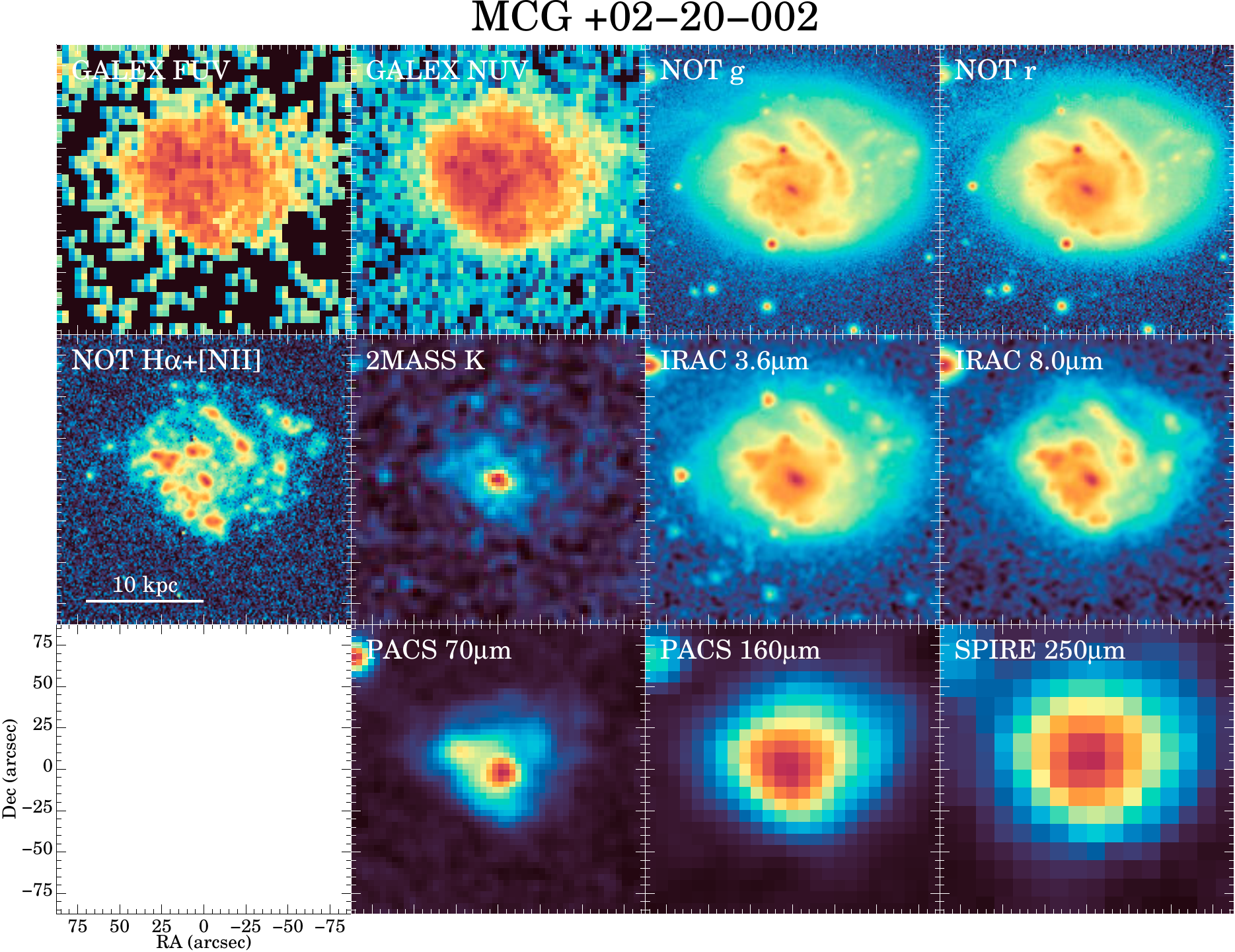}
\includegraphics[width=.8\textwidth]{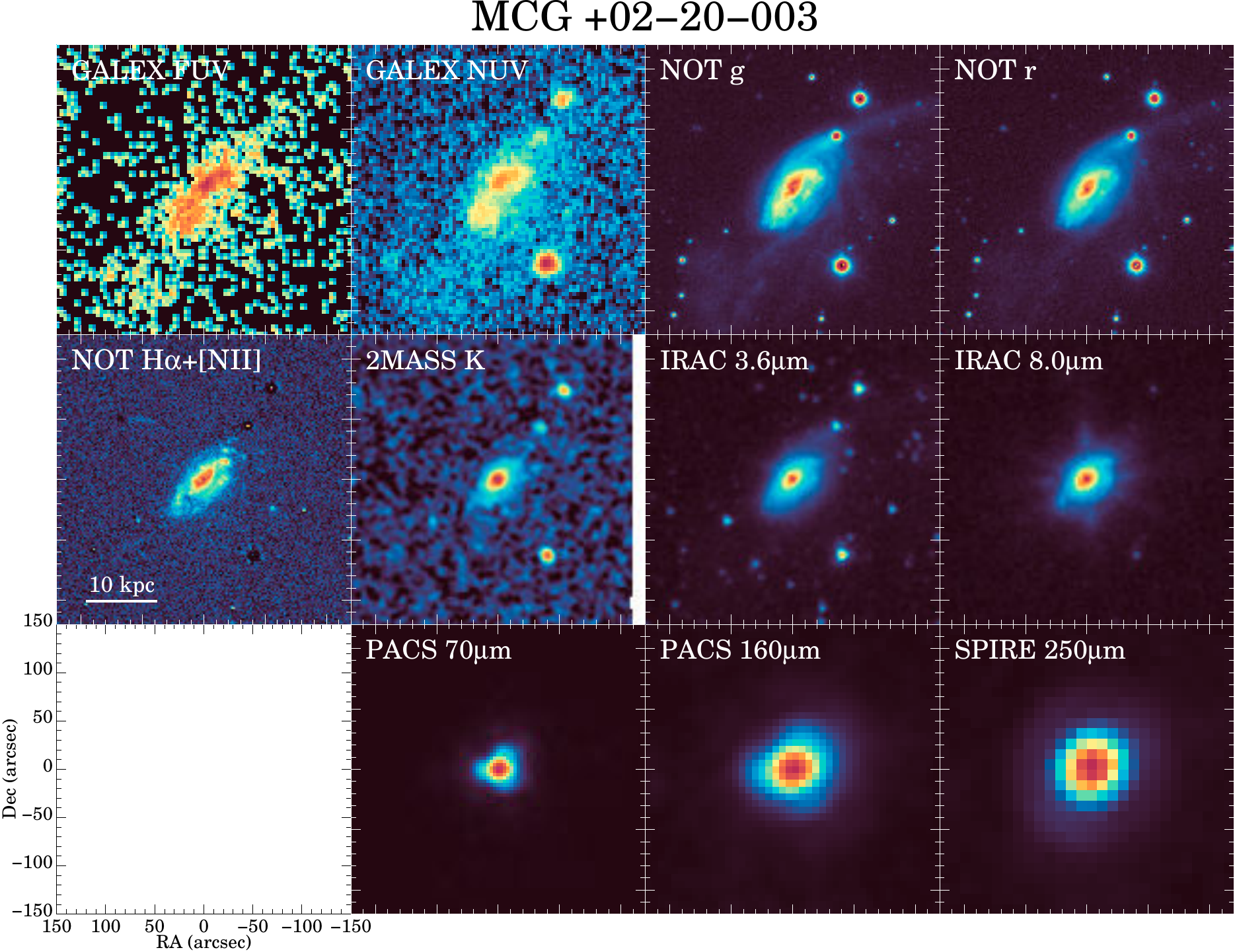}
\caption{Continued.}
\end{figure*}
\clearpage

\begin{figure*}[!h]
\centering
\addtocounter{figure}{-1}
\includegraphics[width=.8\textwidth]{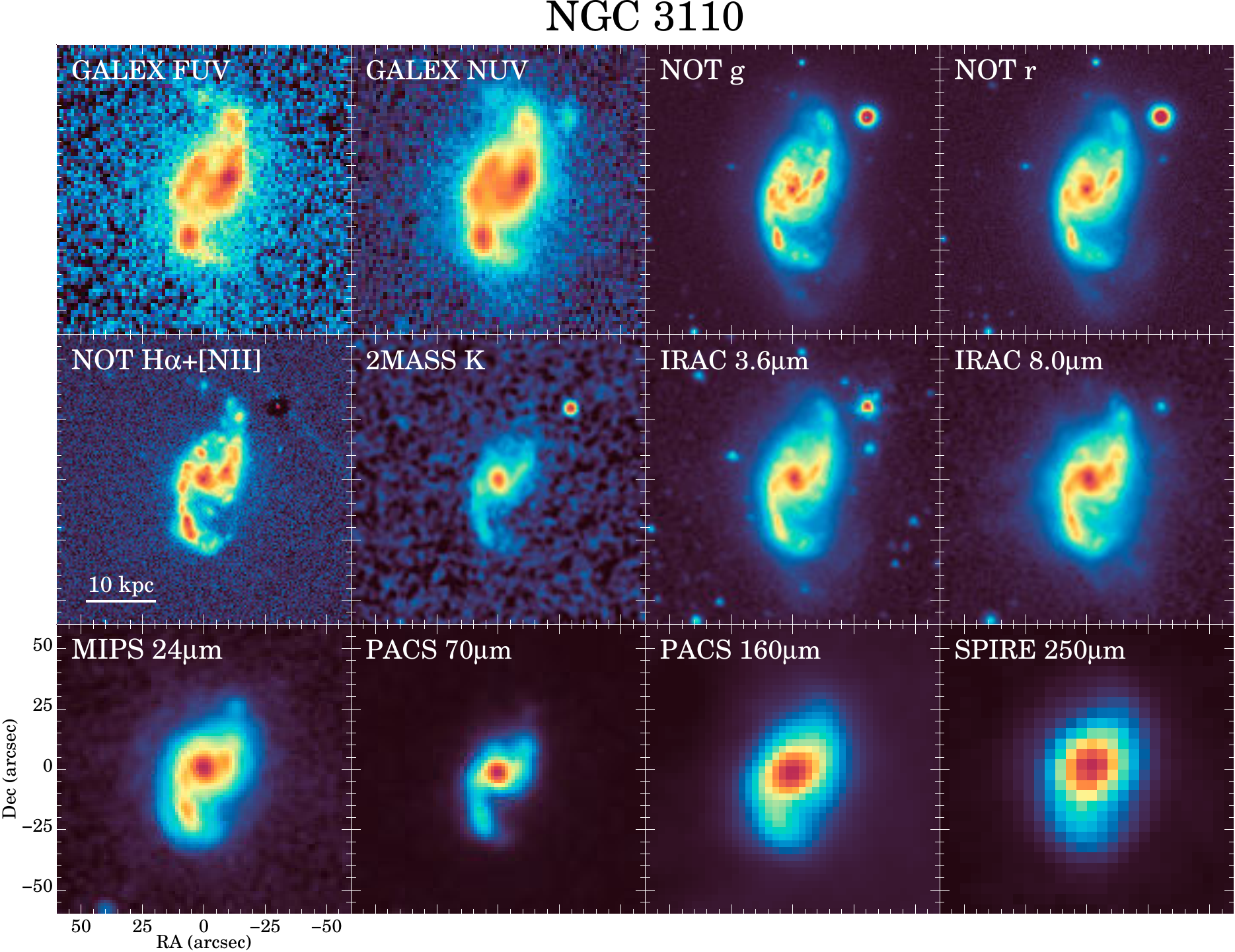}
\includegraphics[width=.8\textwidth]{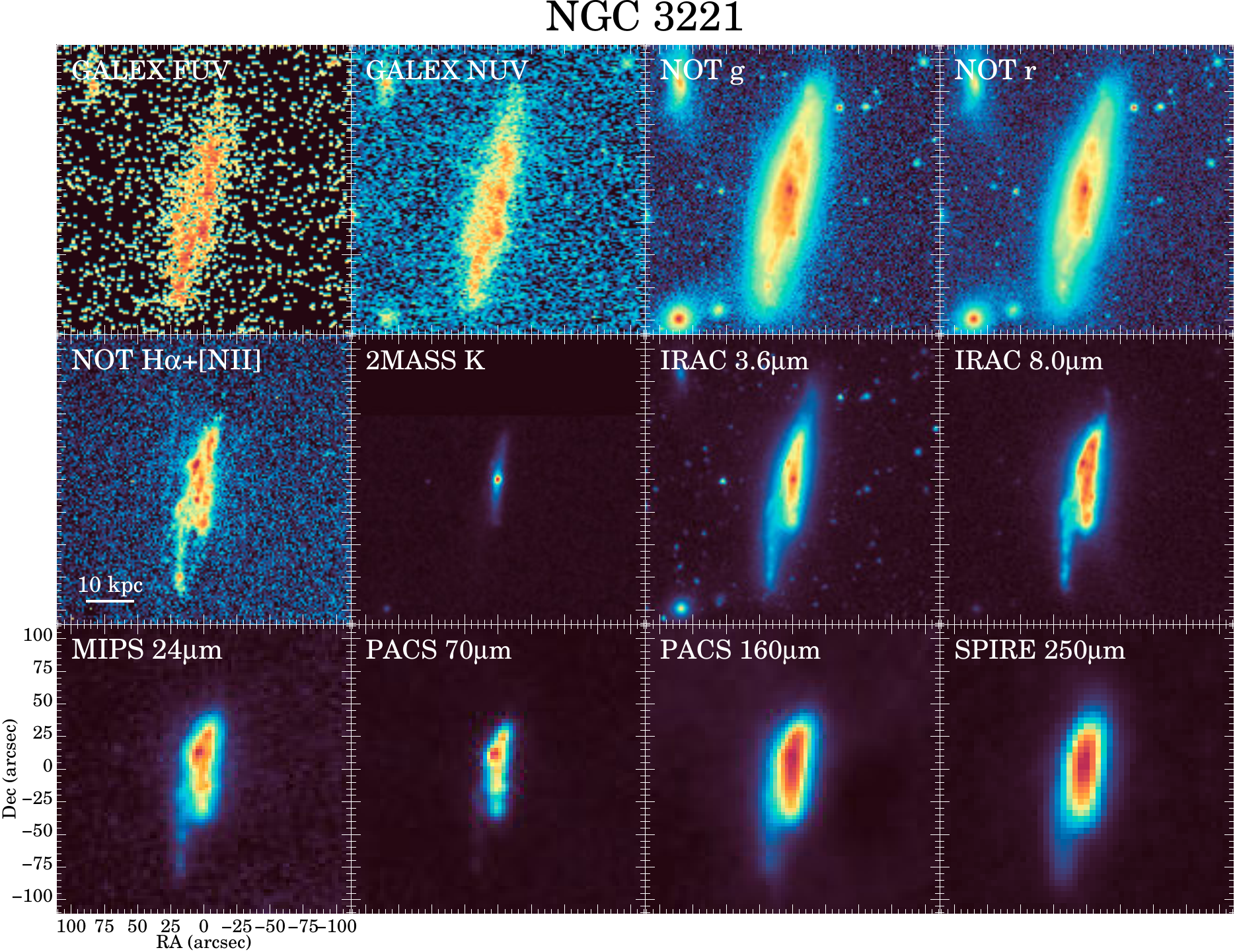}
\caption{Continued.}
\end{figure*}
\clearpage

\begin{figure*}[!h]
\centering
\addtocounter{figure}{-1}
\includegraphics[width=.8\textwidth]{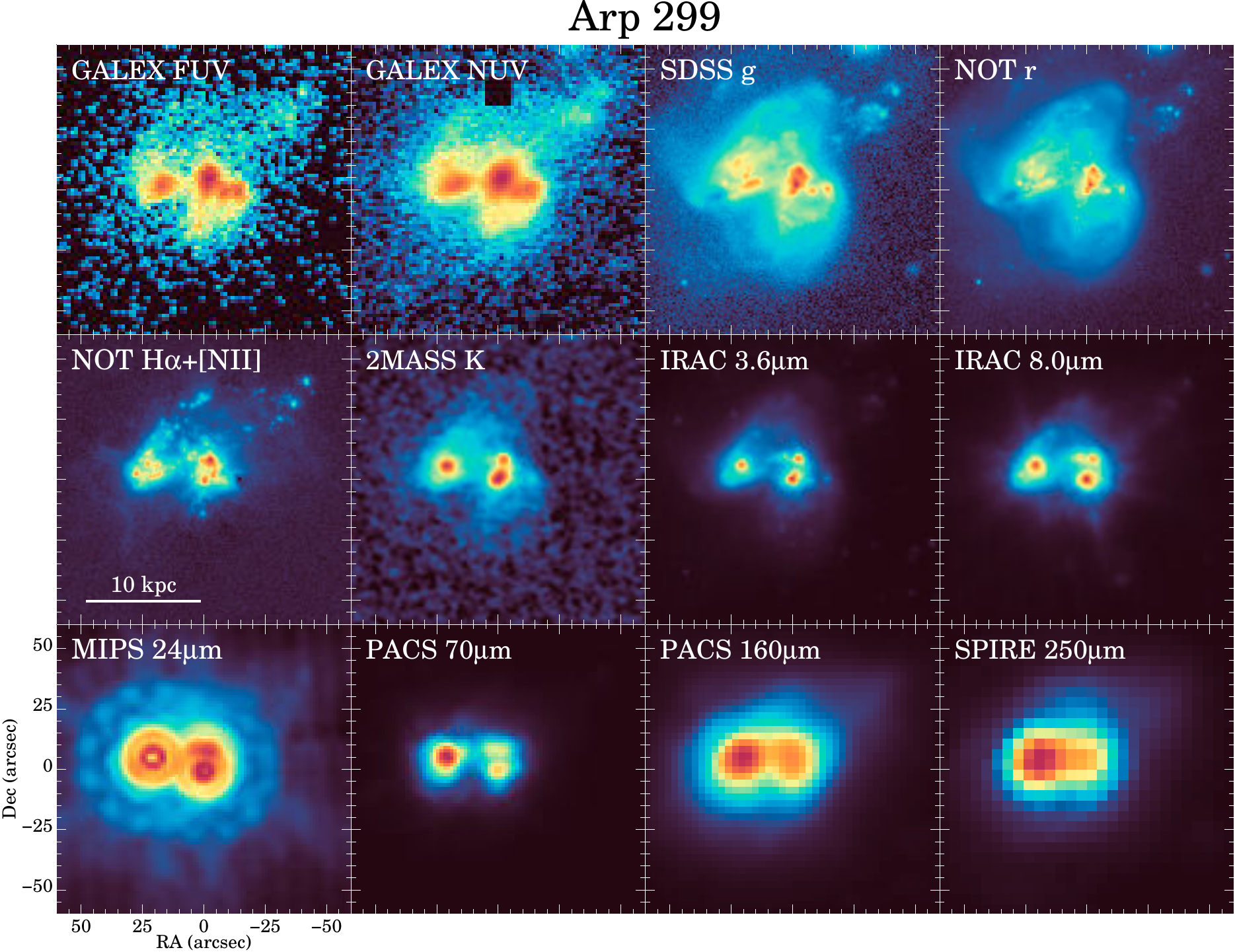}
\includegraphics[width=.8\textwidth]{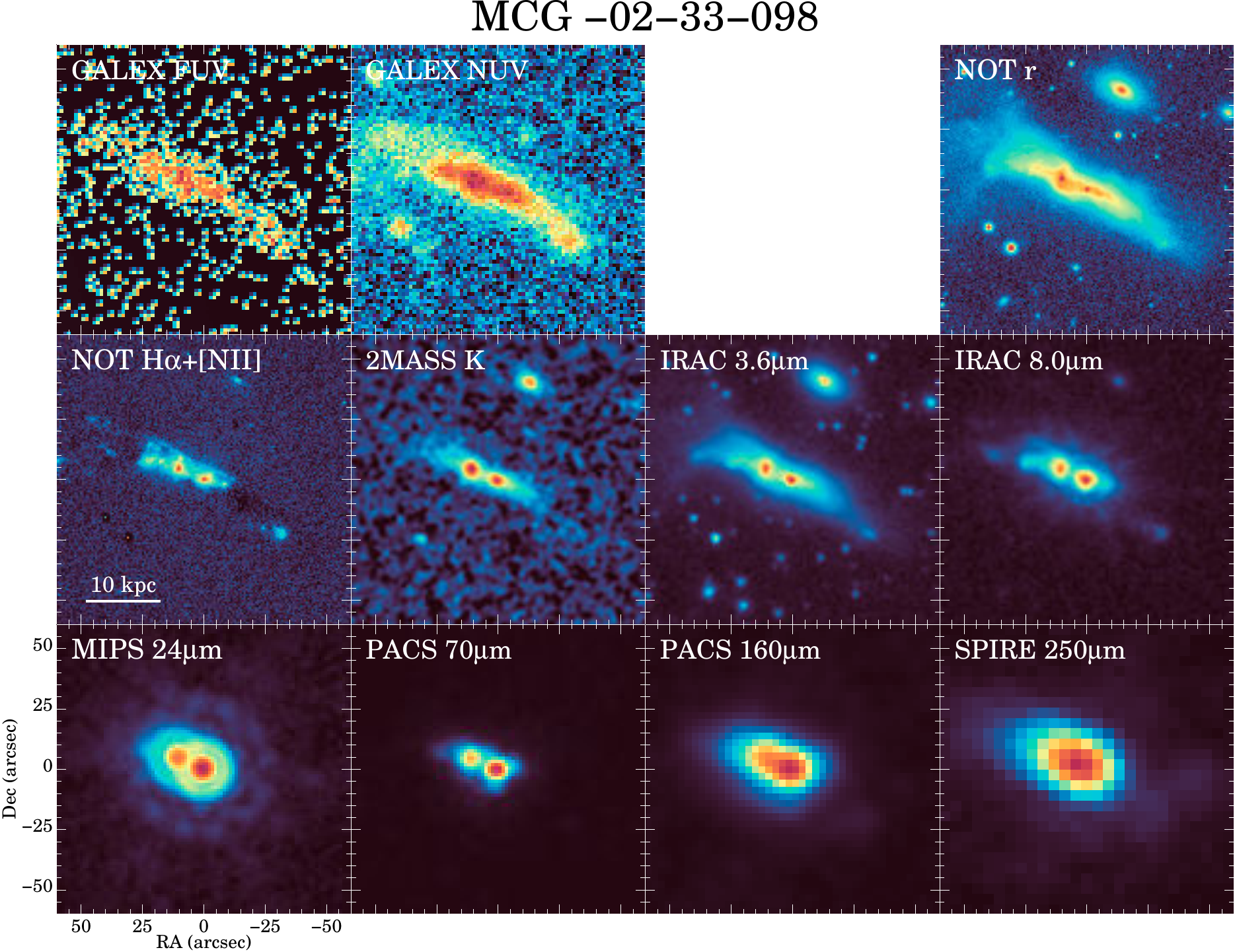}
\caption{Continued.}
\end{figure*}
\clearpage

\begin{figure*}[!h]
\centering
\addtocounter{figure}{-1}
\includegraphics[width=.8\textwidth]{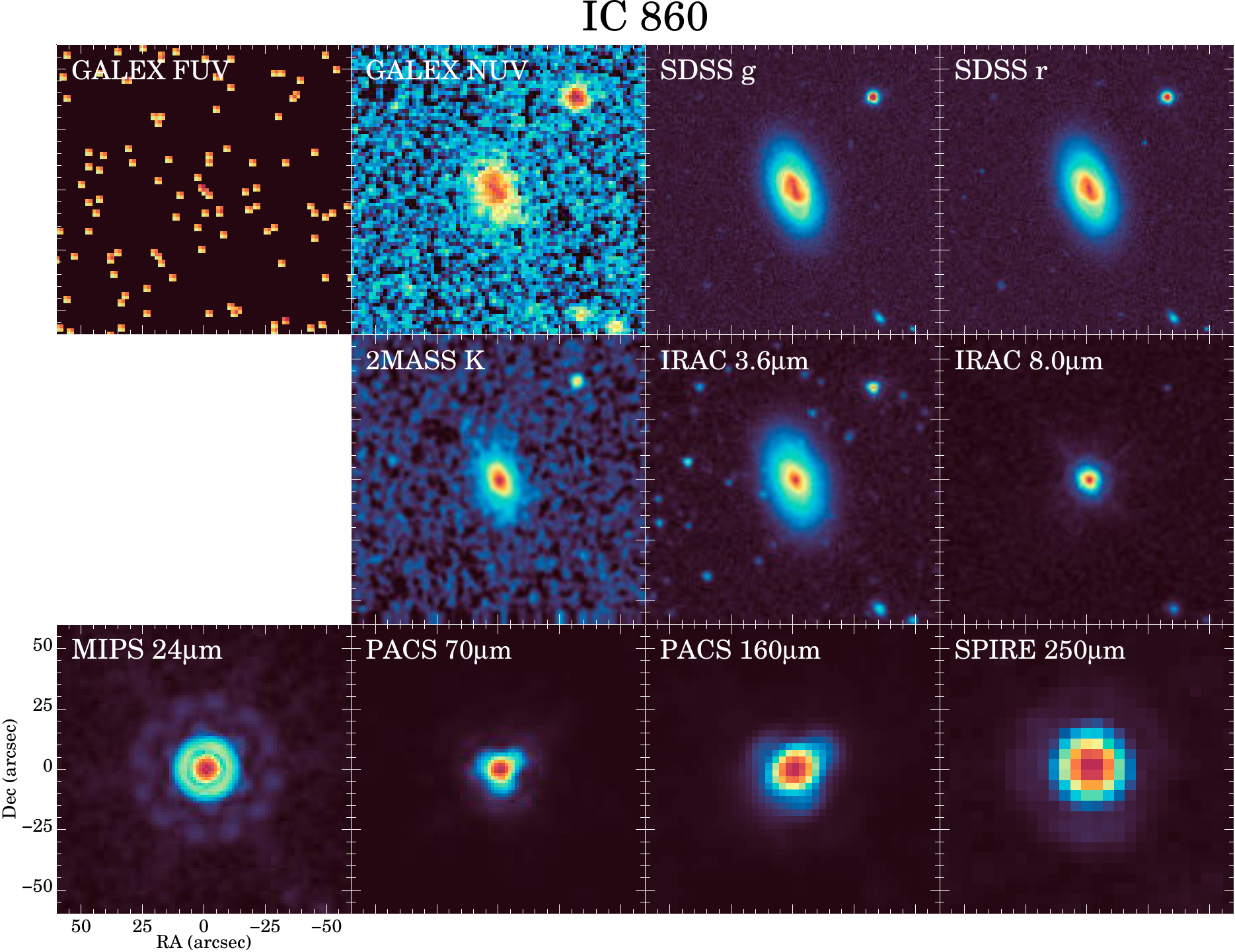}
\includegraphics[width=.8\textwidth]{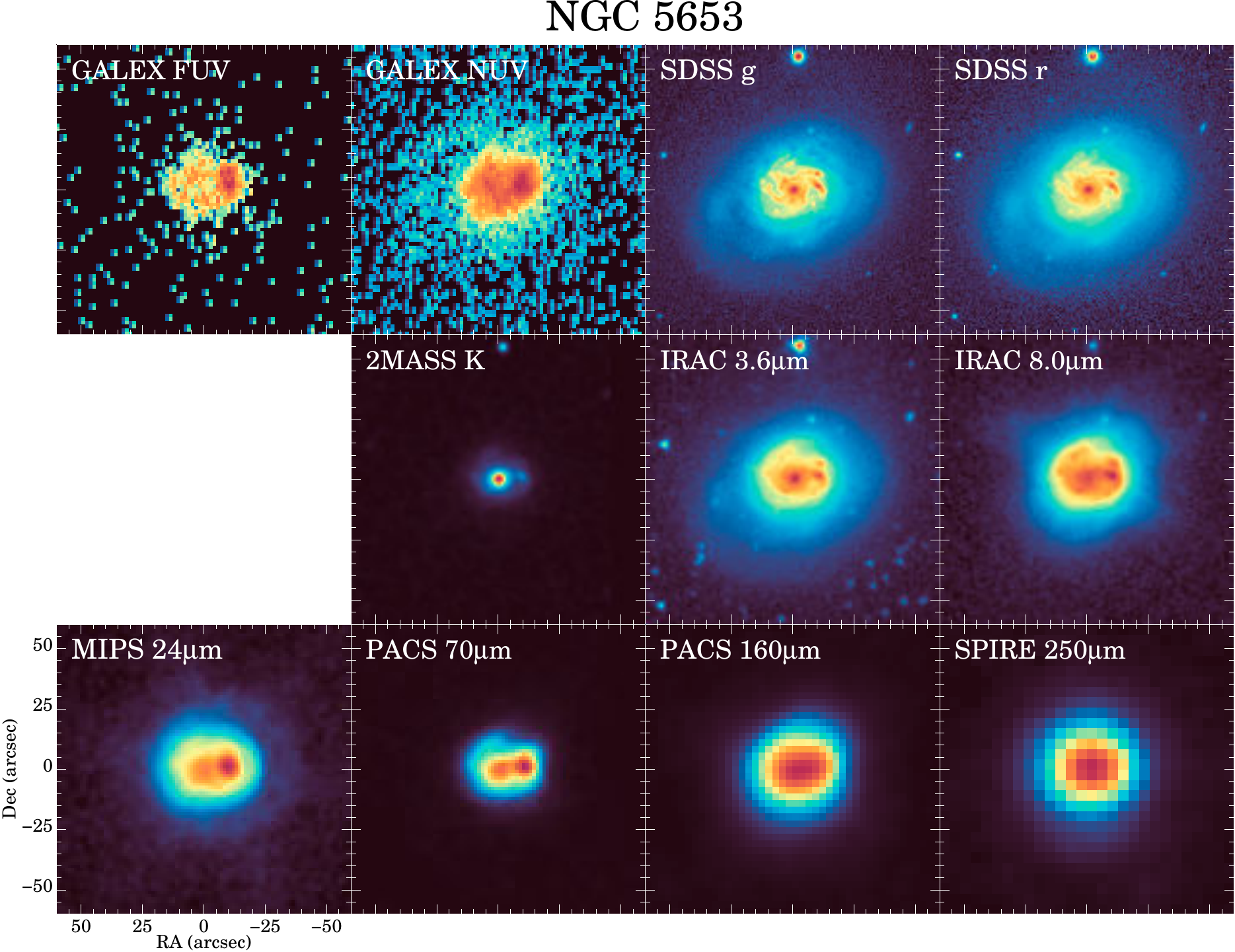}
\caption{Continued.}
\end{figure*}
\clearpage

\begin{figure*}[!h]
\centering
\addtocounter{figure}{-1}
\includegraphics[width=.8\textwidth]{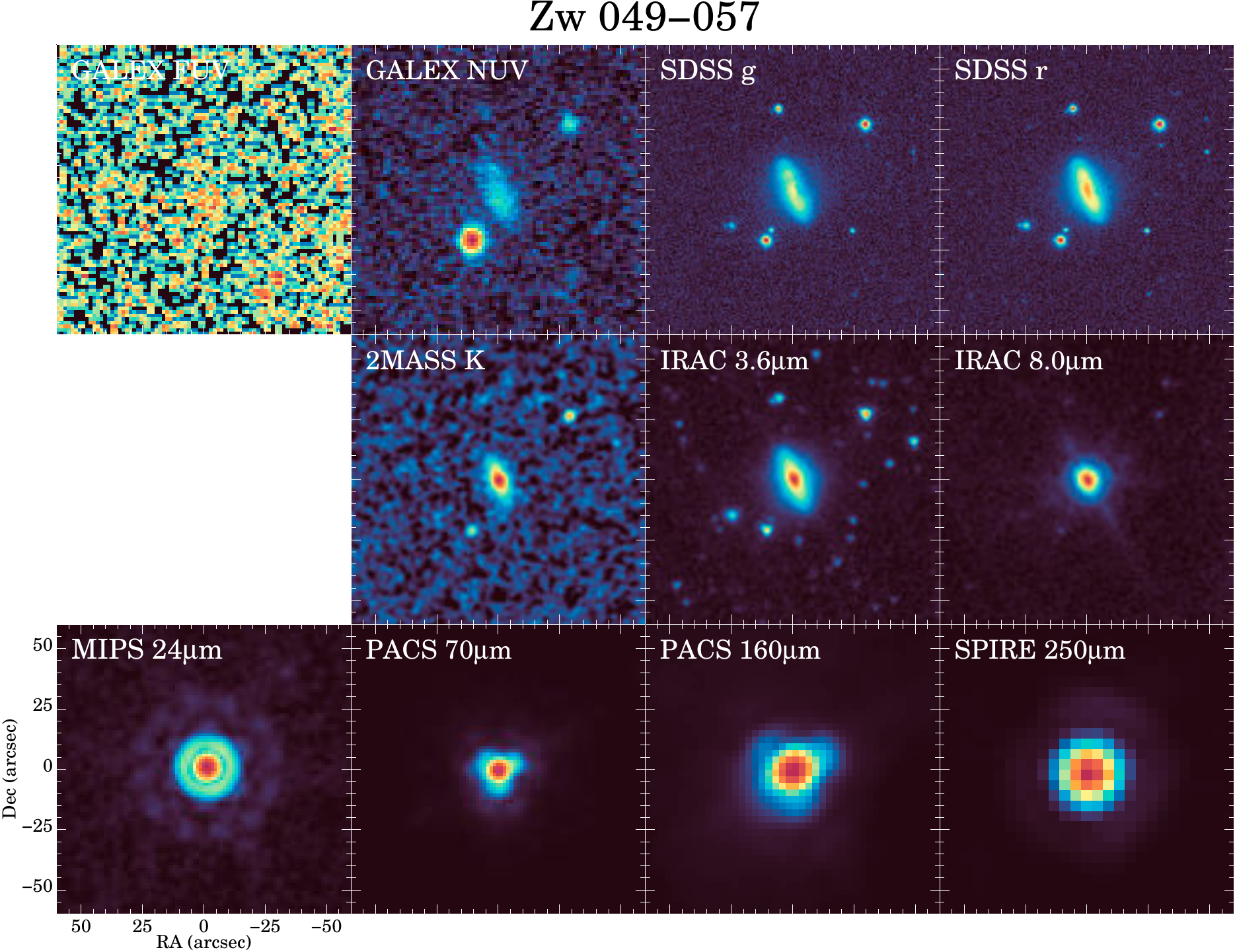}
\includegraphics[width=.8\textwidth]{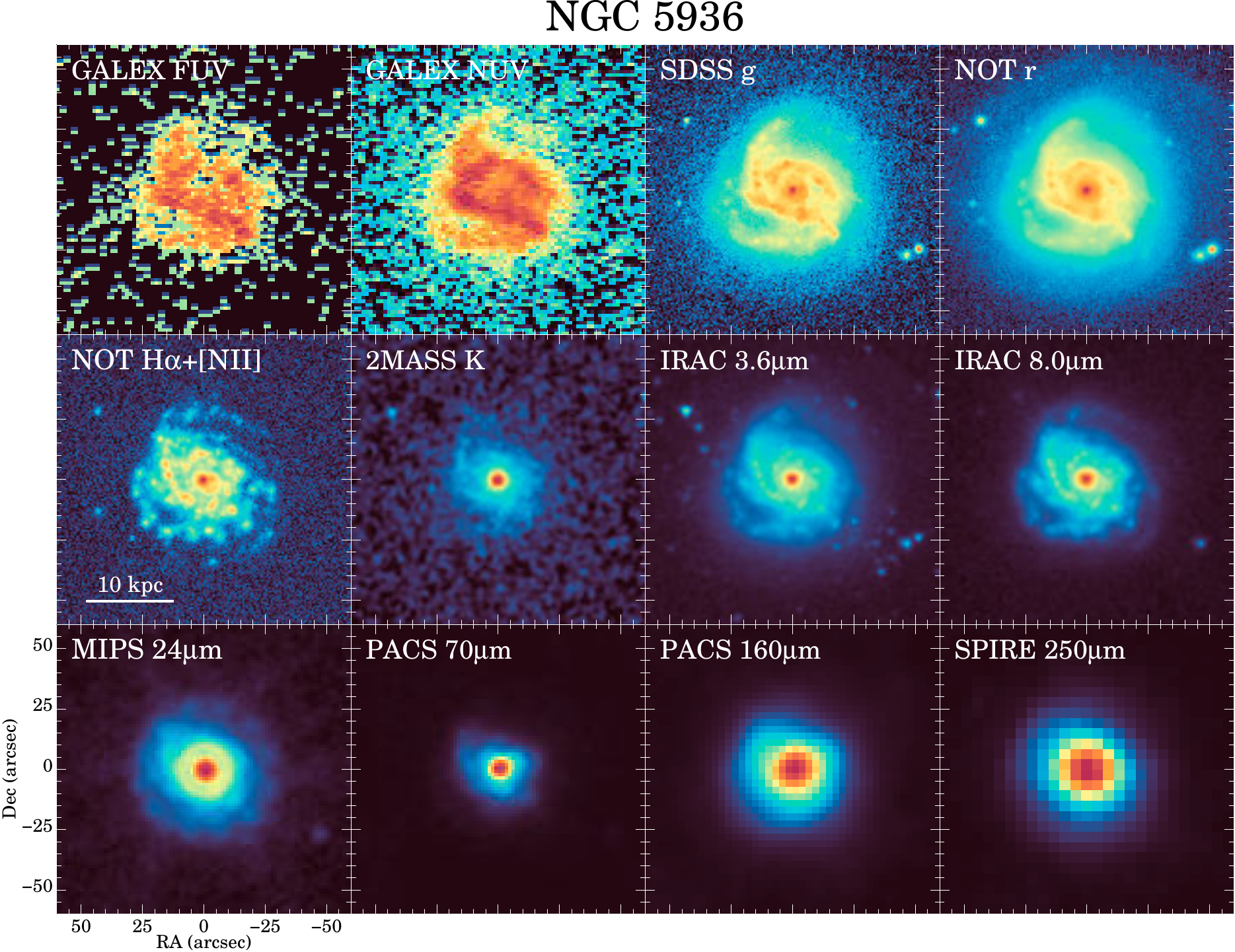}
\caption{Continued.}
\end{figure*}
\clearpage

\begin{figure*}[!h]
\centering
\addtocounter{figure}{-1}
\includegraphics[width=.8\textwidth]{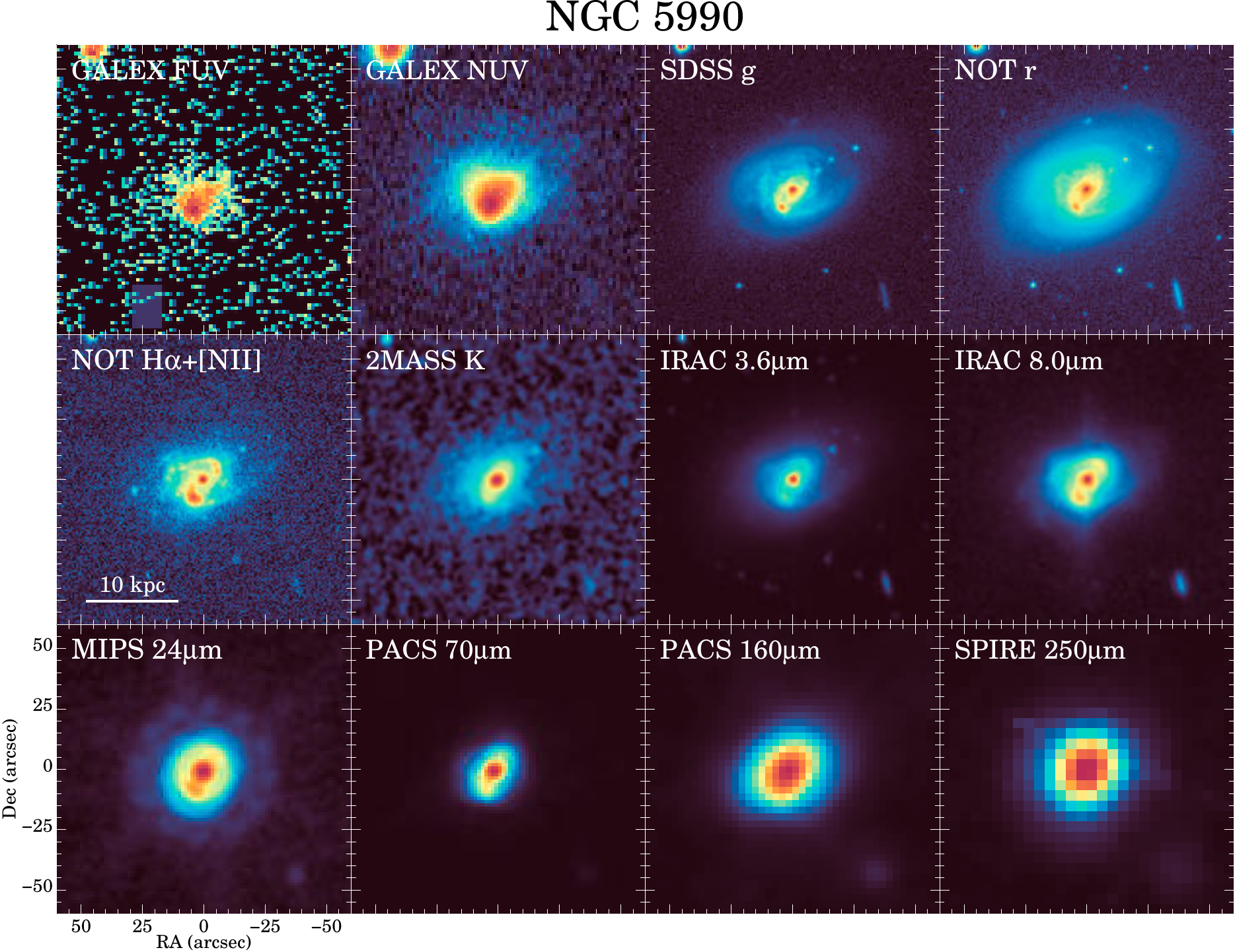}
\includegraphics[width=.8\textwidth]{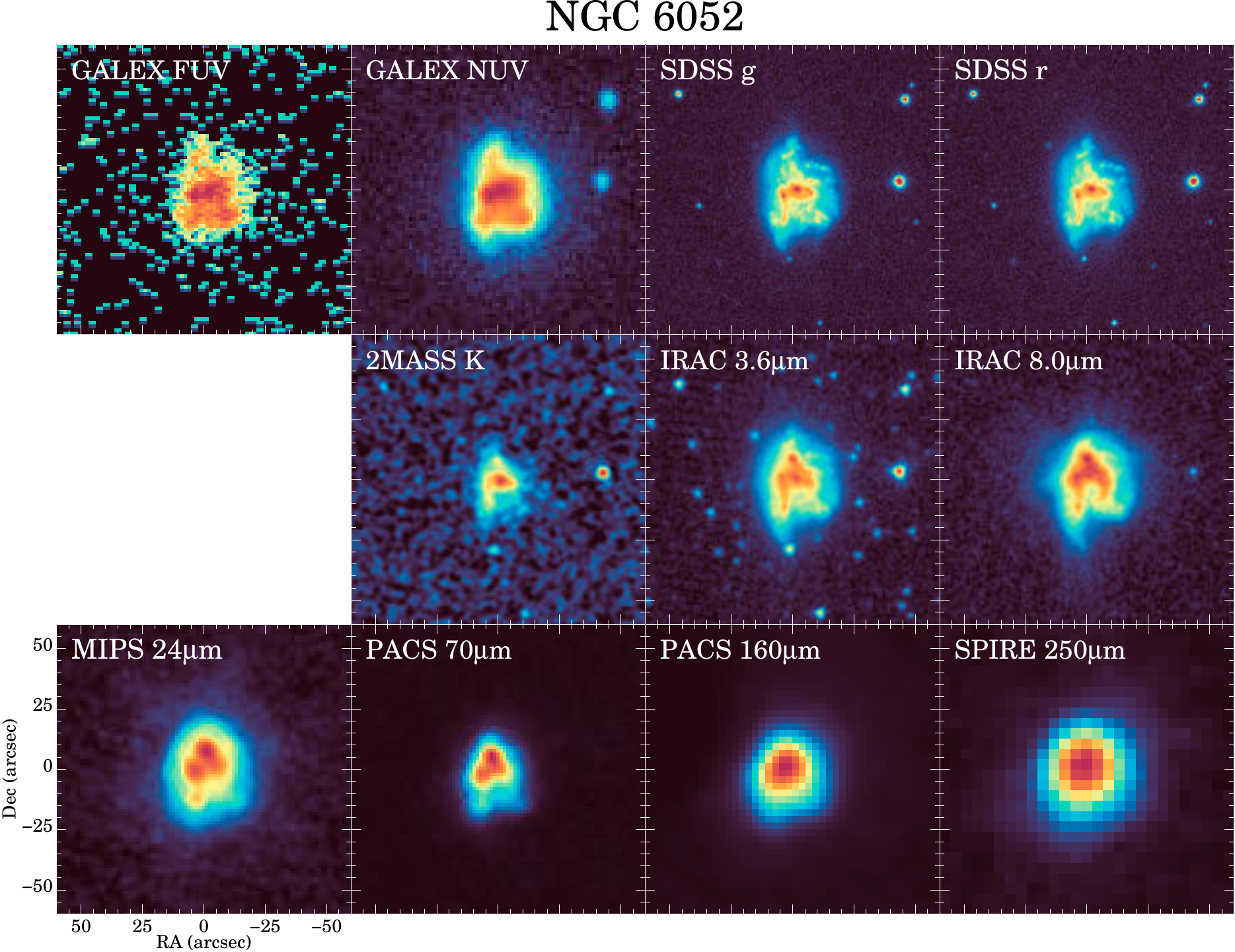}
\caption{Continued.}
\end{figure*}
\clearpage

\begin{figure*}[!h]
\centering
\addtocounter{figure}{-1}
\includegraphics[width=.8\textwidth]{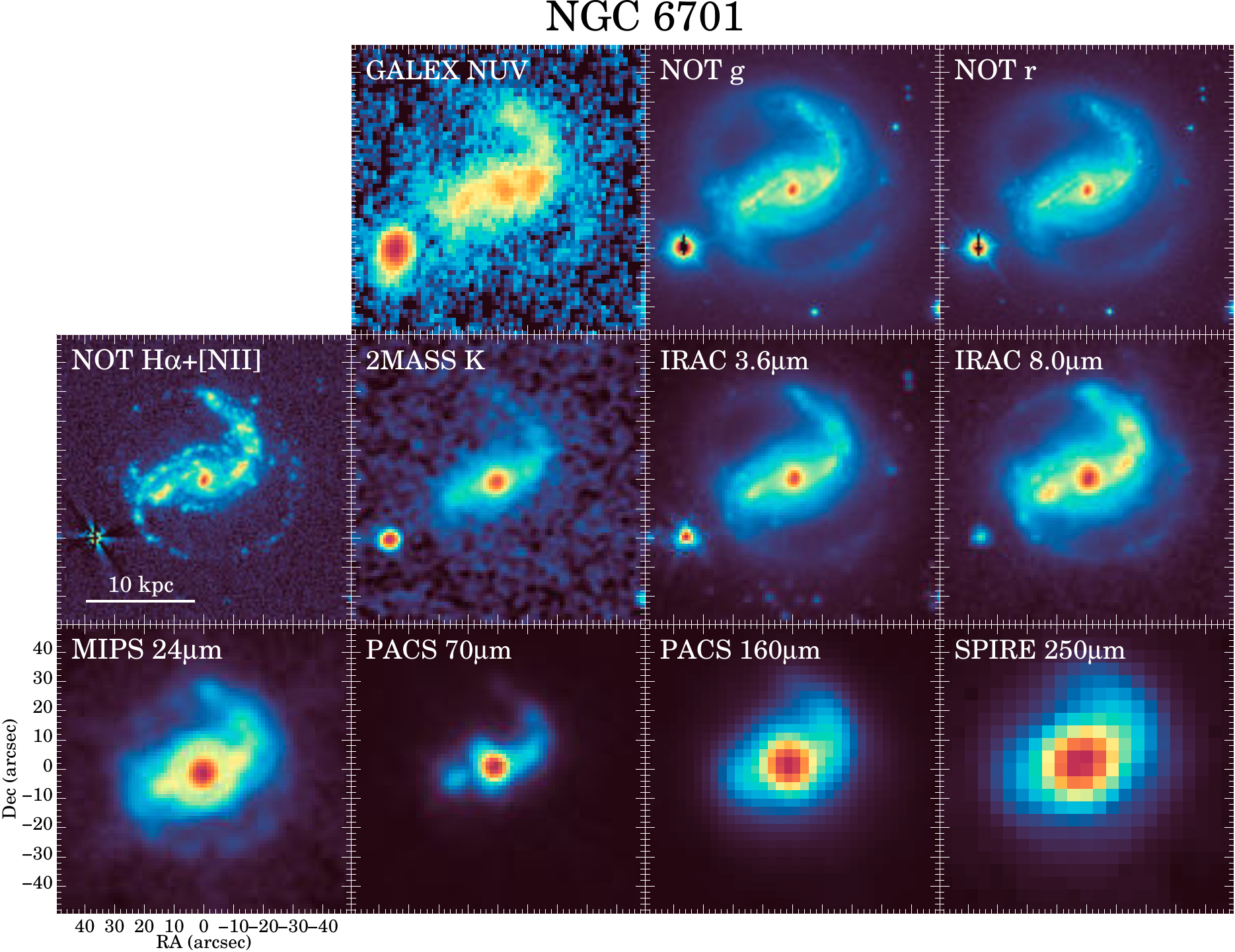}
\includegraphics[width=.8\textwidth]{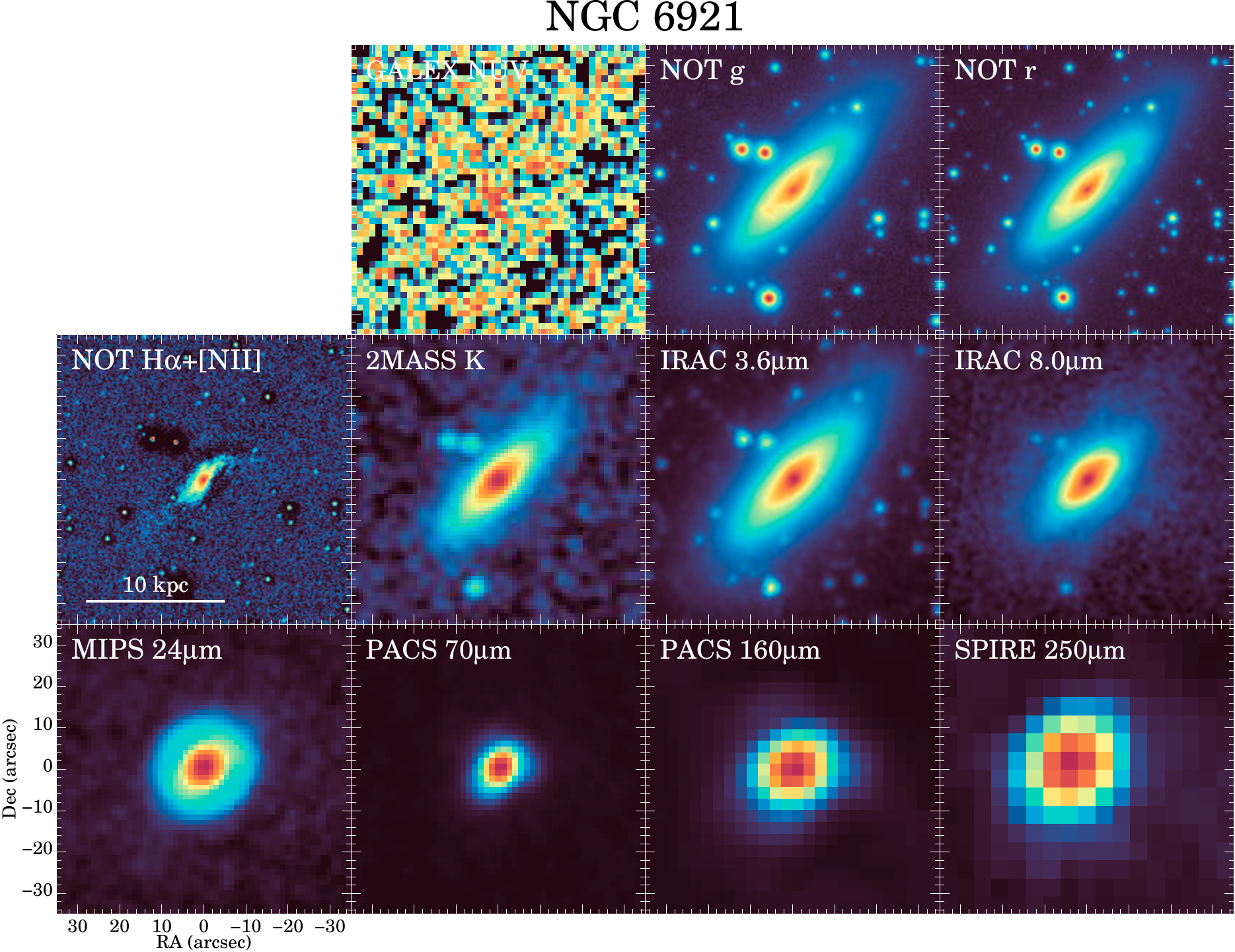}
\caption{Continued.}
\end{figure*}
\clearpage

\begin{figure*}[!h]
\centering
\addtocounter{figure}{-1}
\includegraphics[width=.8\textwidth]{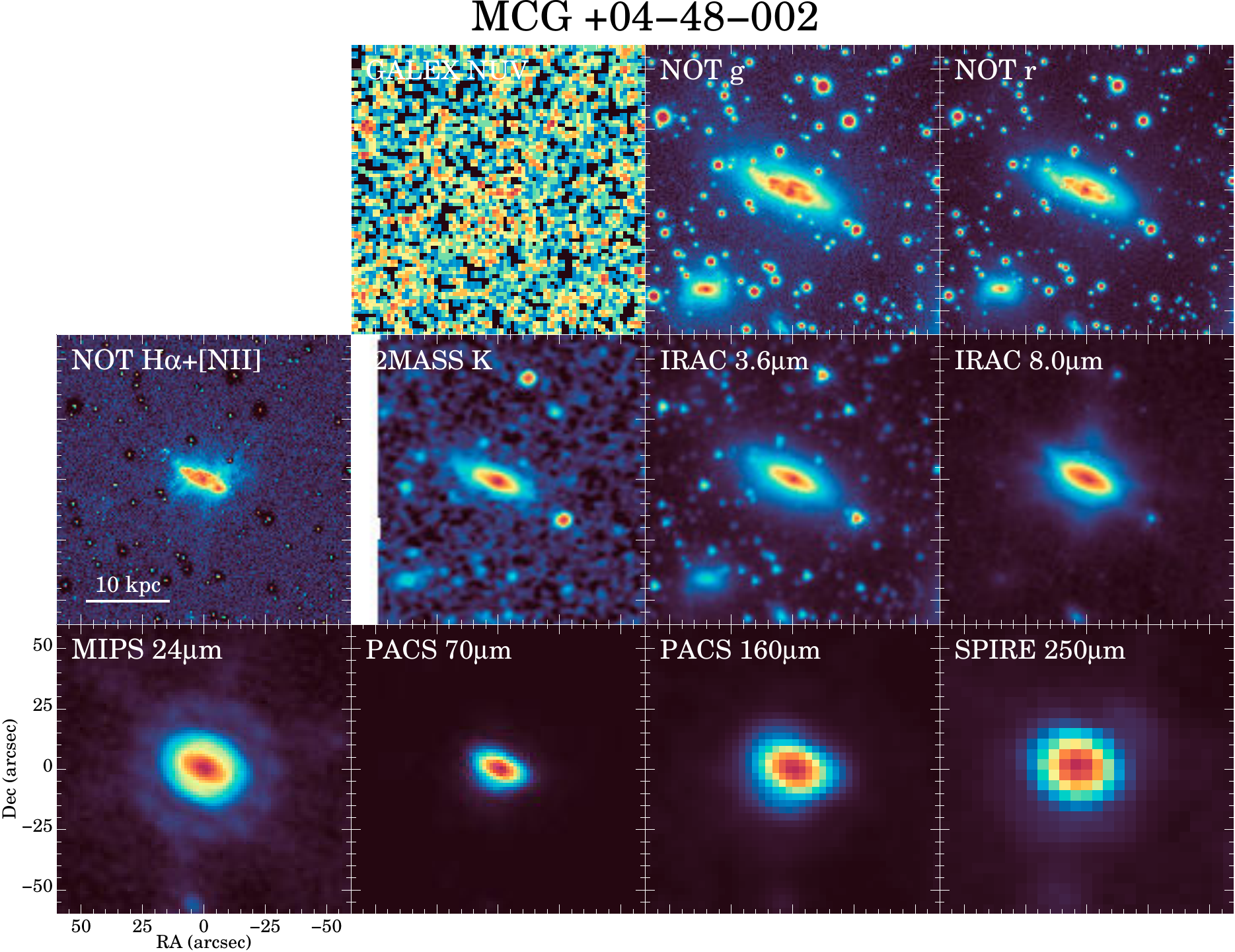}
\includegraphics[width=.8\textwidth]{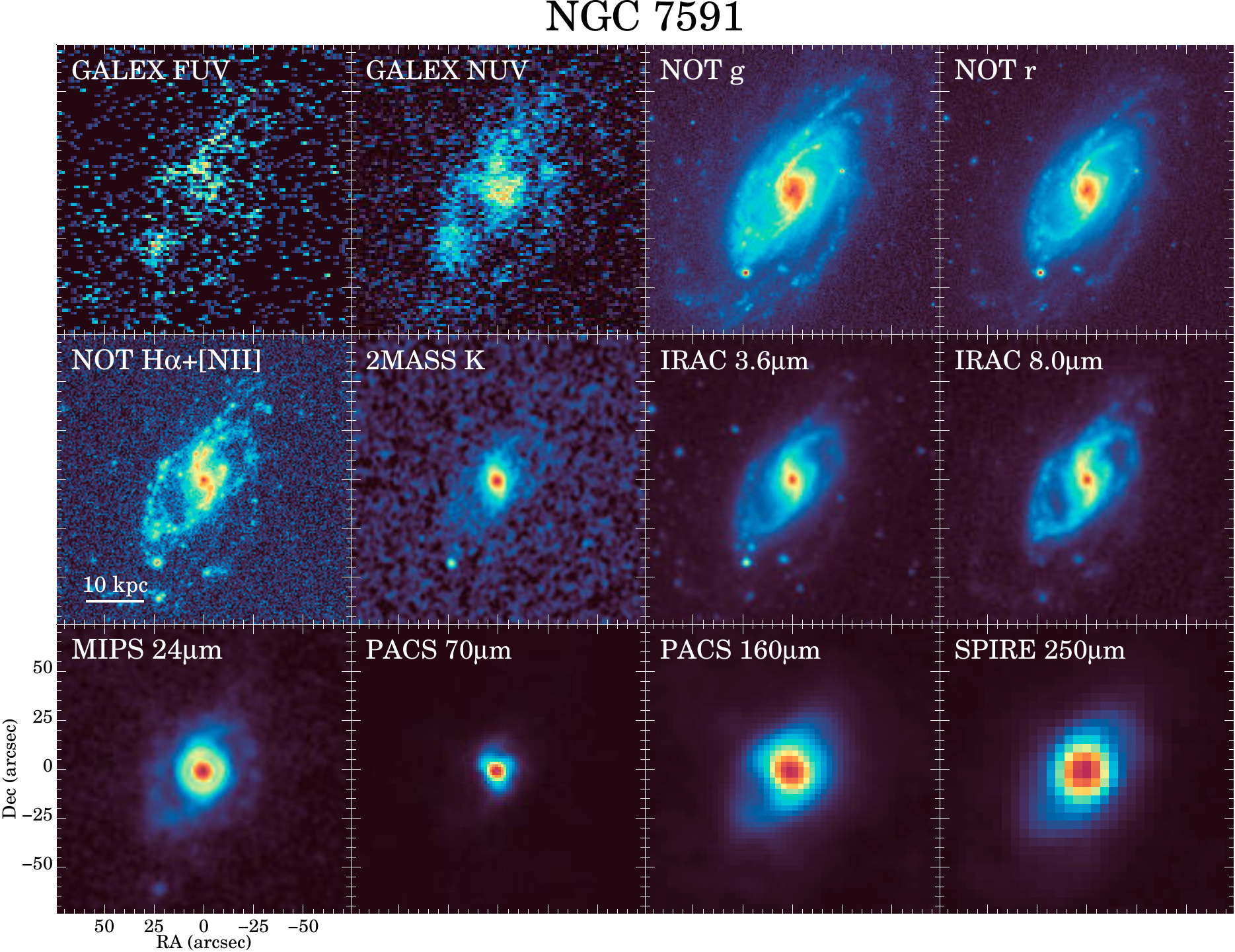}
\caption{Continued.}
\end{figure*}
\clearpage

\begin{figure*}[!h]
\centering
\addtocounter{figure}{-1}
\includegraphics[width=.8\textwidth]{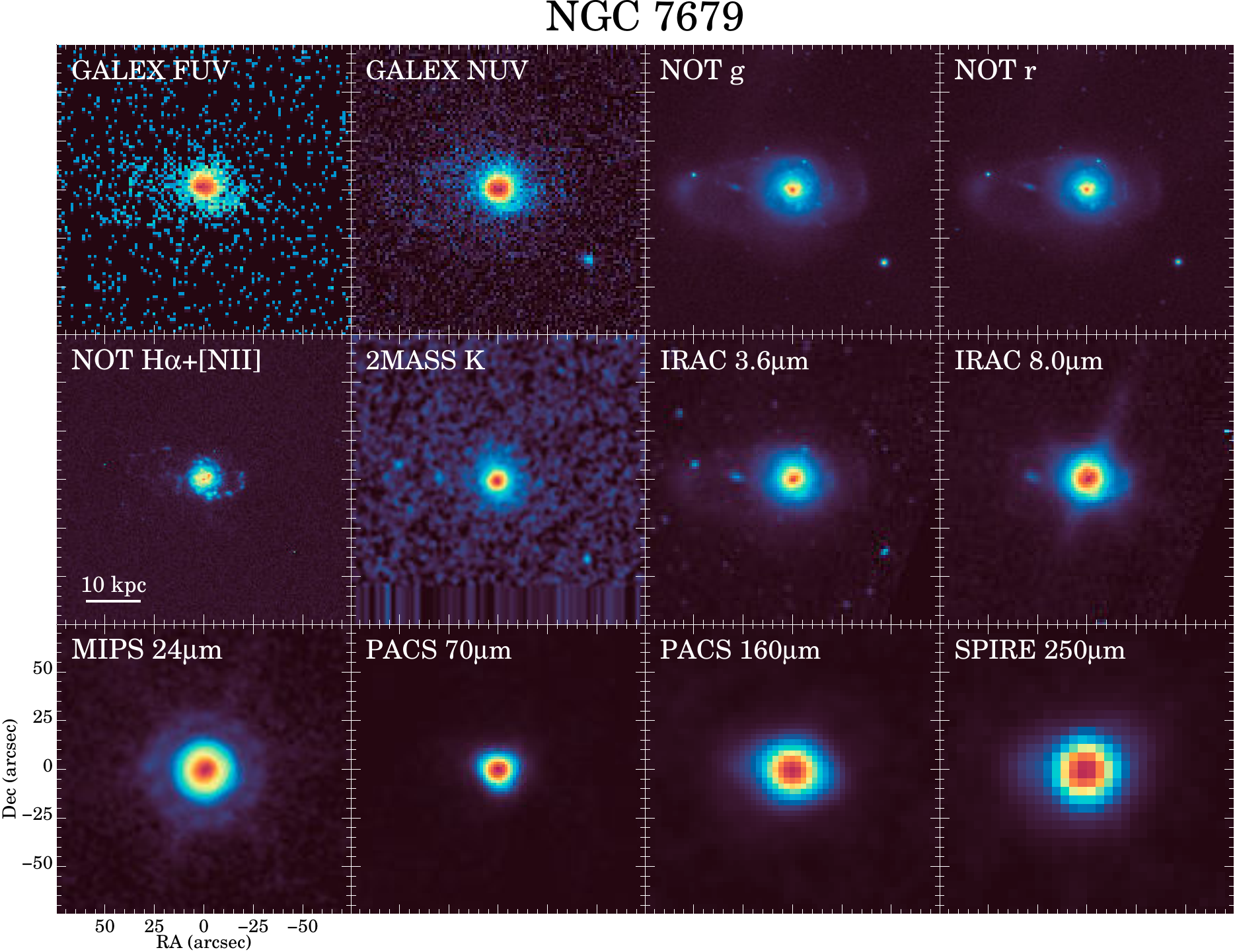}
\includegraphics[width=.8\textwidth]{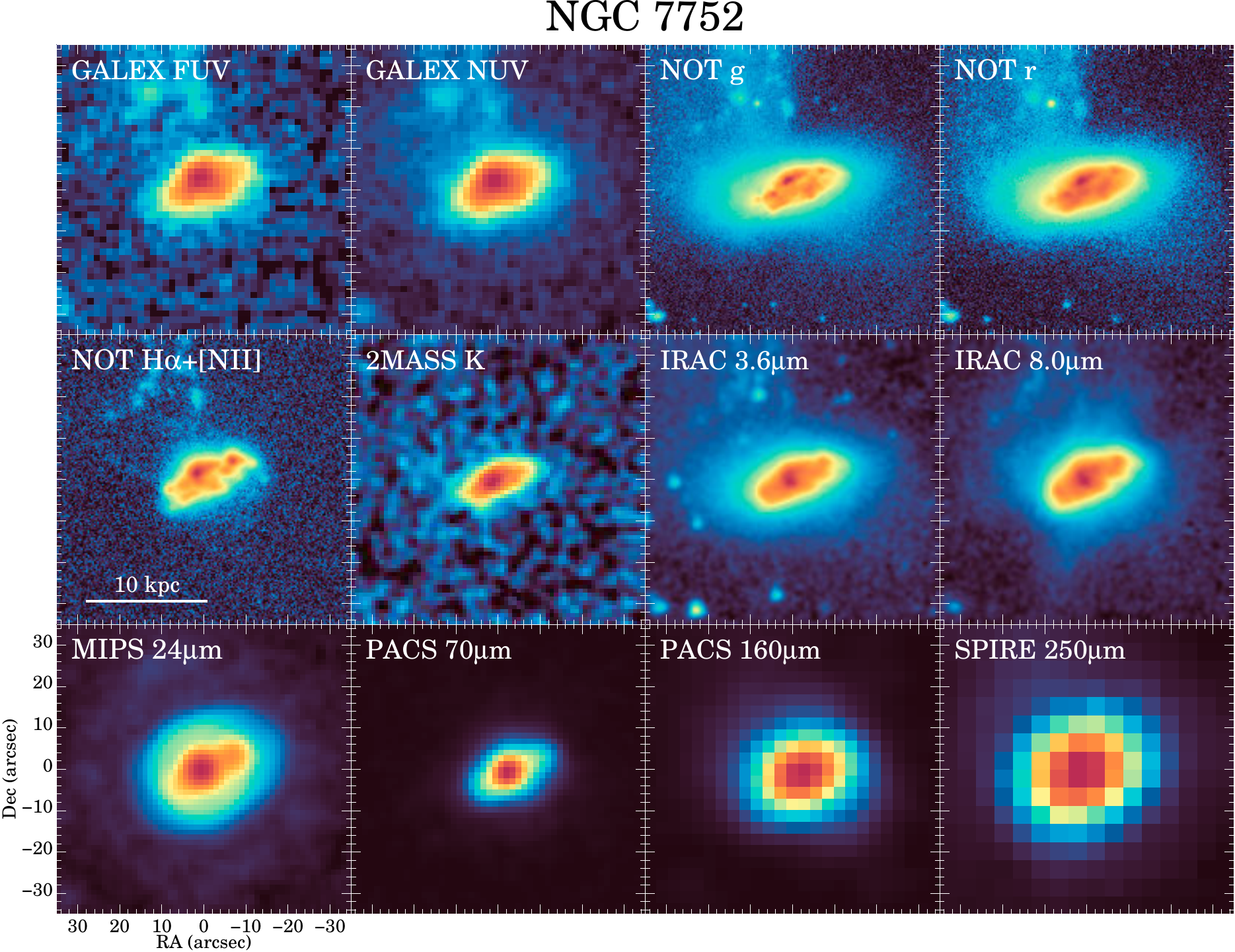}
\caption{Continued.}
\end{figure*}
\clearpage

\begin{figure*}[!h]
\centering
\addtocounter{figure}{-1}
\includegraphics[width=.8\textwidth]{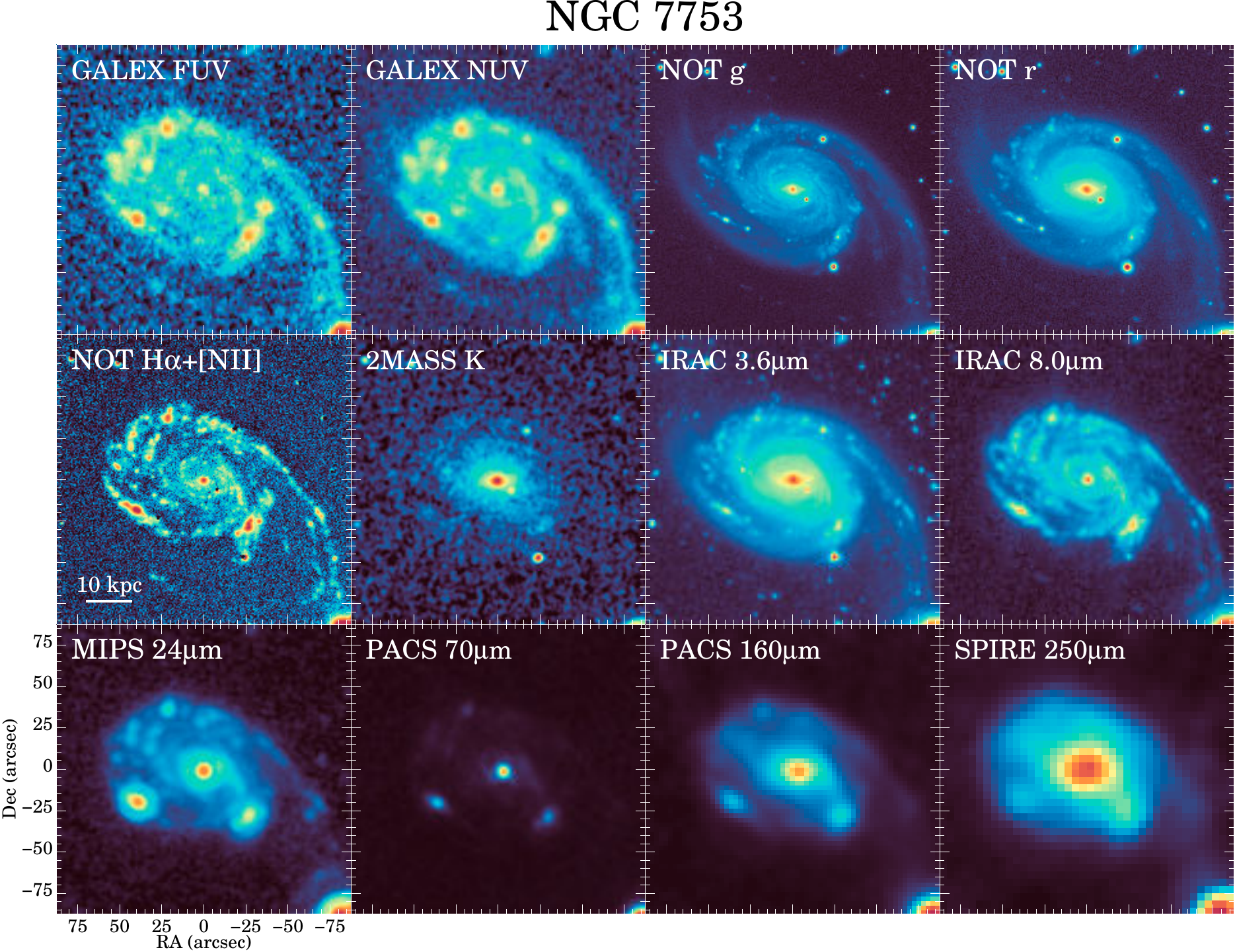}
\includegraphics[width=.8\textwidth]{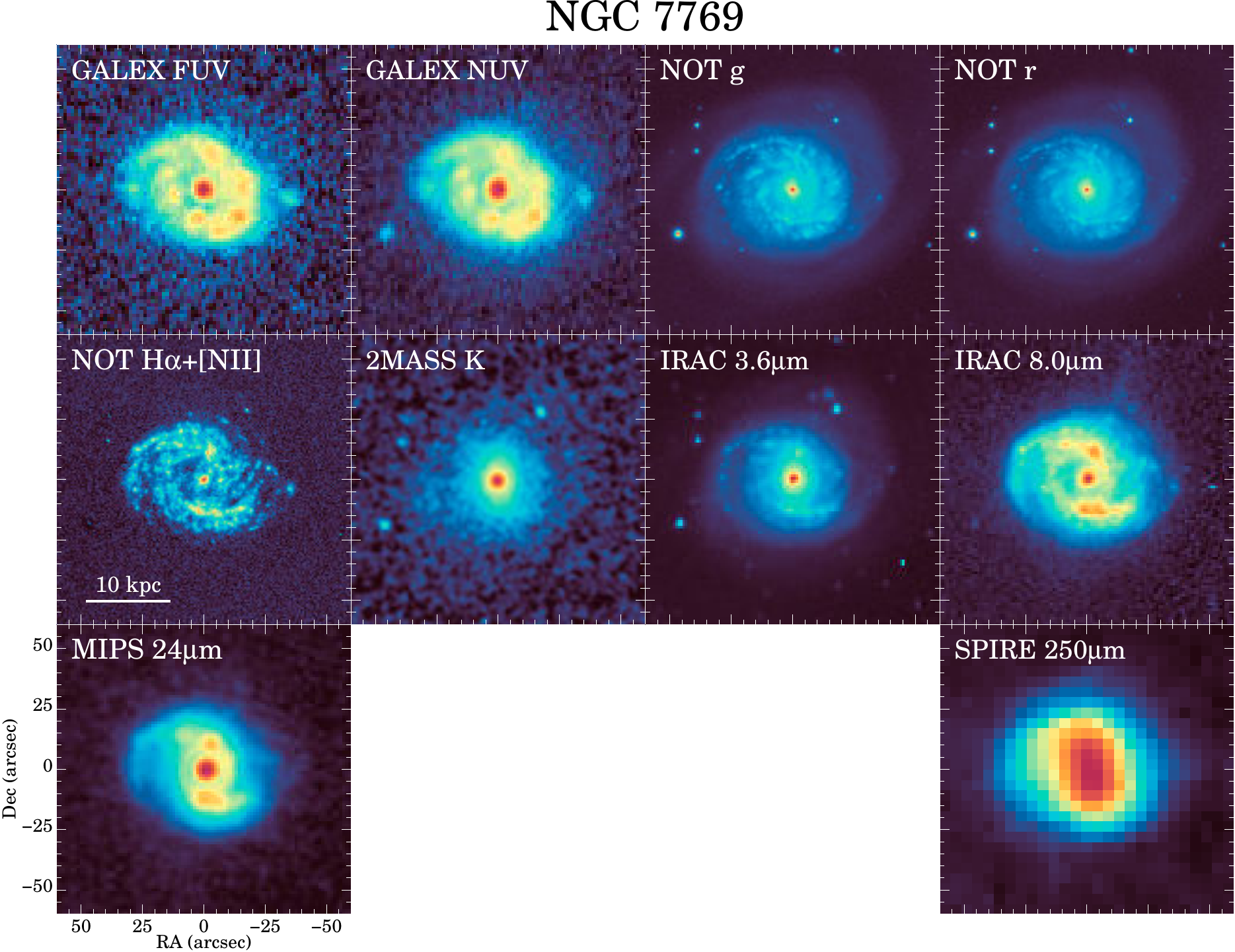}
\caption{Continued.}
\end{figure*}
\clearpage

\begin{figure*}[!h]
\centering
\addtocounter{figure}{-1}
\includegraphics[width=.8\textwidth]{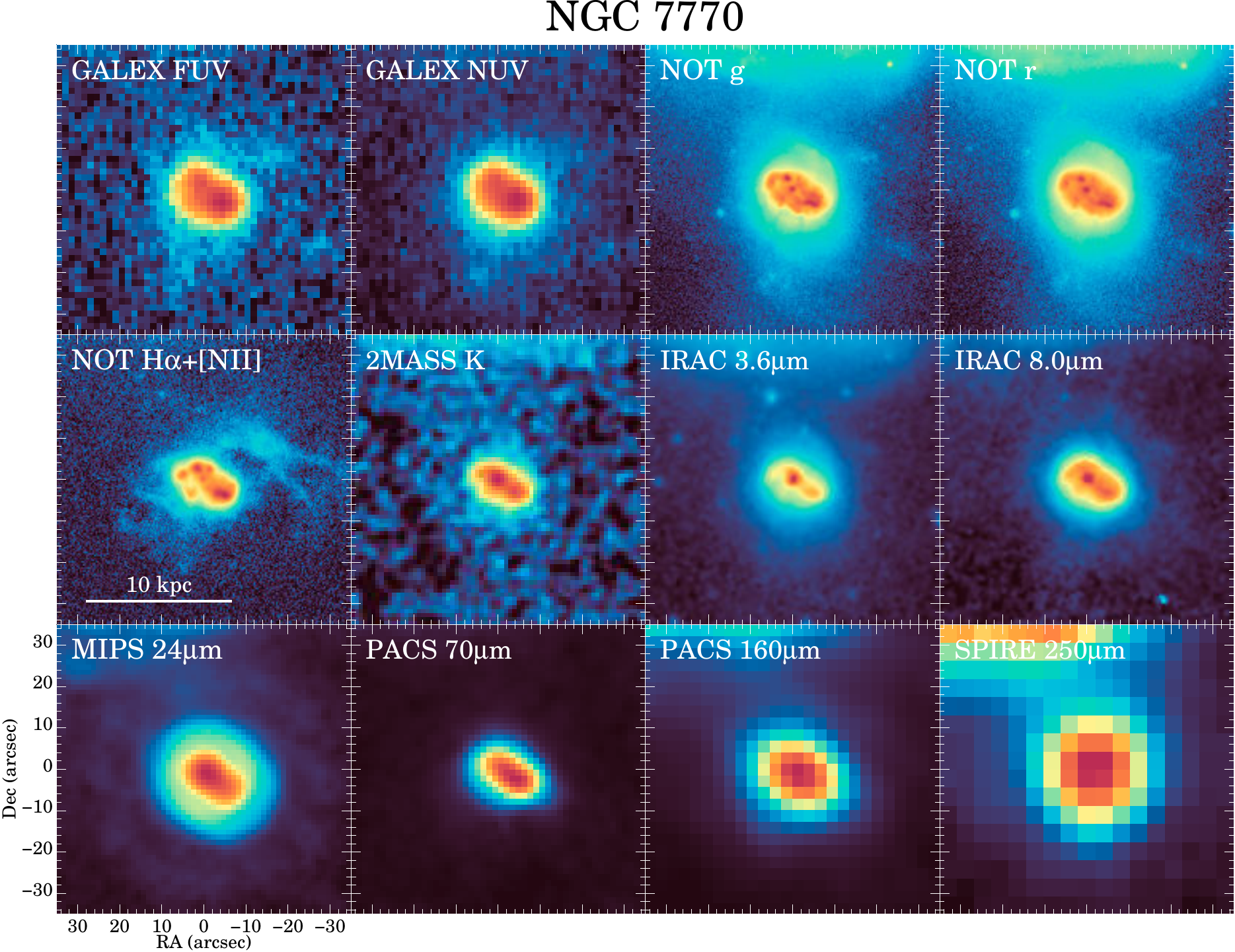}
\includegraphics[width=.8\textwidth]{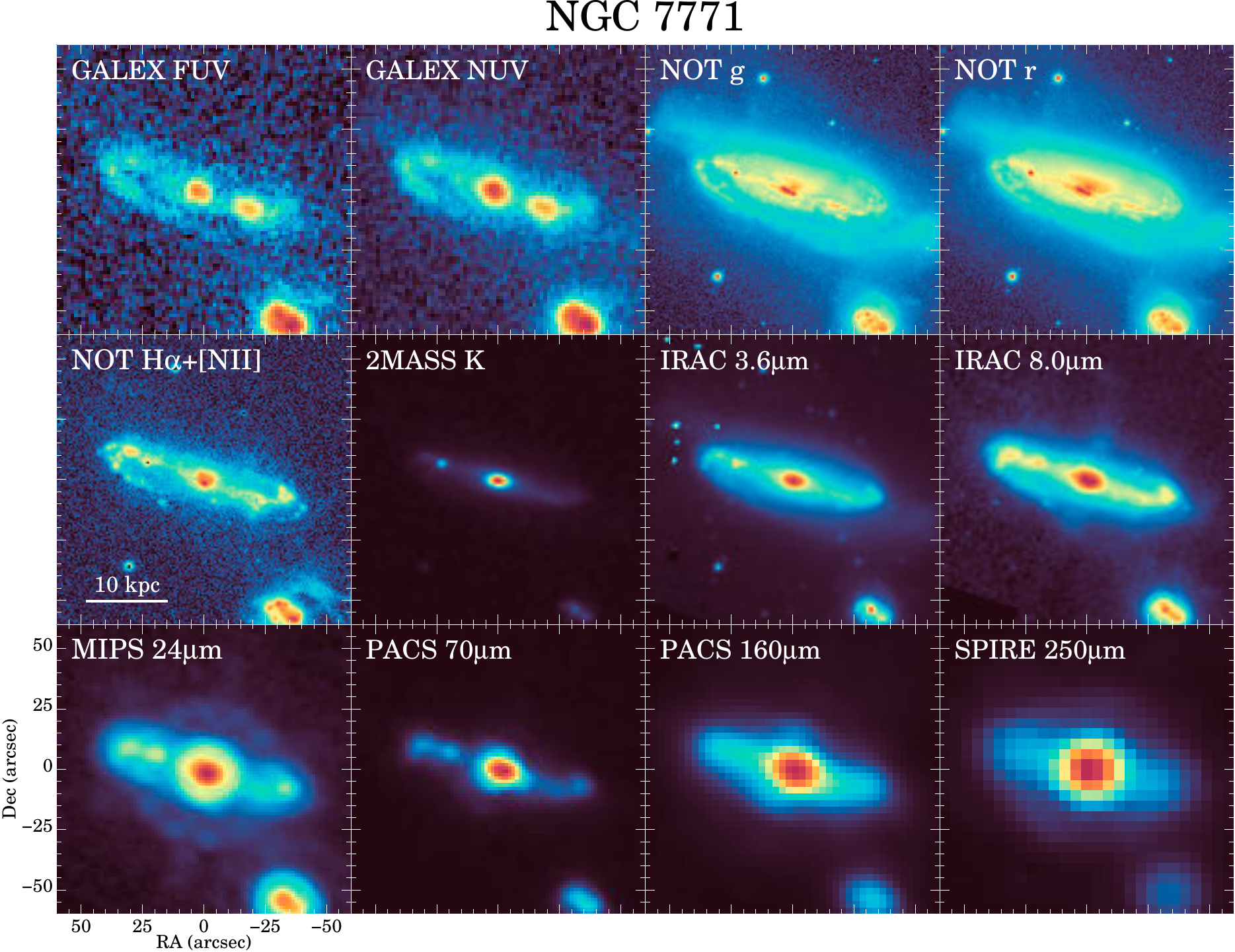}
\caption{Continued.}
\end{figure*}
\clearpage

\section{Best fitting model results to the SED}\label{apx:bestfit}

\begin{figure}[!h]
\centering
\includegraphics[width=0.9\textwidth]{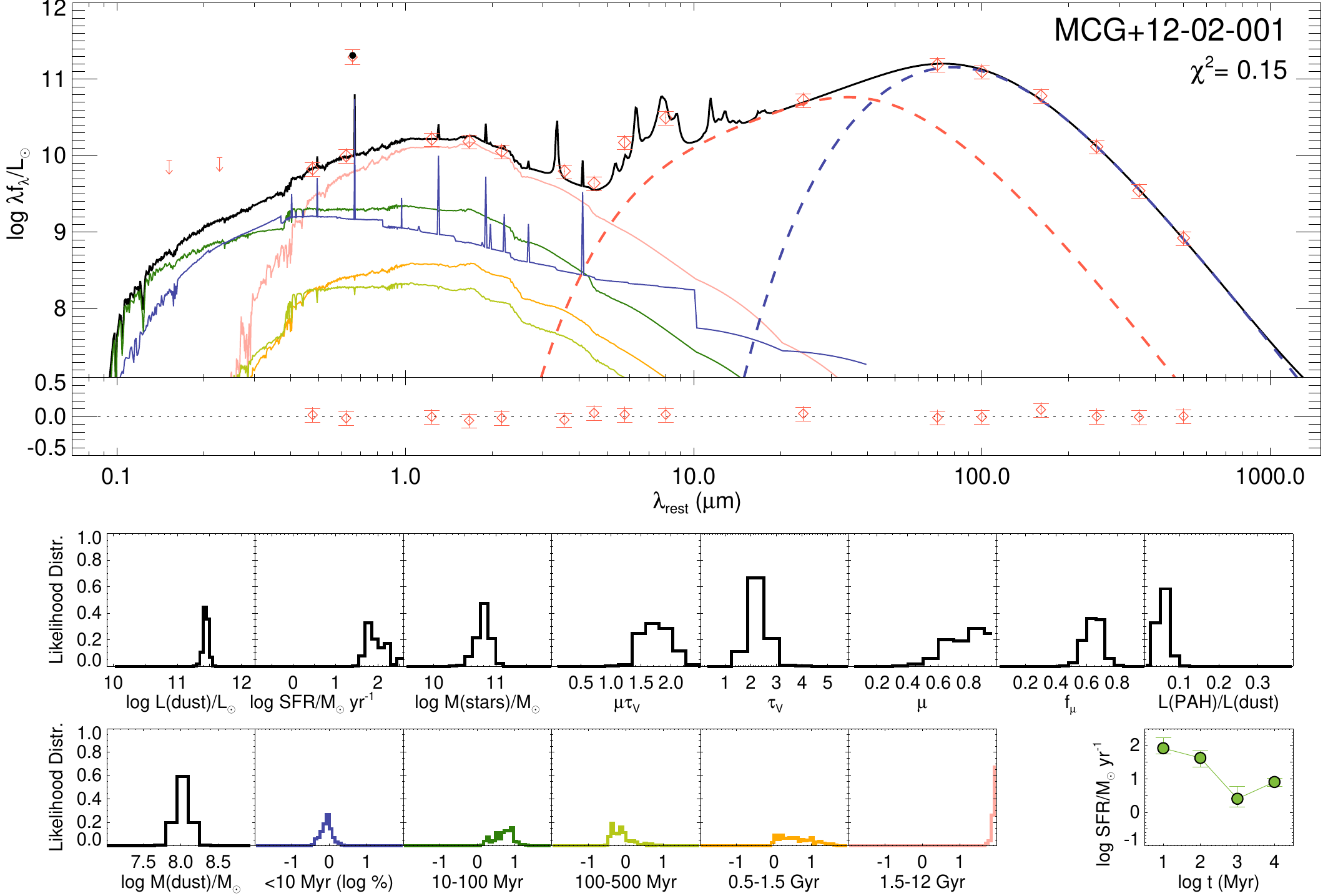}
\includegraphics[width=0.9\textwidth]{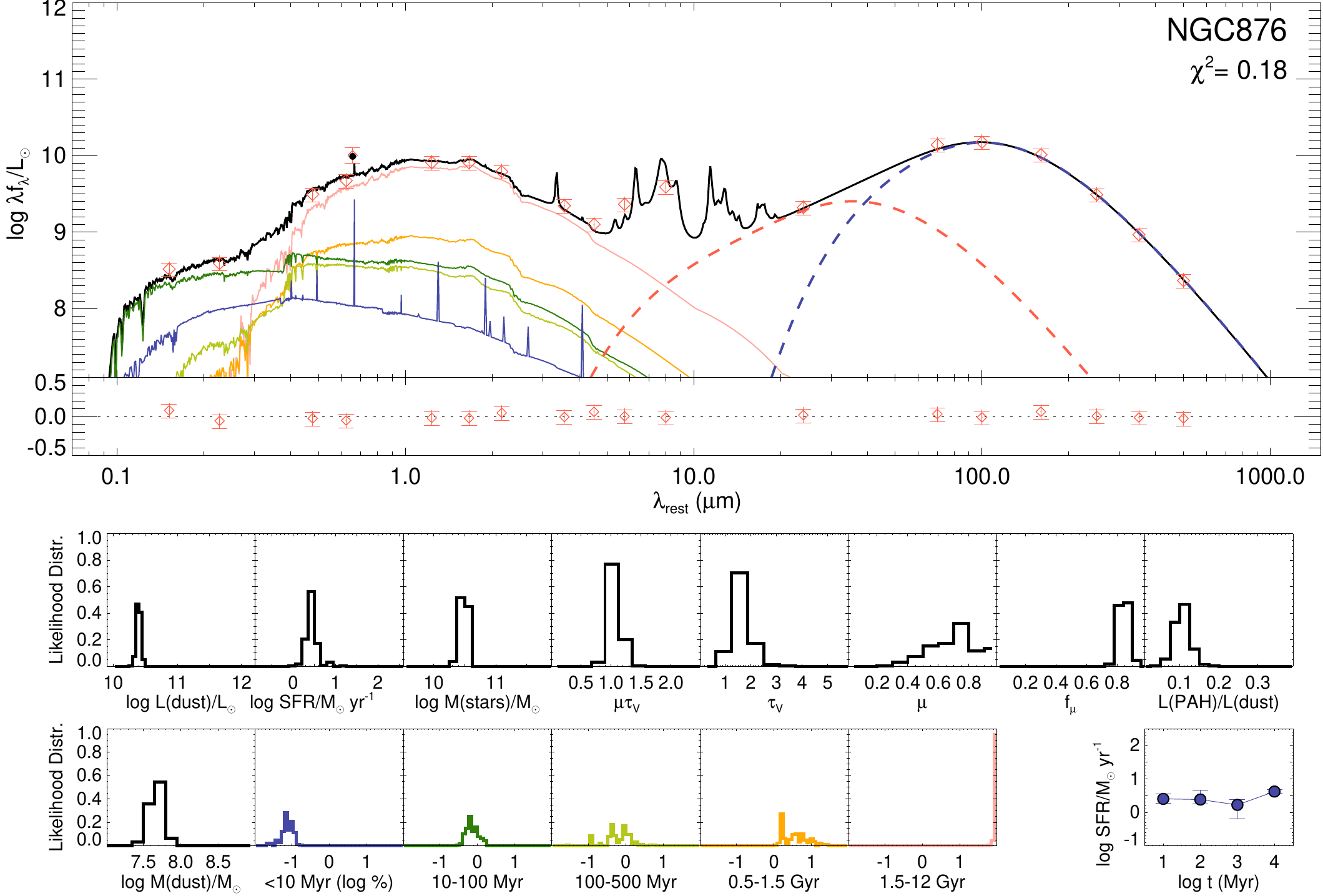}
\caption{Same as Figure \ref{fig:sedfitting}.}
\end{figure}
\clearpage
\addtocounter{figure}{-1}
\begin{figure*}
\centering
\includegraphics[width=0.9\textwidth]{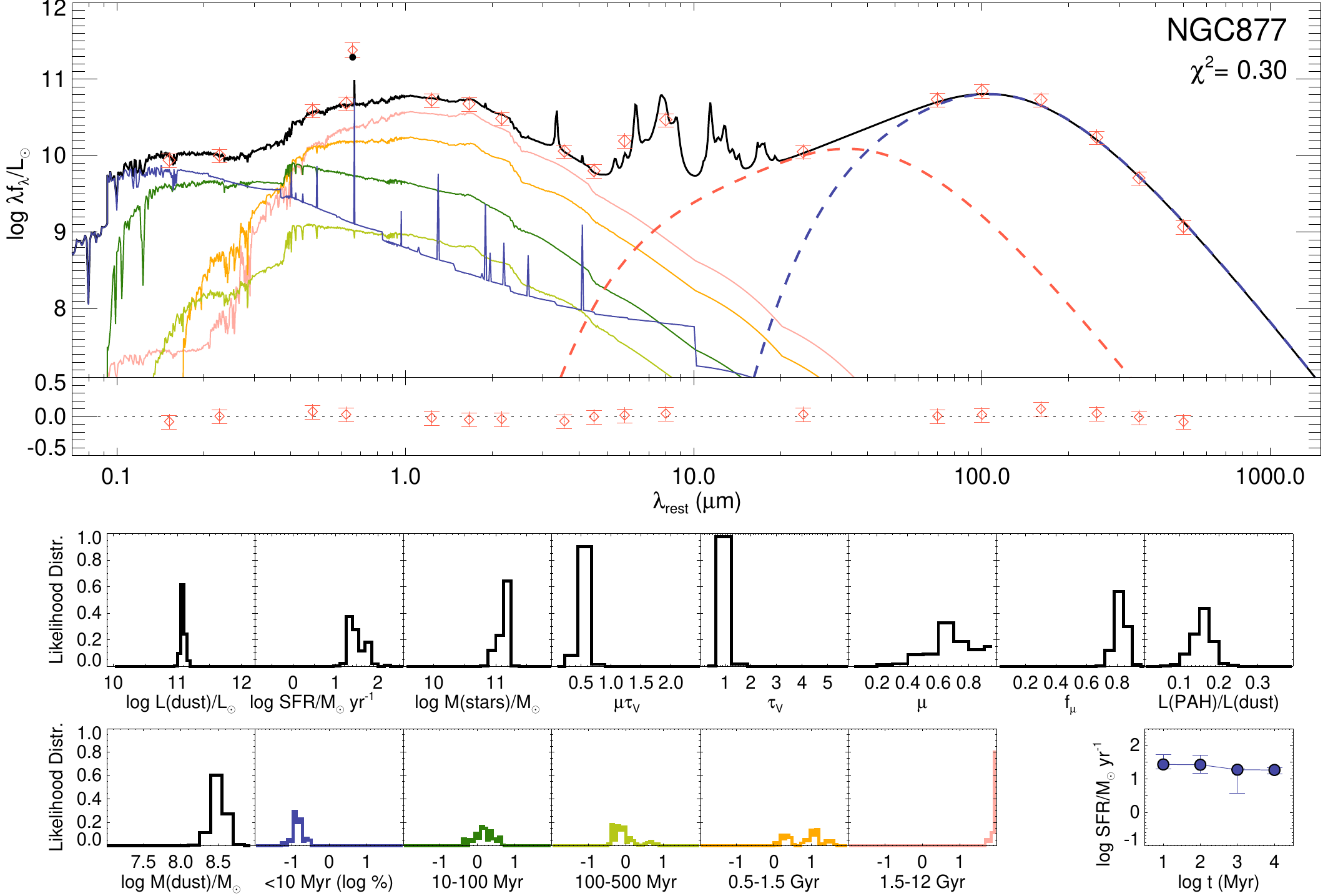}
\includegraphics[width=0.9\textwidth]{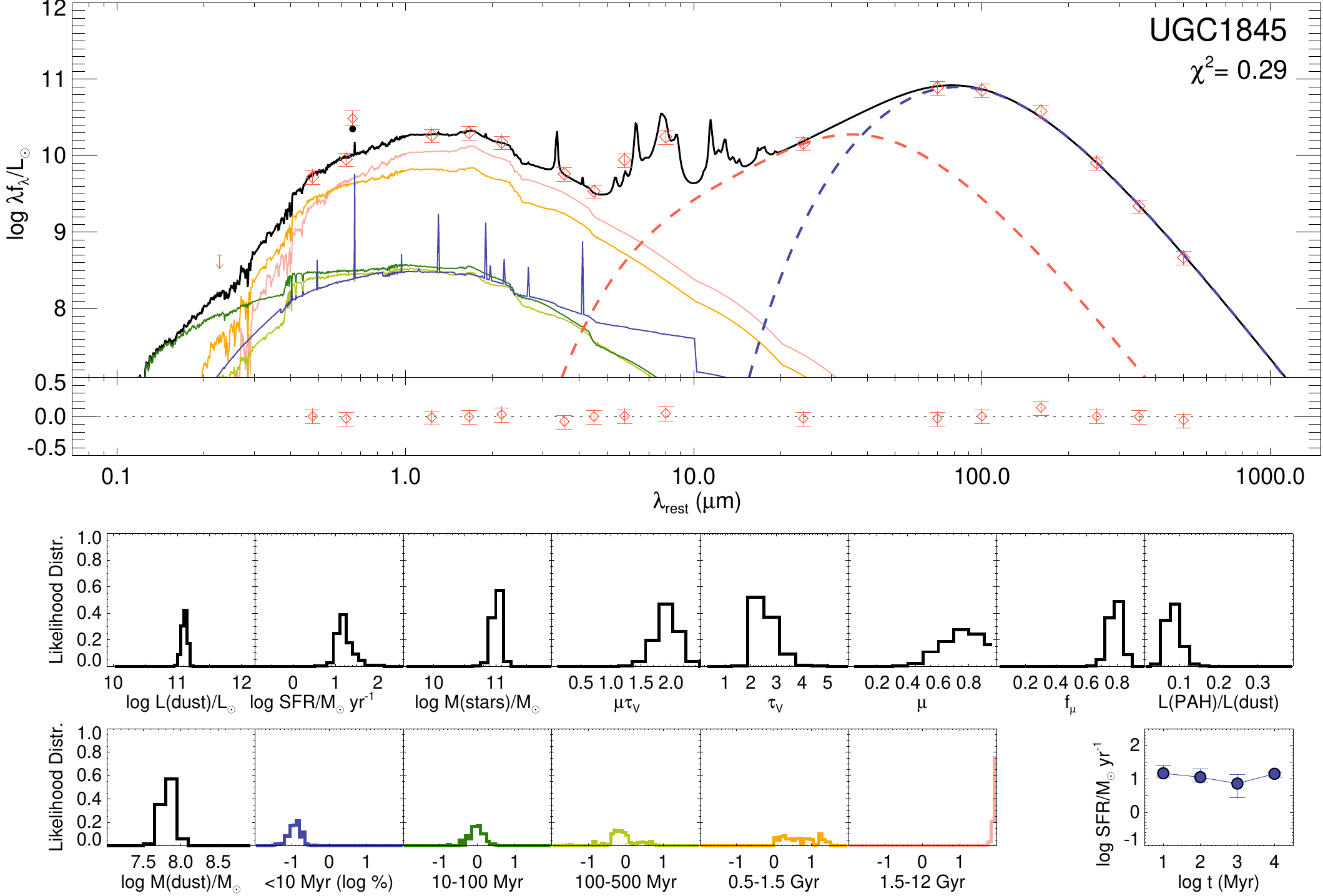}
\caption{Continued.}
\end{figure*}
\clearpage
\addtocounter{figure}{-1}
\begin{figure*}
\centering
\includegraphics[width=0.9\textwidth]{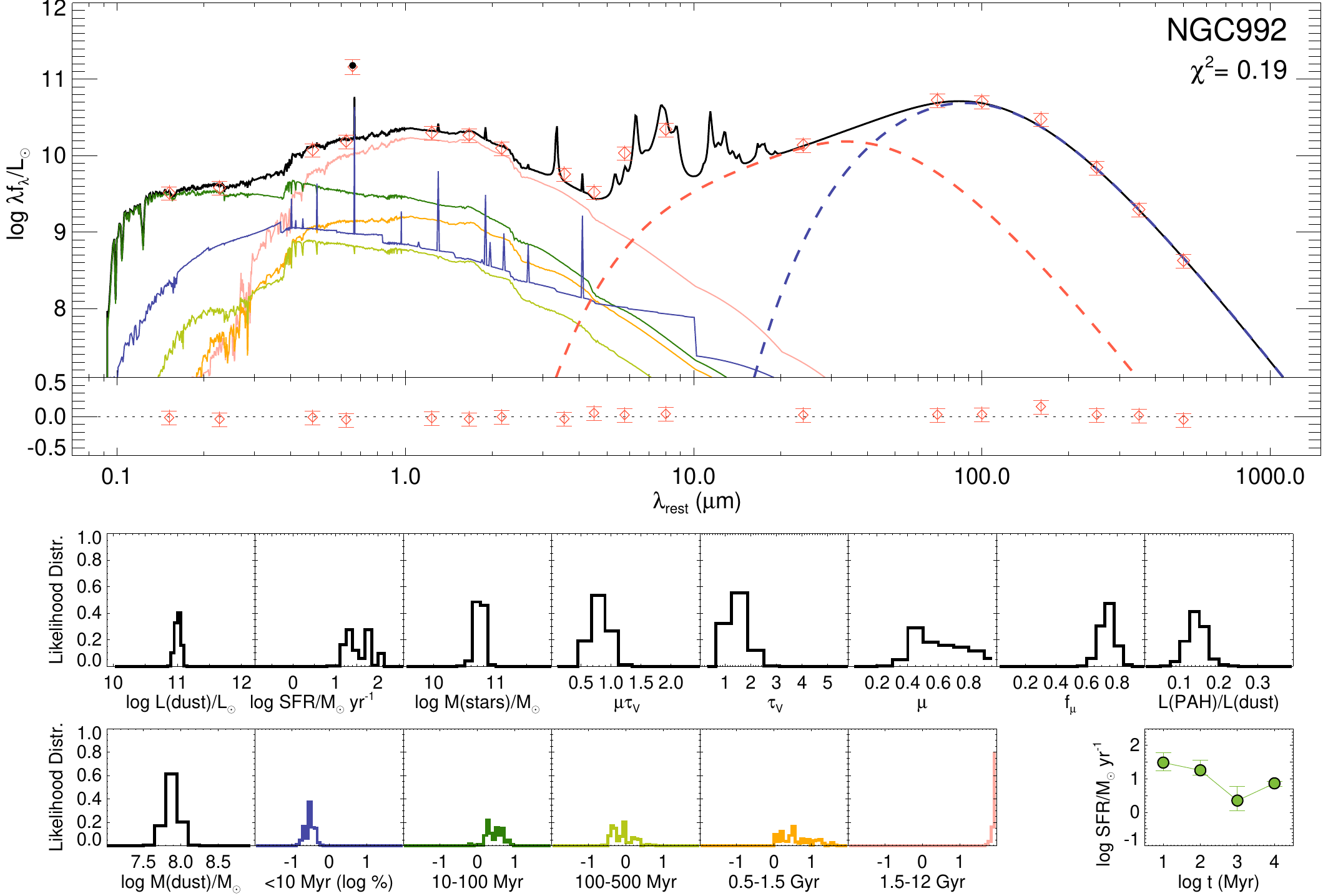}
\includegraphics[width=0.9\textwidth]{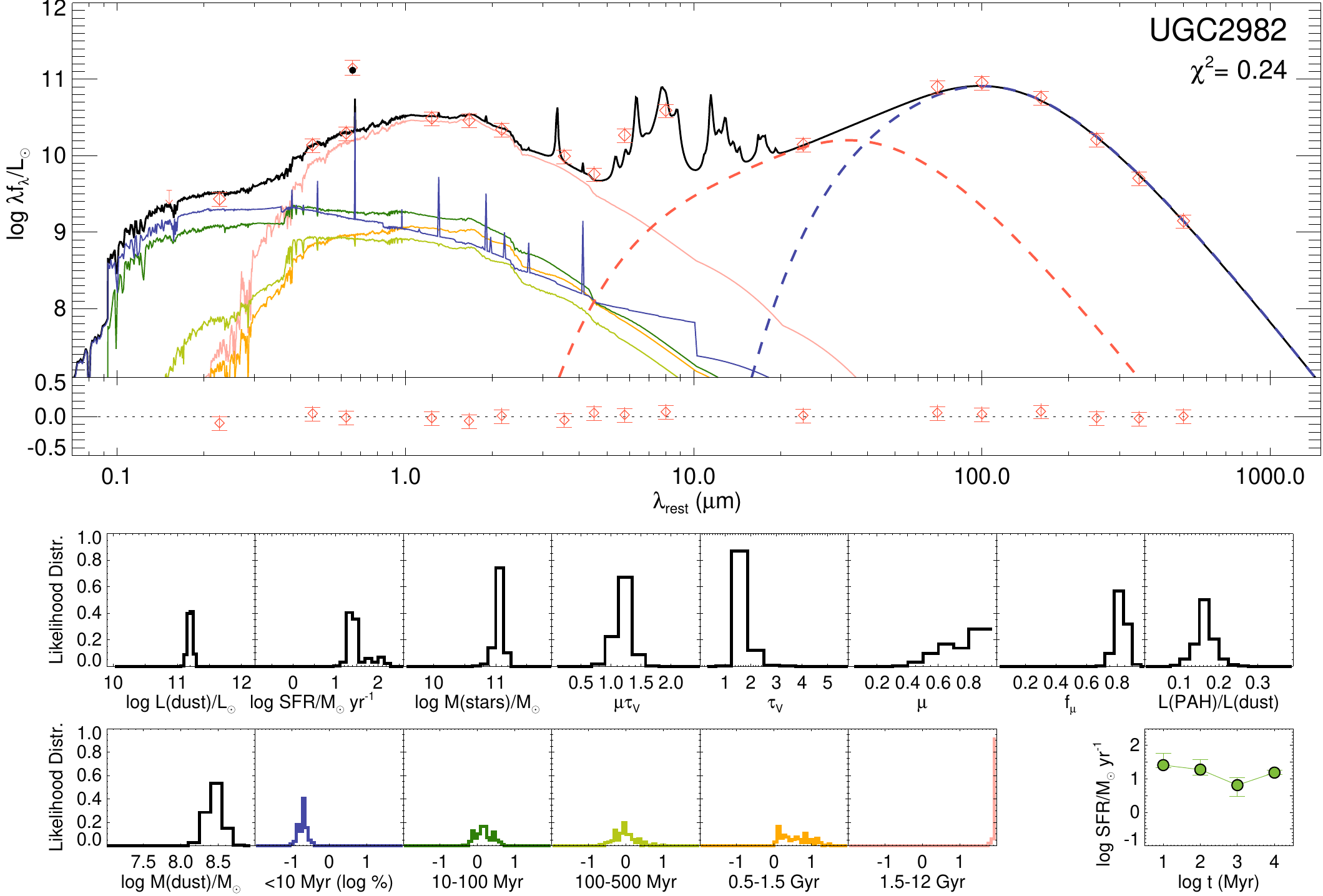}
\caption{Continued.}
\end{figure*}
\clearpage
\addtocounter{figure}{-1}
\begin{figure*}
\centering
\includegraphics[width=0.9\textwidth]{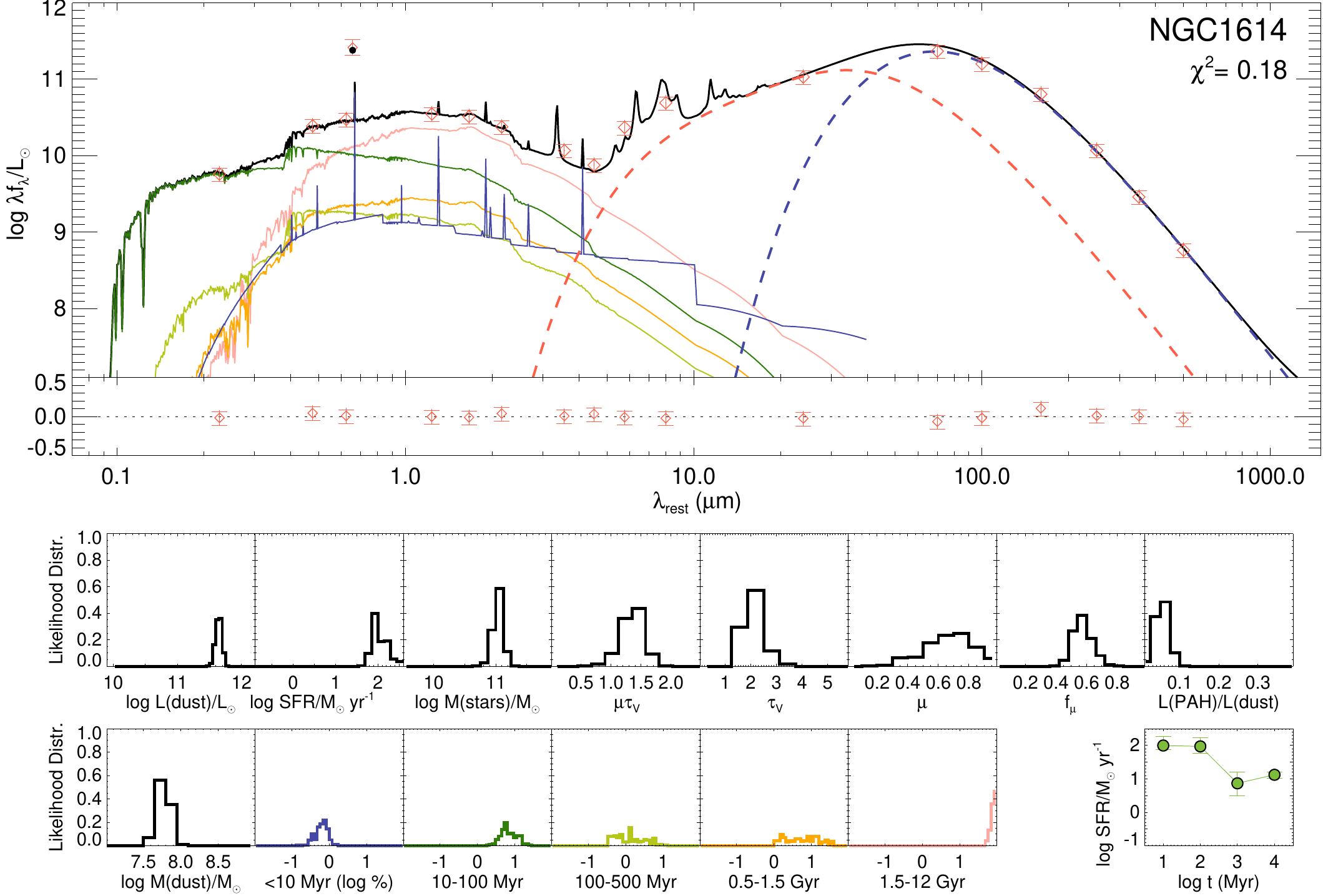}
\includegraphics[width=0.9\textwidth]{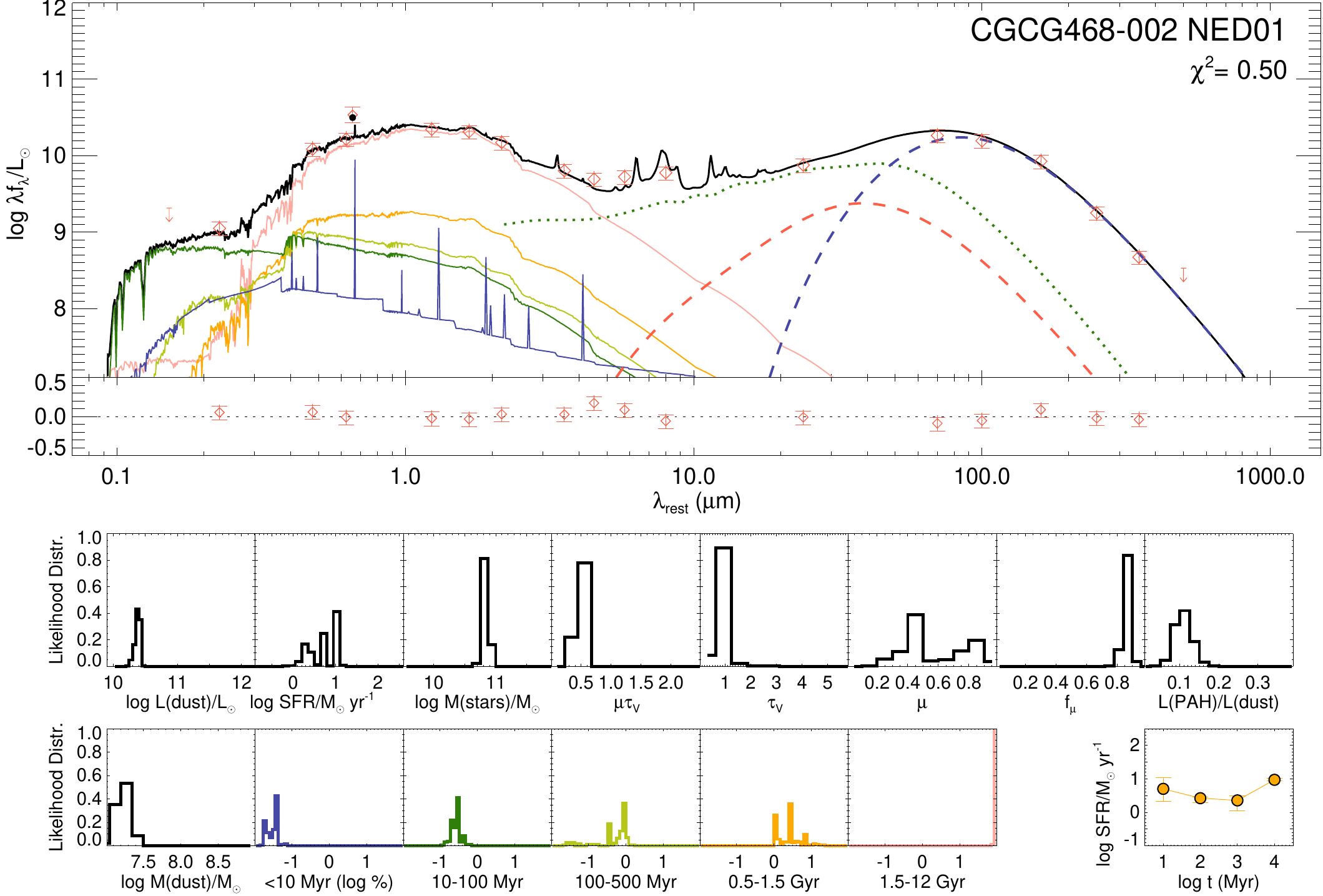}
\caption{Continued.}
\end{figure*}
\clearpage
\addtocounter{figure}{-1}
\begin{figure*}
\centering
\includegraphics[width=0.9\textwidth]{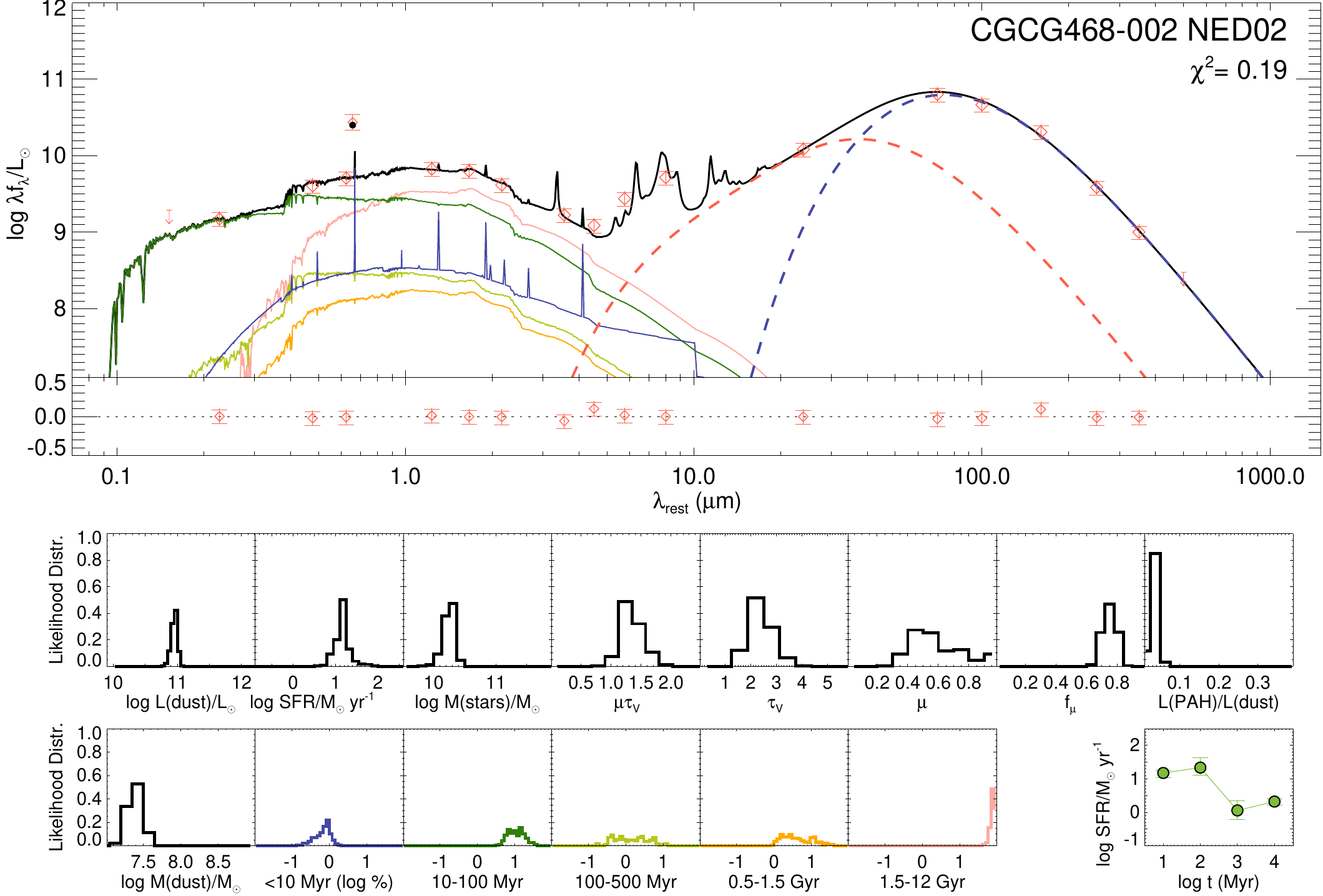}
\includegraphics[width=0.9\textwidth]{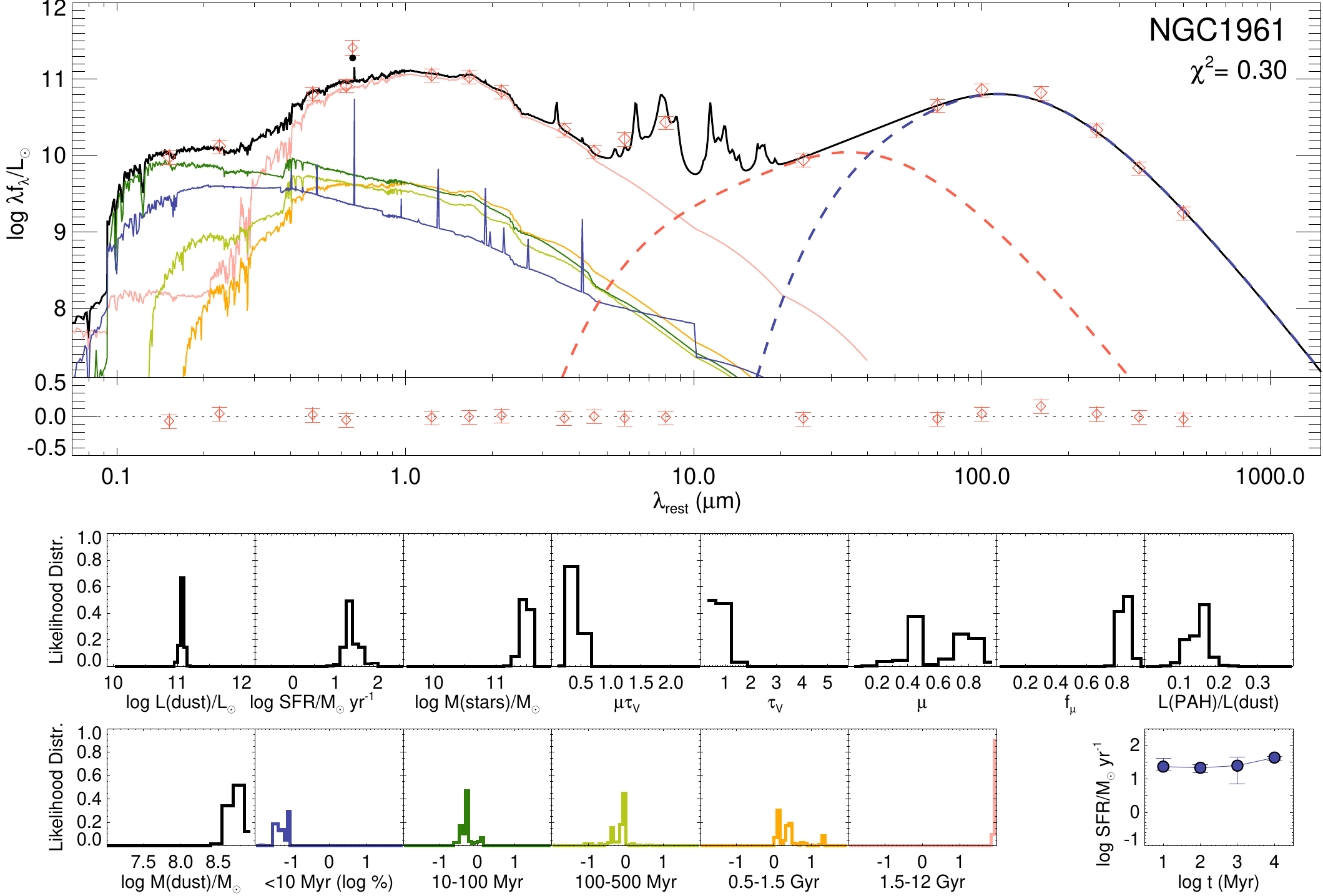}
\caption{Continued.}
\end{figure*}
\clearpage
\addtocounter{figure}{-1}
\begin{figure*}
\centering
\includegraphics[width=0.9\textwidth]{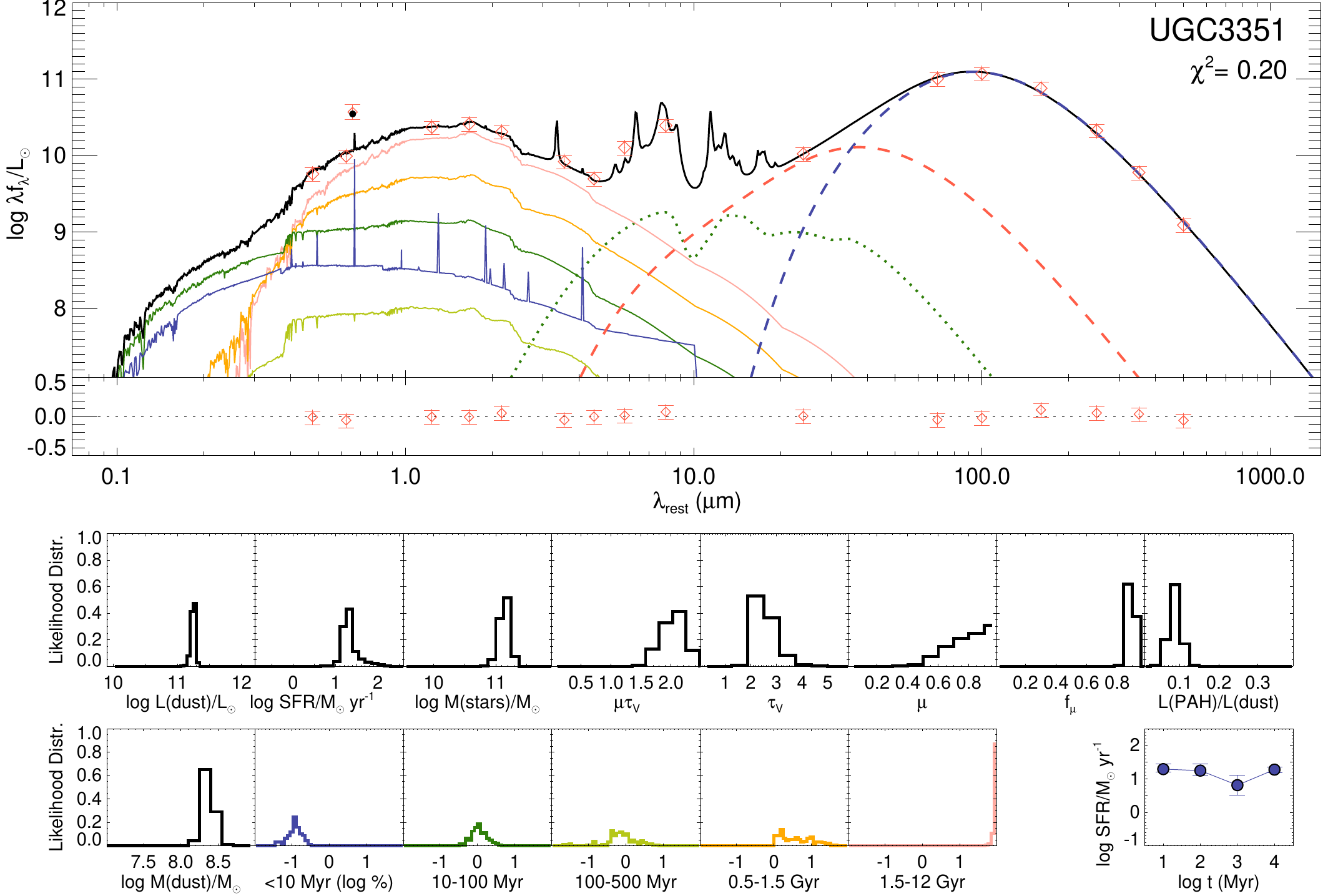}
\includegraphics[width=0.9\textwidth]{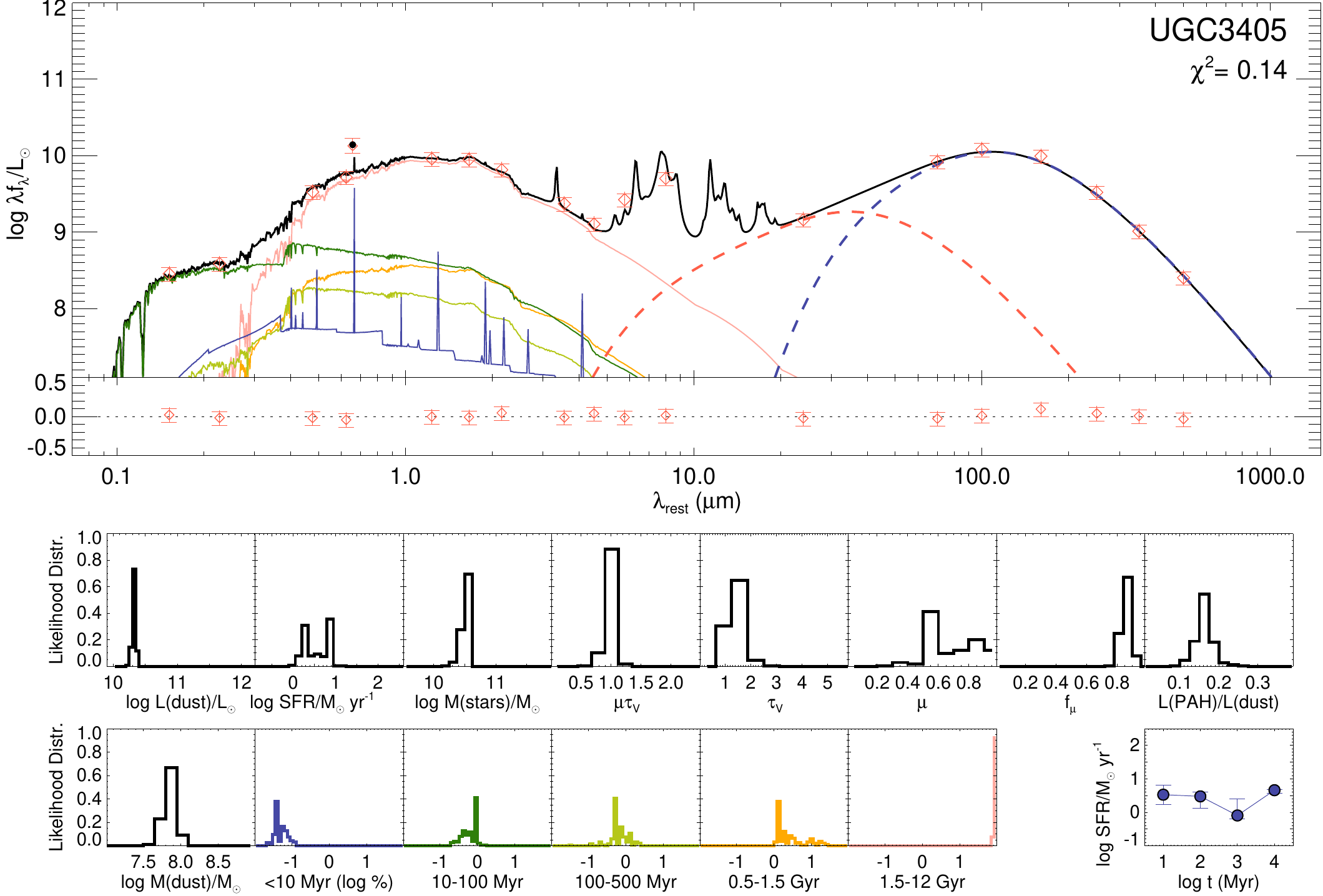}
\caption{Continued.}
\end{figure*}
\clearpage
\addtocounter{figure}{-1}
\begin{figure*}
\centering
\includegraphics[width=0.9\textwidth]{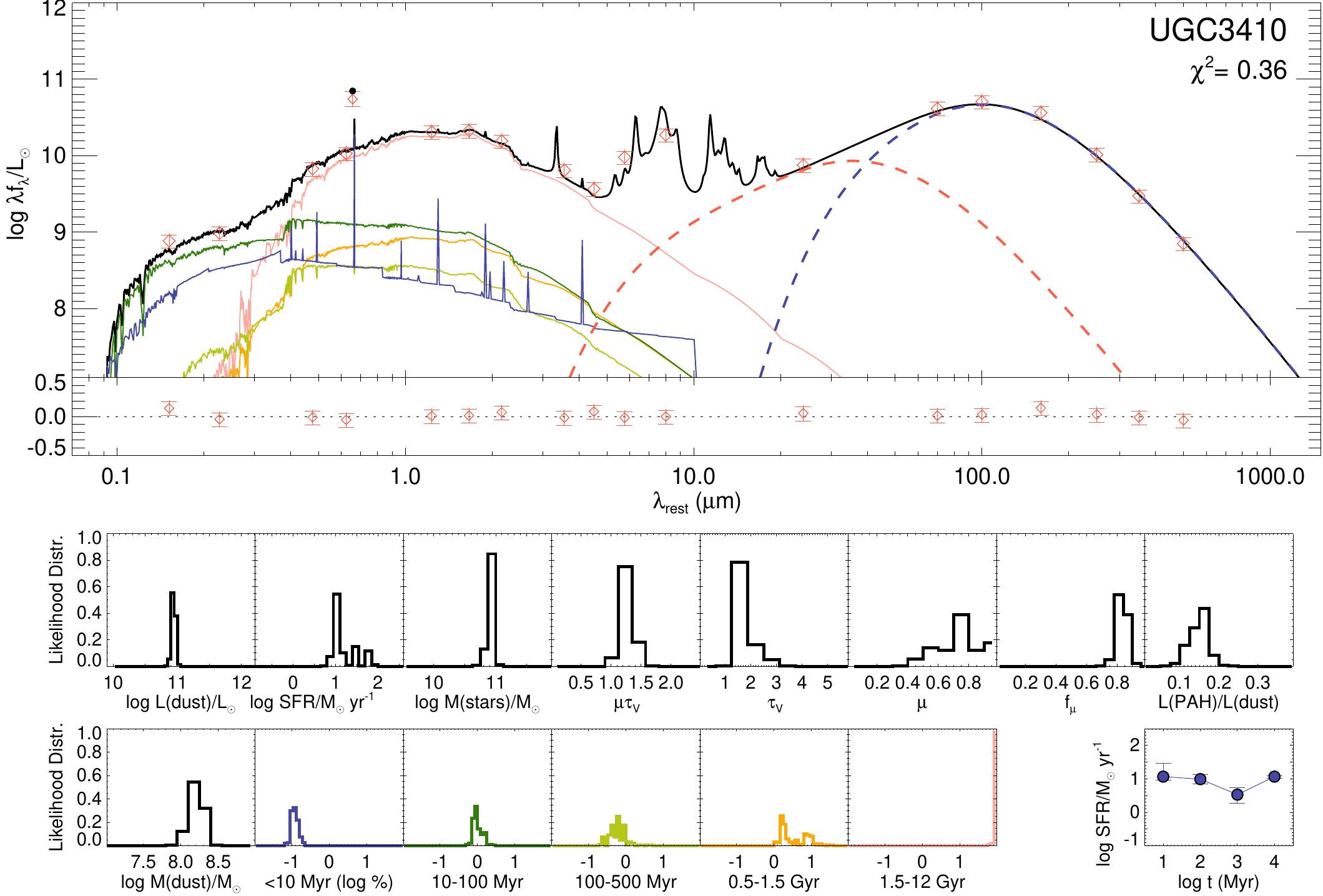}
\includegraphics[width=0.9\textwidth]{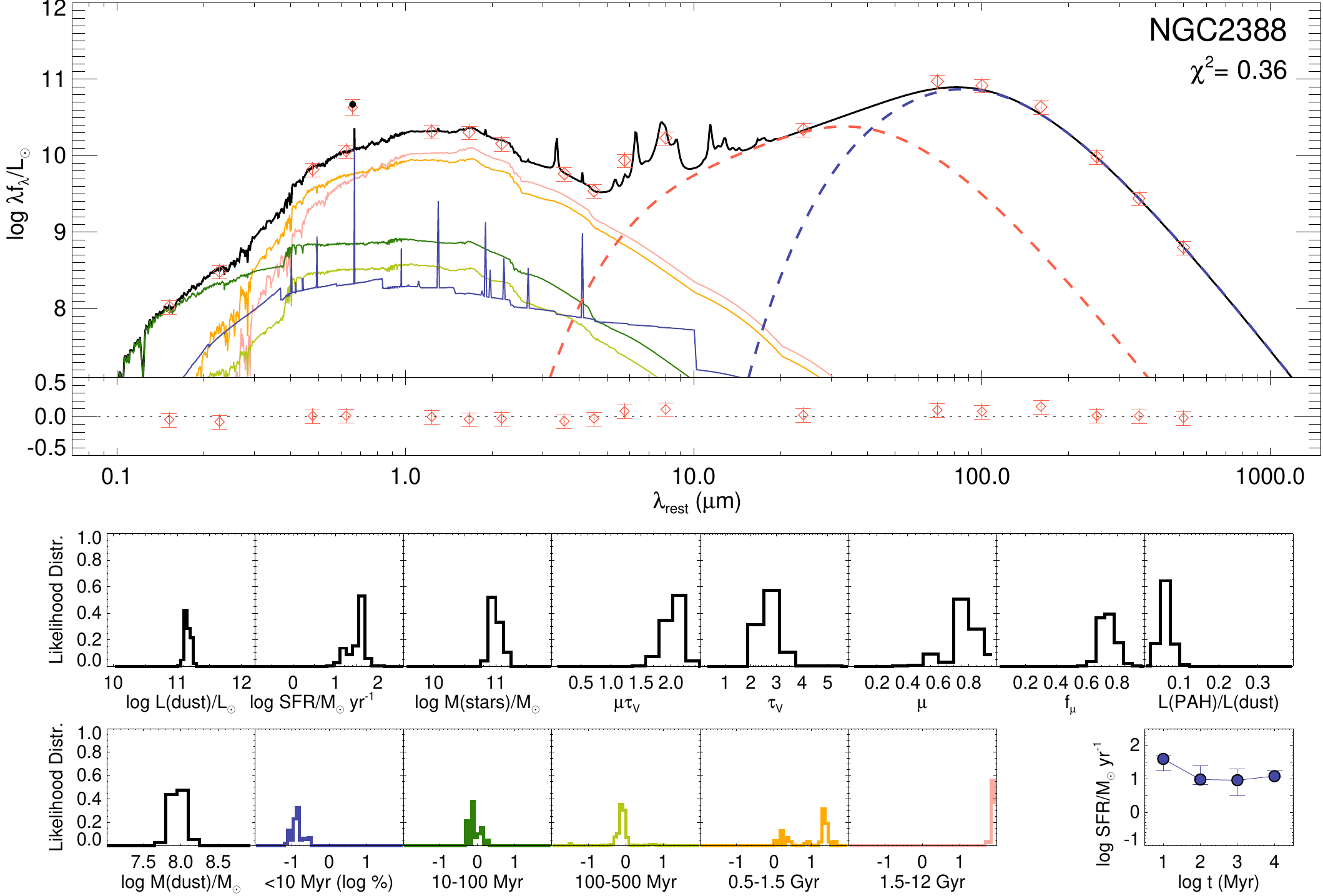}
\caption{Continued.}
\end{figure*}
\clearpage
\addtocounter{figure}{-1}
\begin{figure*}
\centering
\includegraphics[width=0.9\textwidth]{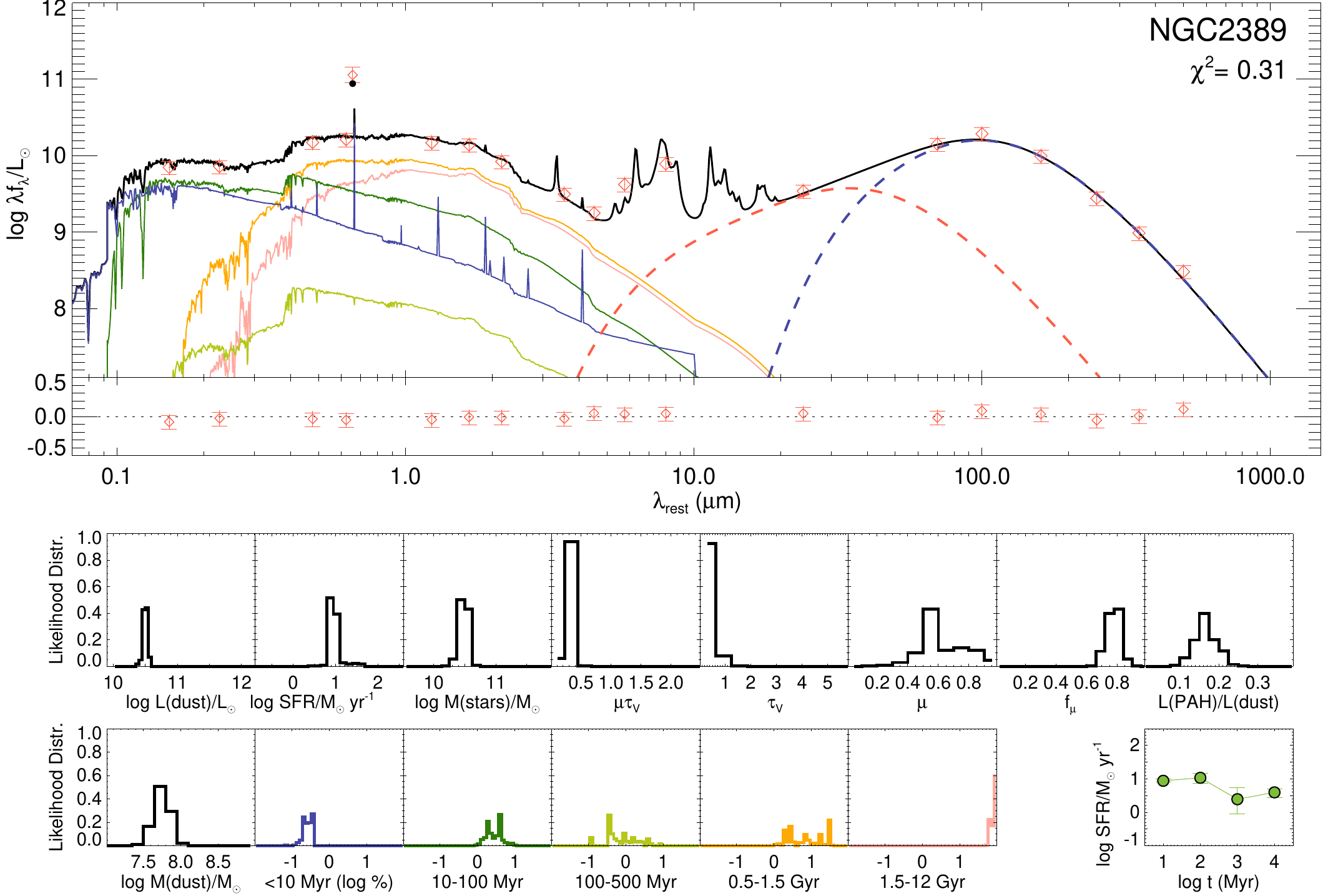}
\includegraphics[width=0.9\textwidth]{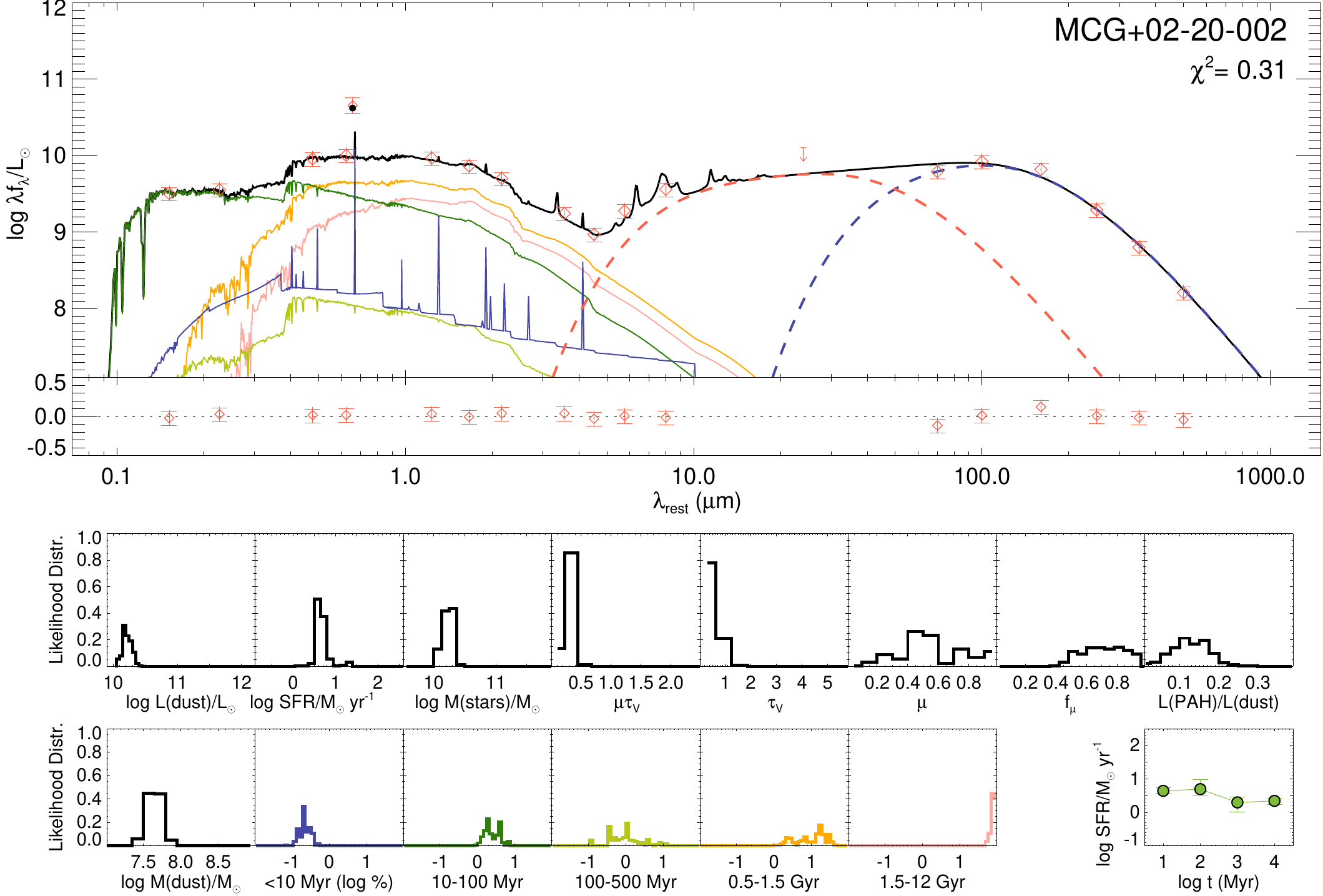}
\caption{Continued.}
\end{figure*}
\clearpage
\addtocounter{figure}{-1}
\begin{figure*}
\centering
\includegraphics[width=0.9\textwidth]{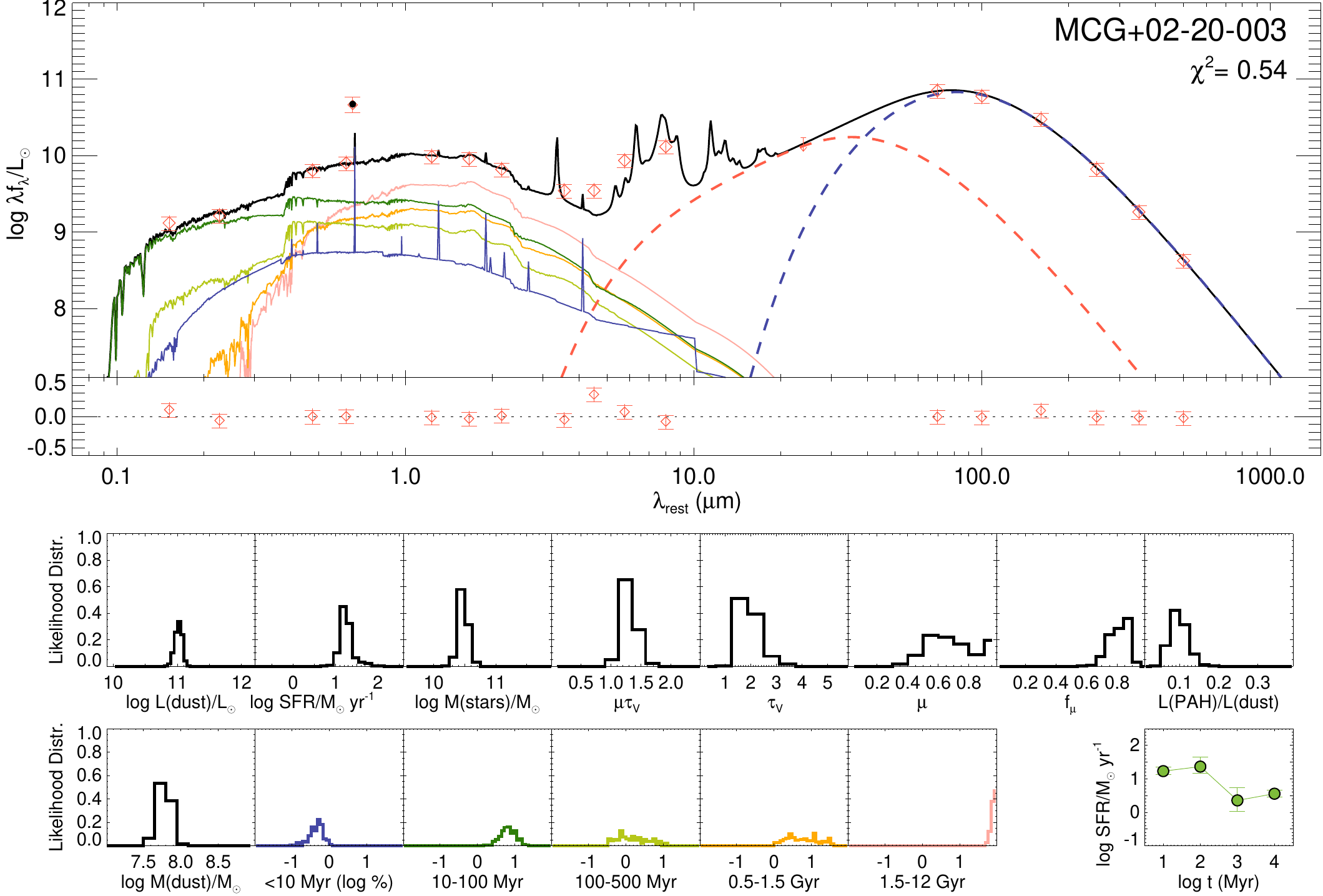}
\includegraphics[width=0.9\textwidth]{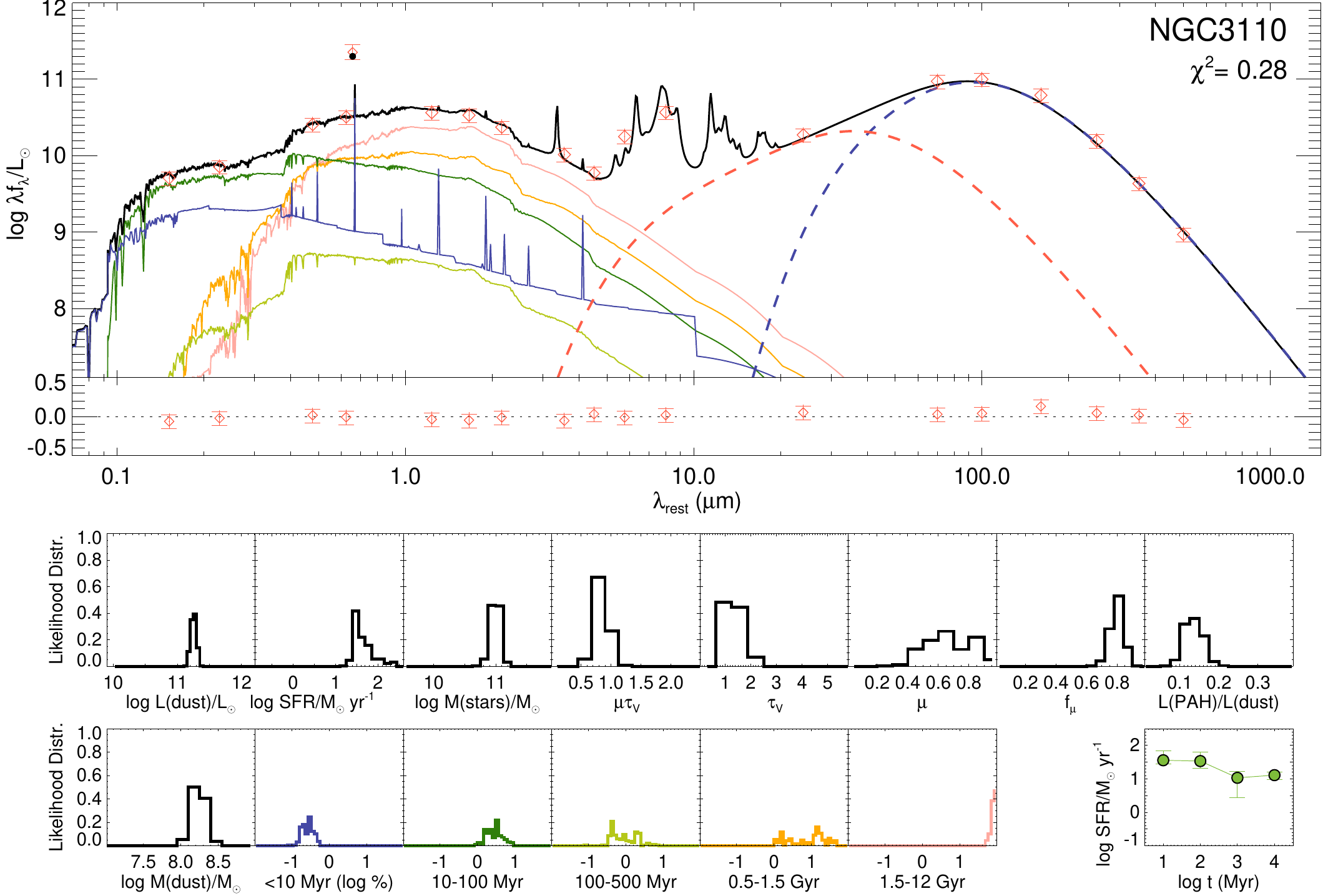}
\caption{Continued.}
\end{figure*}
\clearpage
\addtocounter{figure}{-1}
\begin{figure*}
\centering
\includegraphics[width=0.9\textwidth]{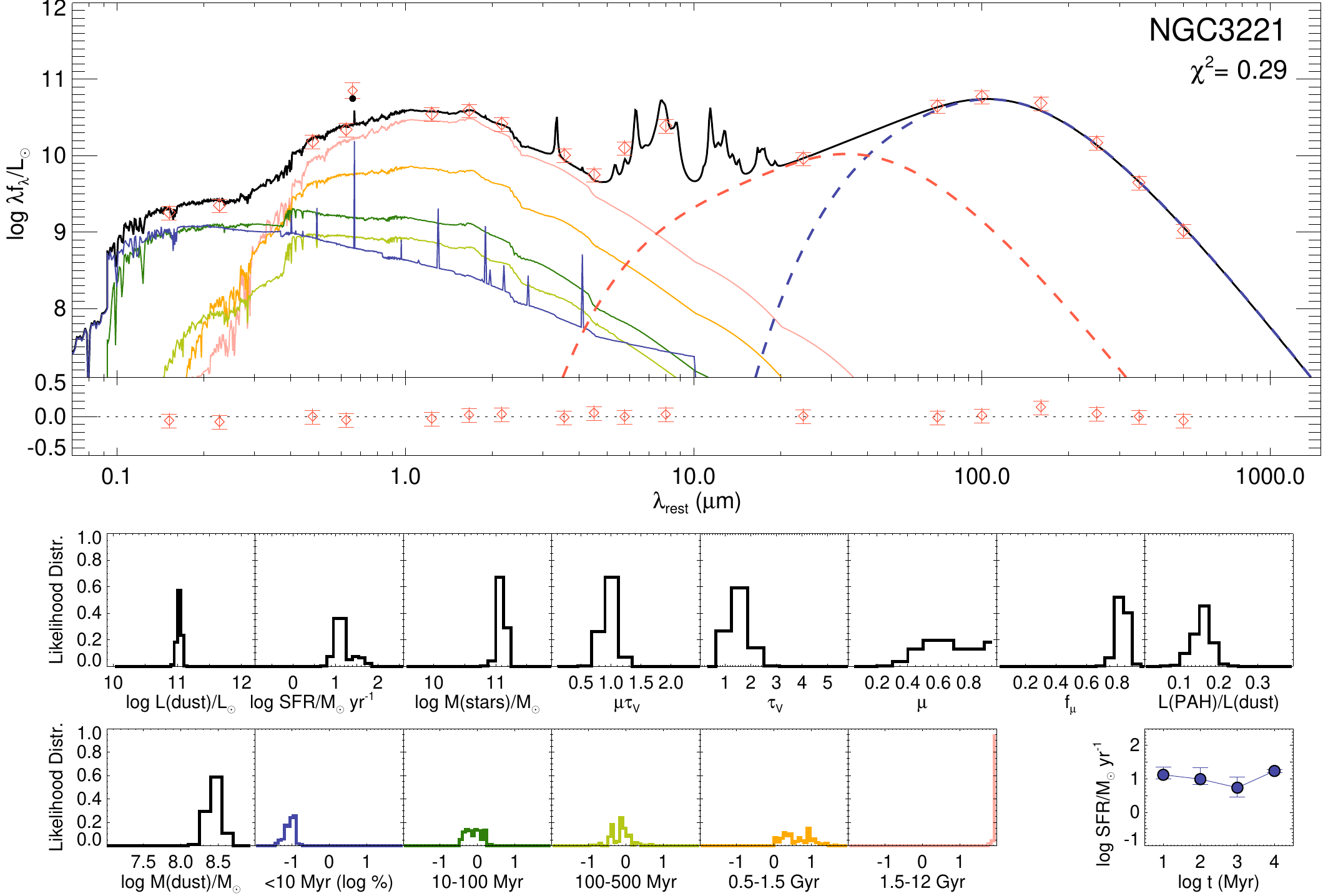}
\includegraphics[width=0.9\textwidth]{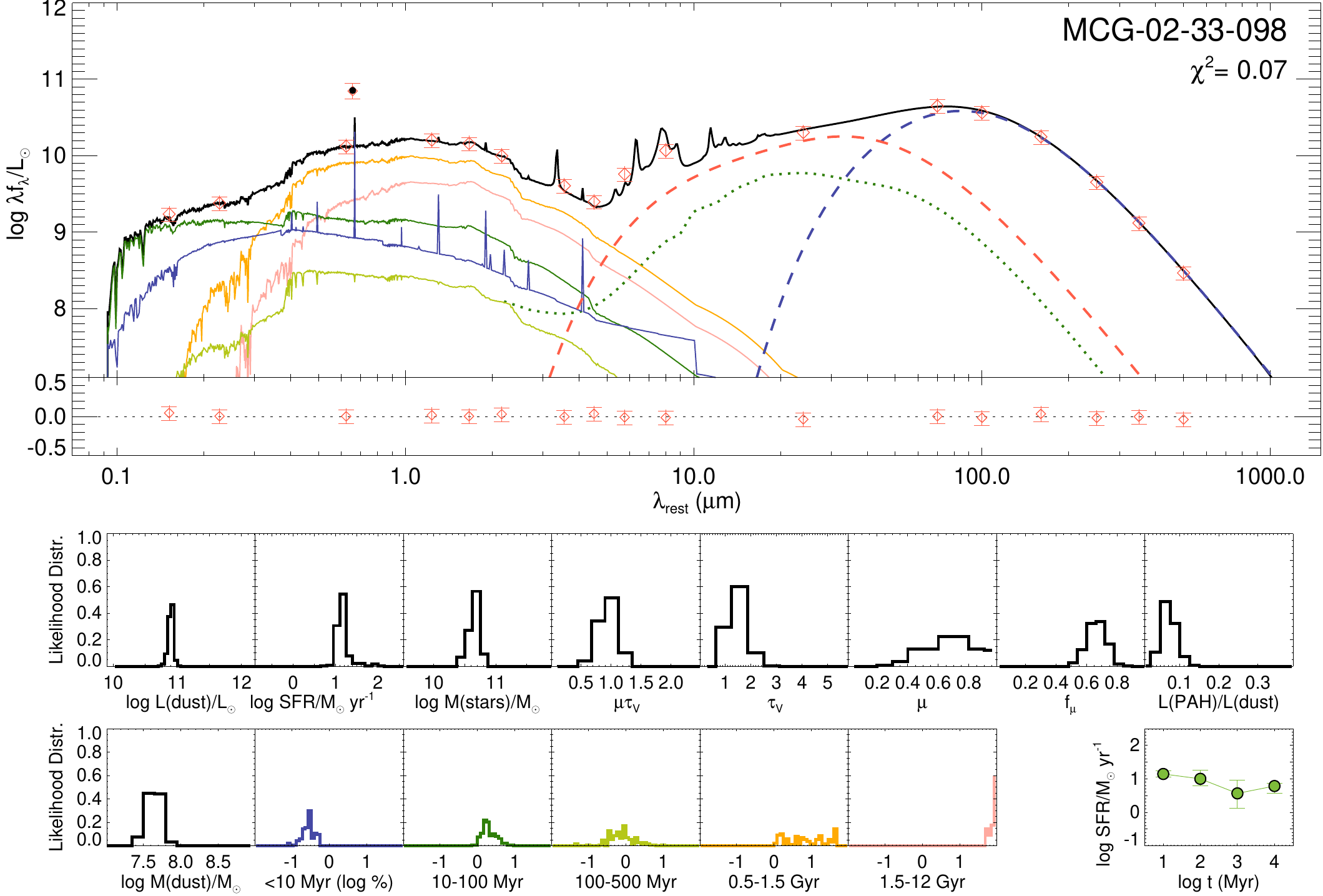}
\caption{Continued.}
\end{figure*}
\clearpage
\addtocounter{figure}{-1}

\begin{figure*}
\centering
\includegraphics[width=0.9\textwidth]{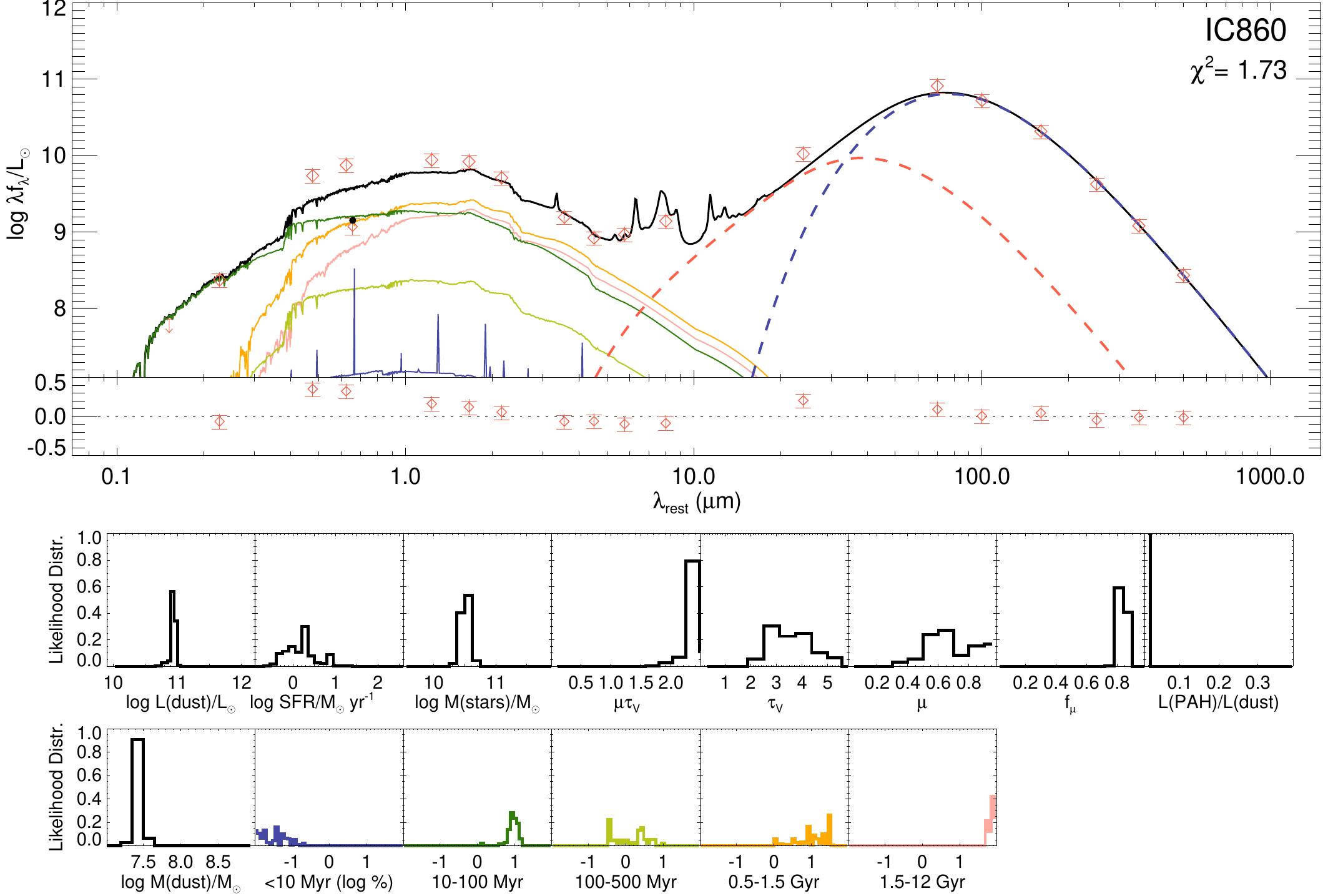}
\includegraphics[width=0.9\textwidth]{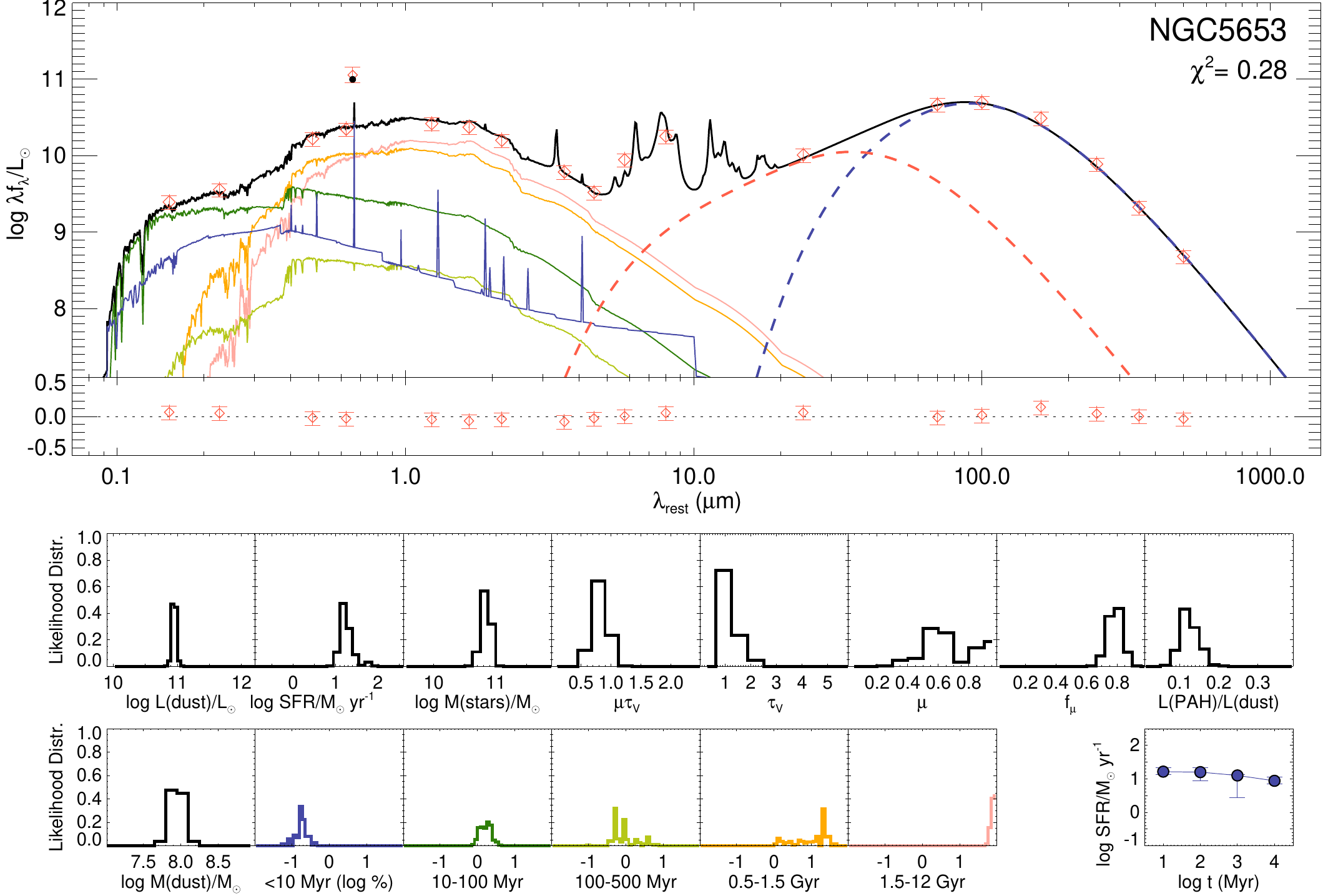}
\caption{Continued.}
\end{figure*}
\clearpage
\addtocounter{figure}{-1}
\begin{figure*}
\centering
\includegraphics[width=0.9\textwidth]{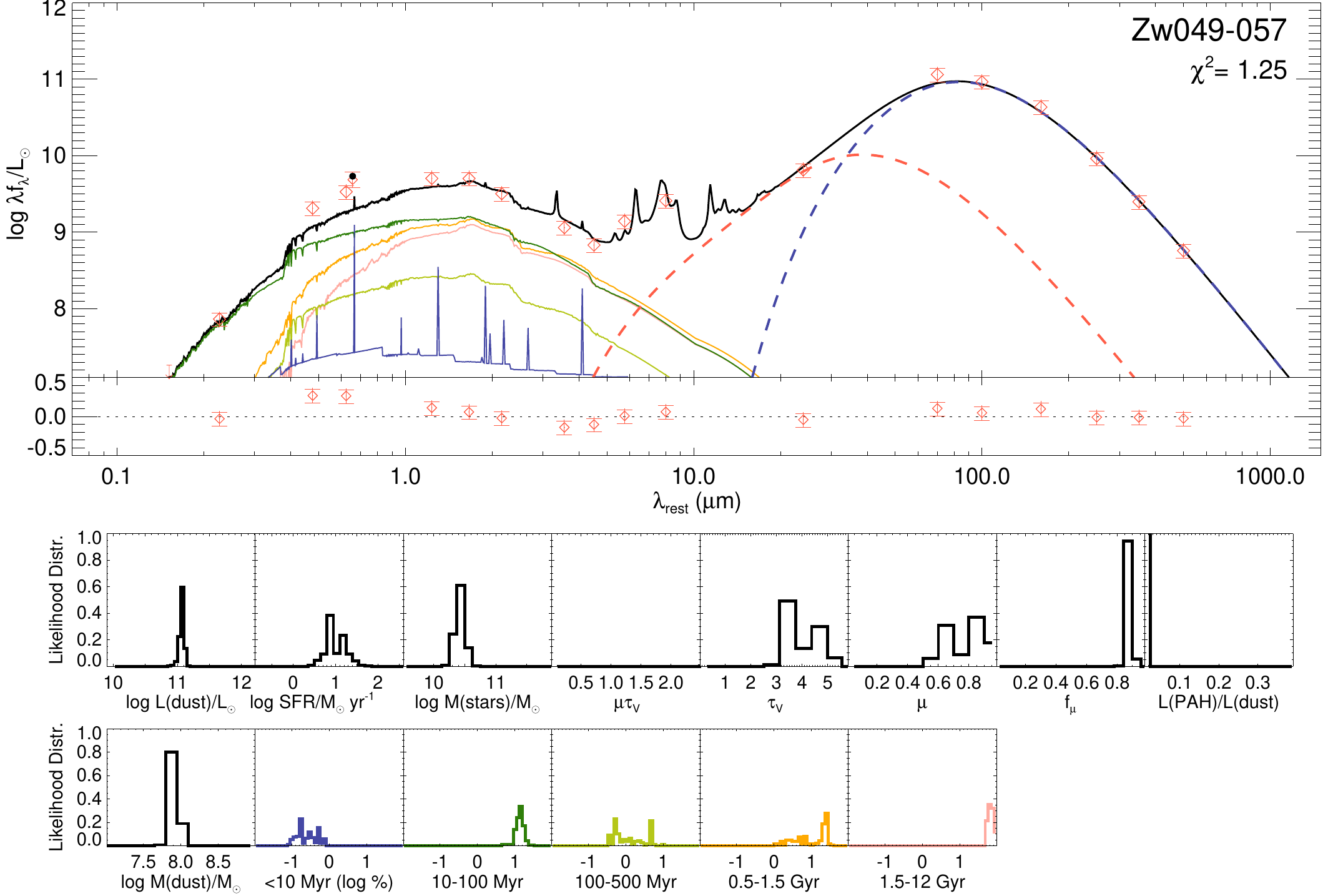}
\includegraphics[width=0.9\textwidth]{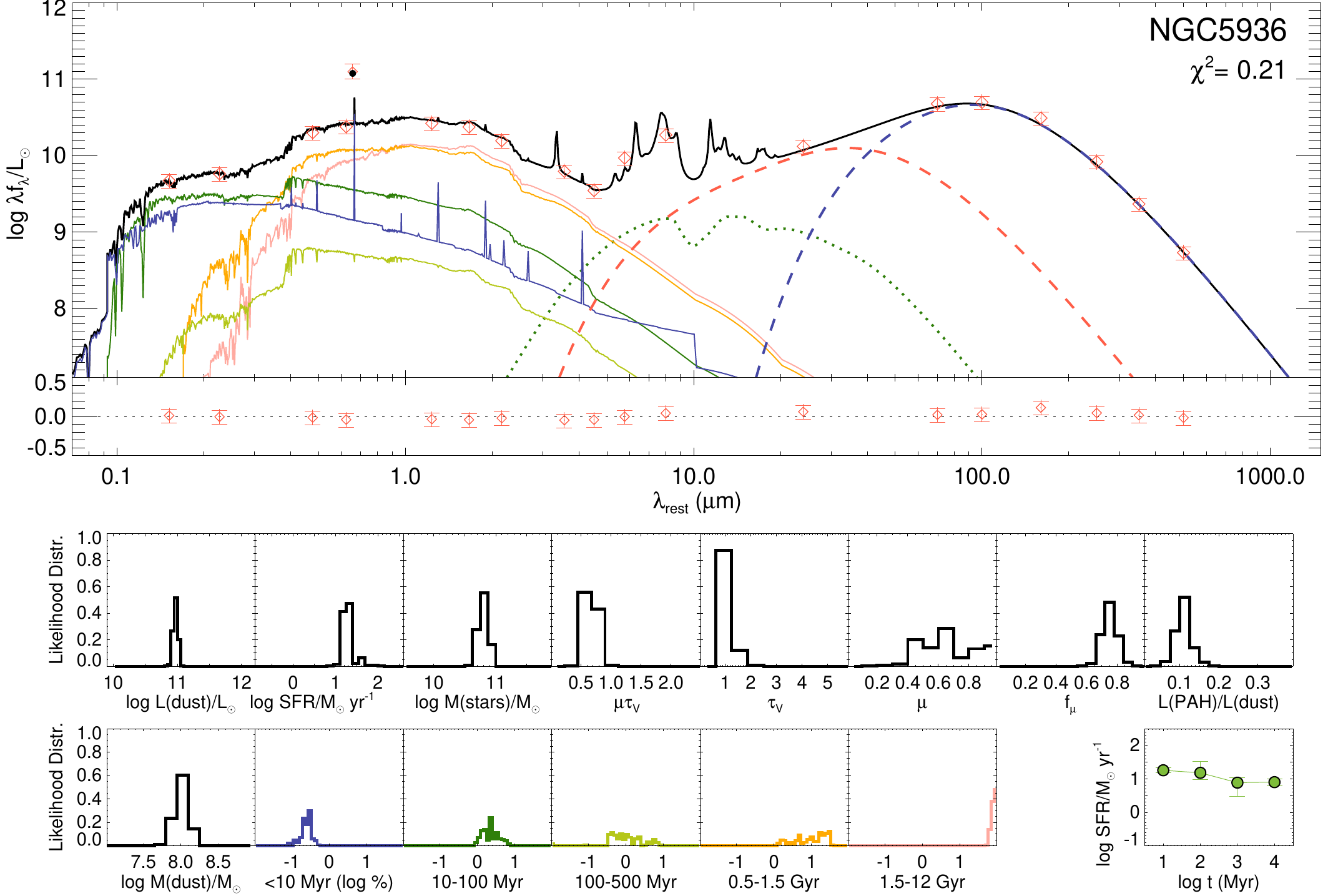}
\caption{Continued.}
\end{figure*}
\clearpage
\addtocounter{figure}{-1}
\begin{figure*}
\centering
\includegraphics[width=0.9\textwidth]{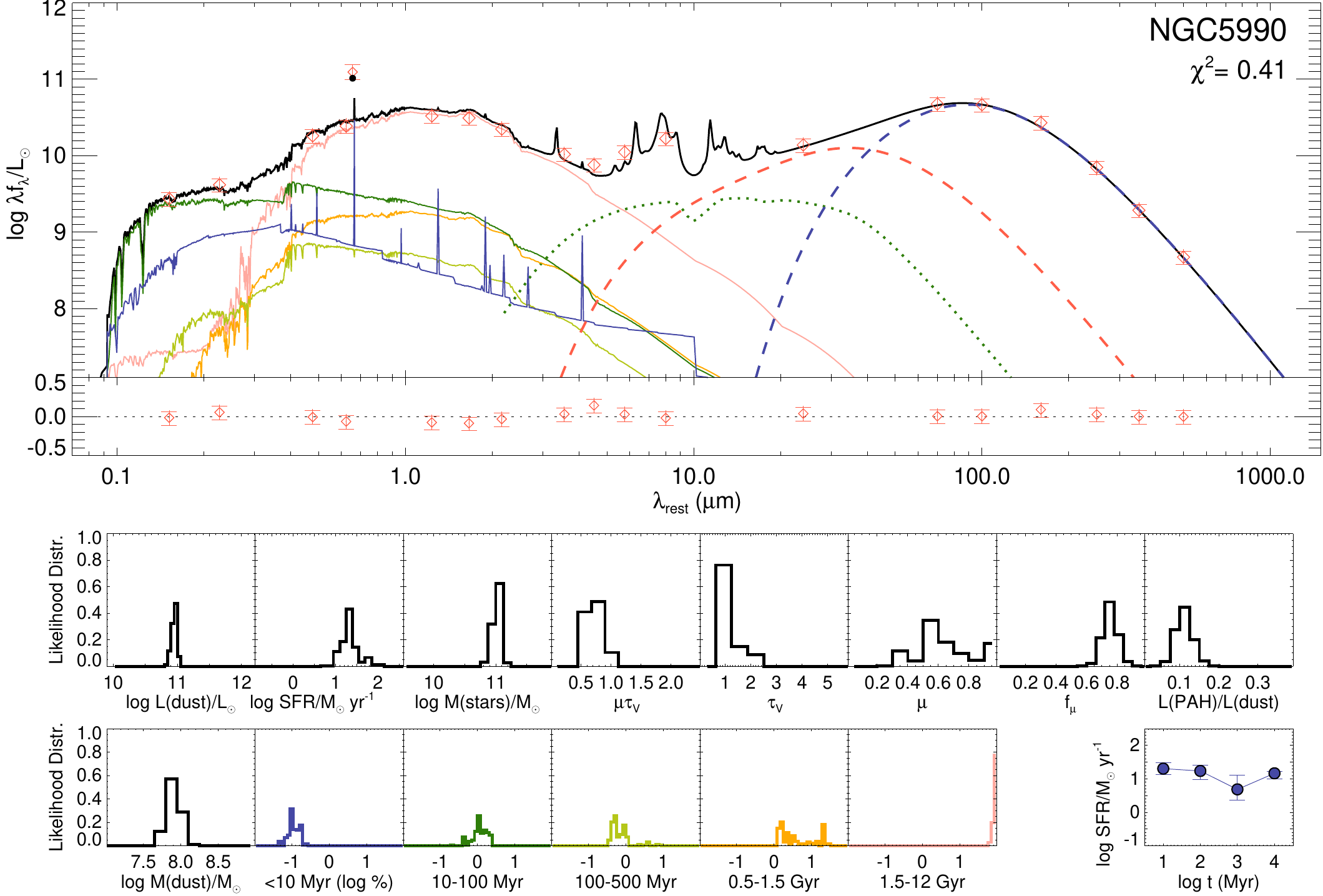}
\includegraphics[width=0.9\textwidth]{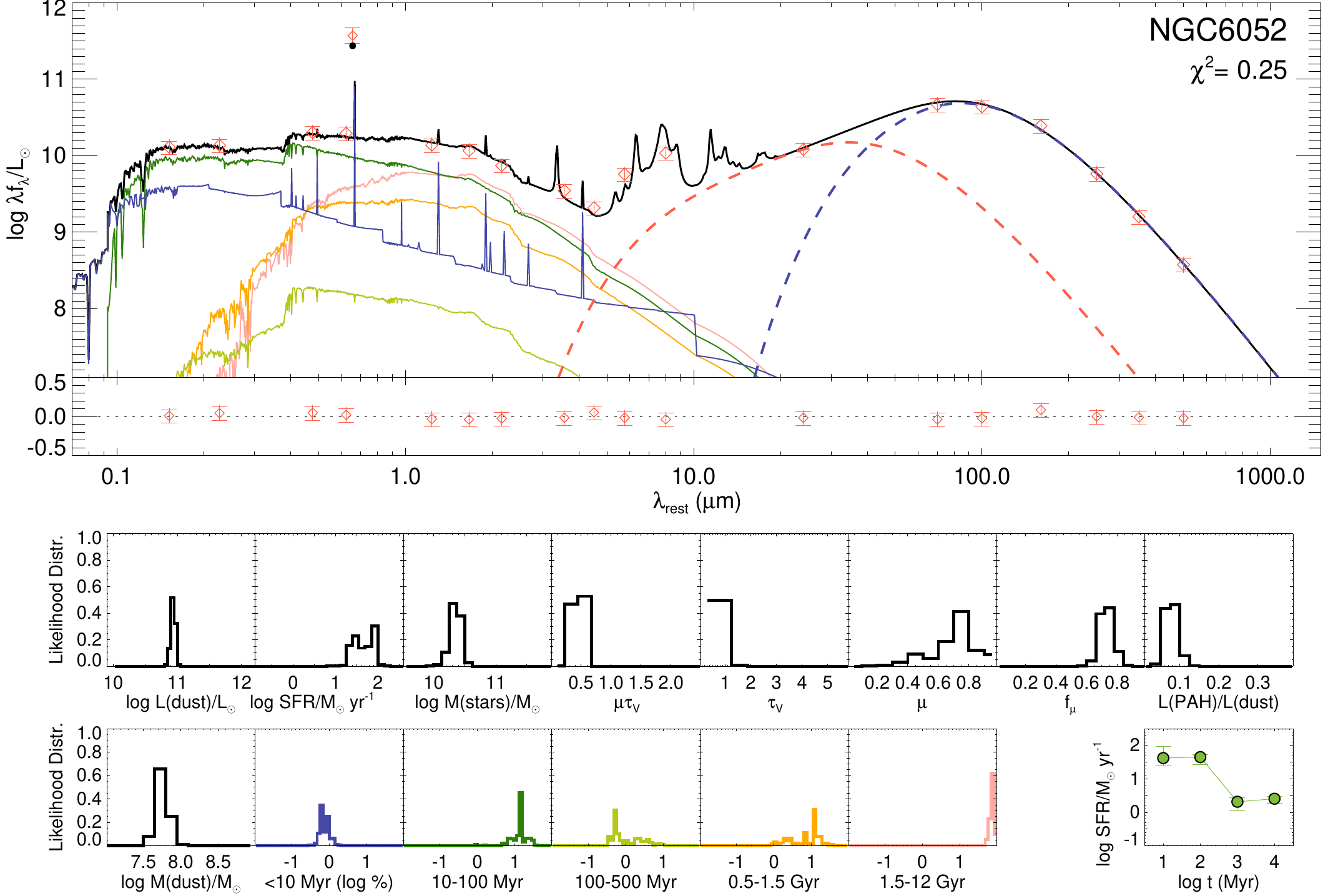}
\caption{Continued.}
\end{figure*}
\clearpage
\addtocounter{figure}{-1}
\begin{figure*}
\centering
\includegraphics[width=0.9\textwidth]{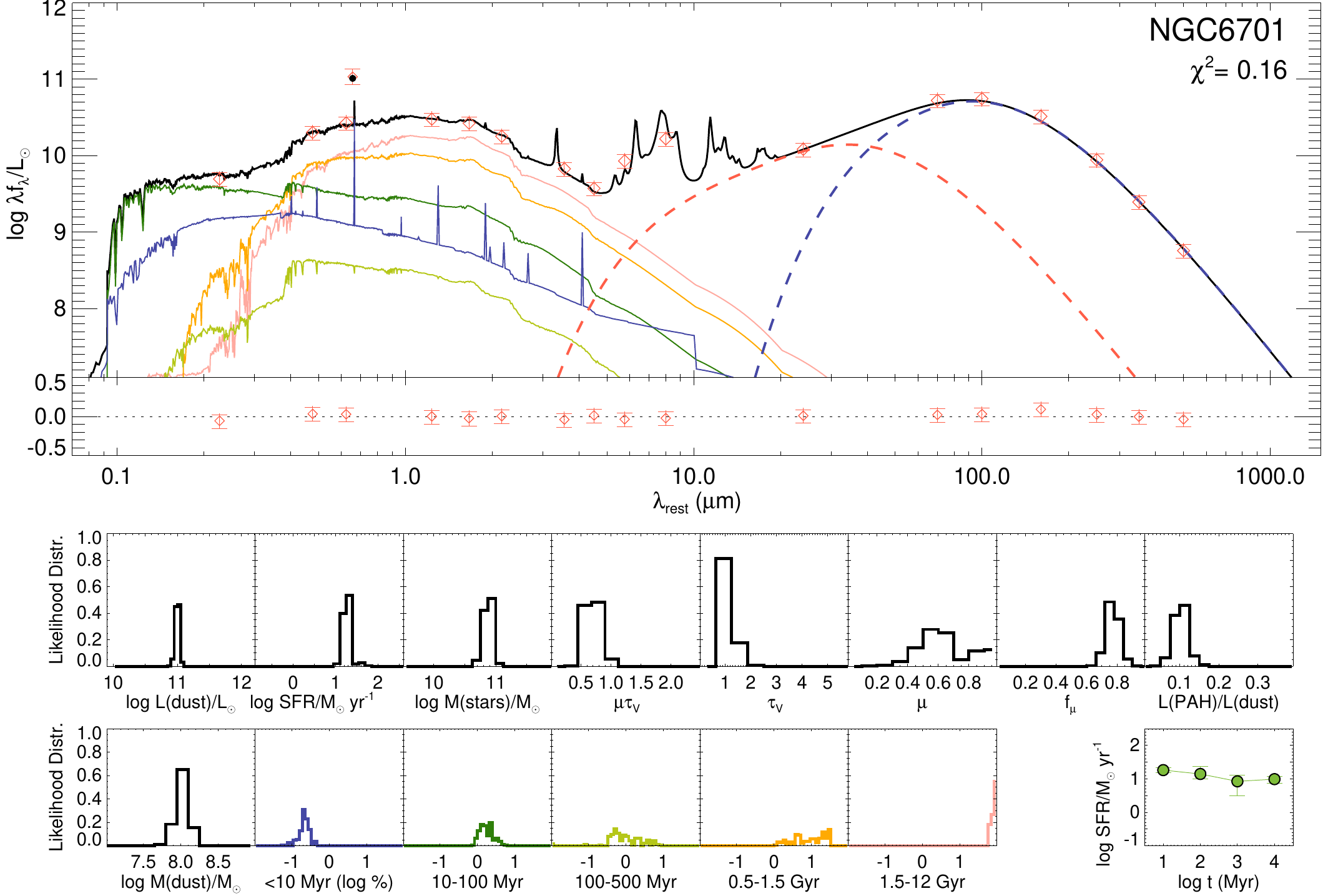}
\includegraphics[width=0.9\textwidth]{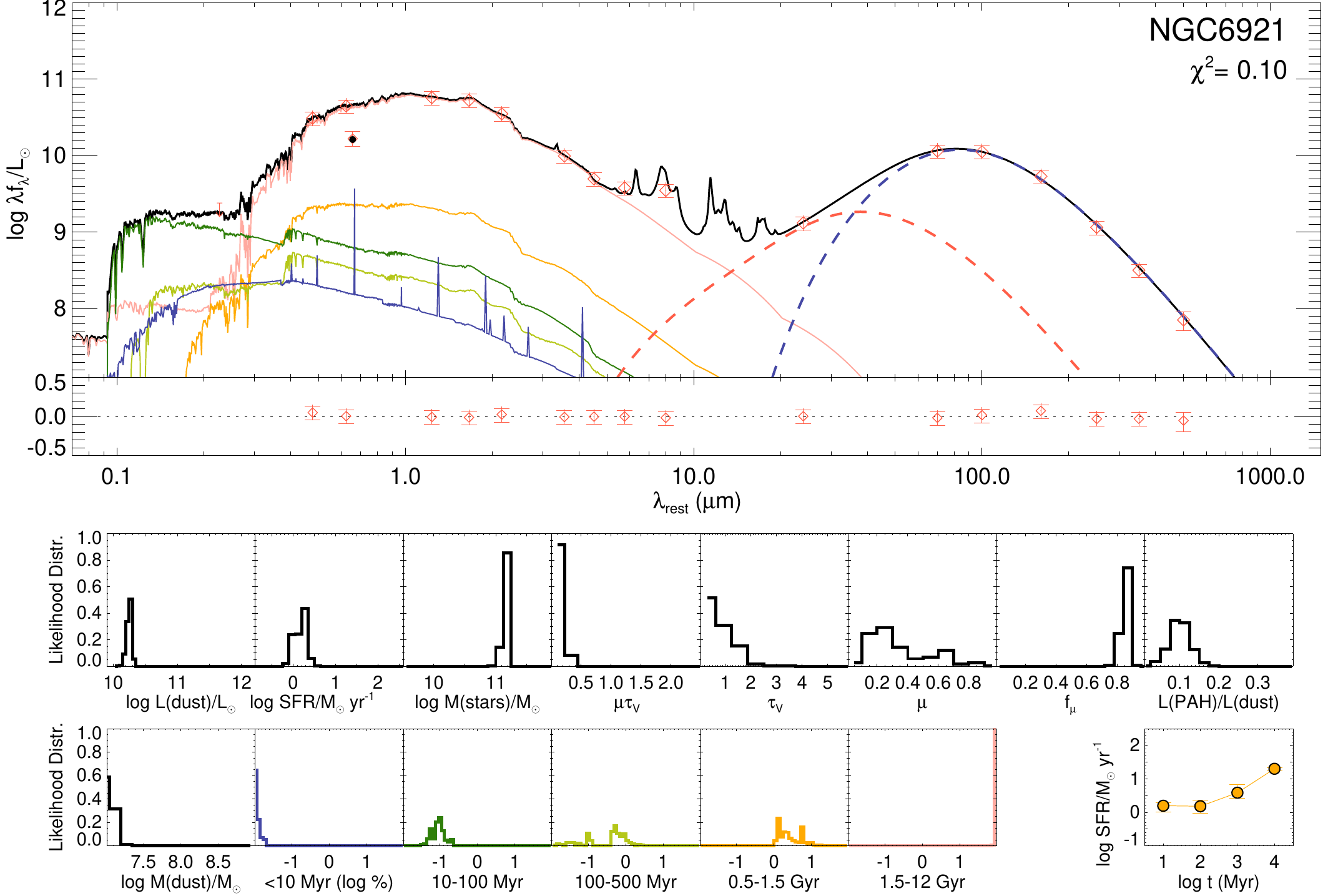}
\caption{Continued.}
\end{figure*}
\clearpage
\addtocounter{figure}{-1}
\begin{figure*}
\centering
\includegraphics[width=0.9\textwidth]{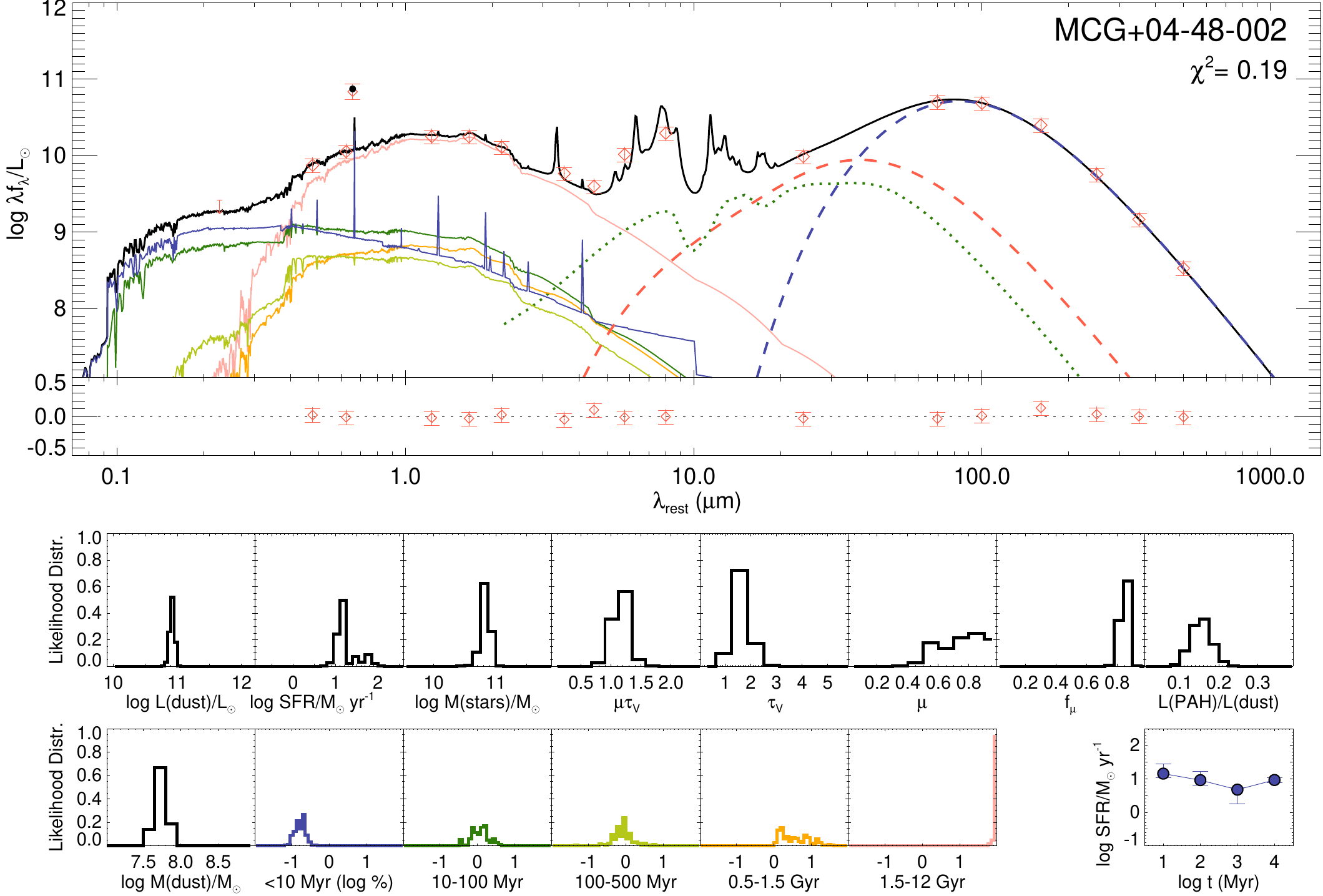}
\includegraphics[width=0.9\textwidth]{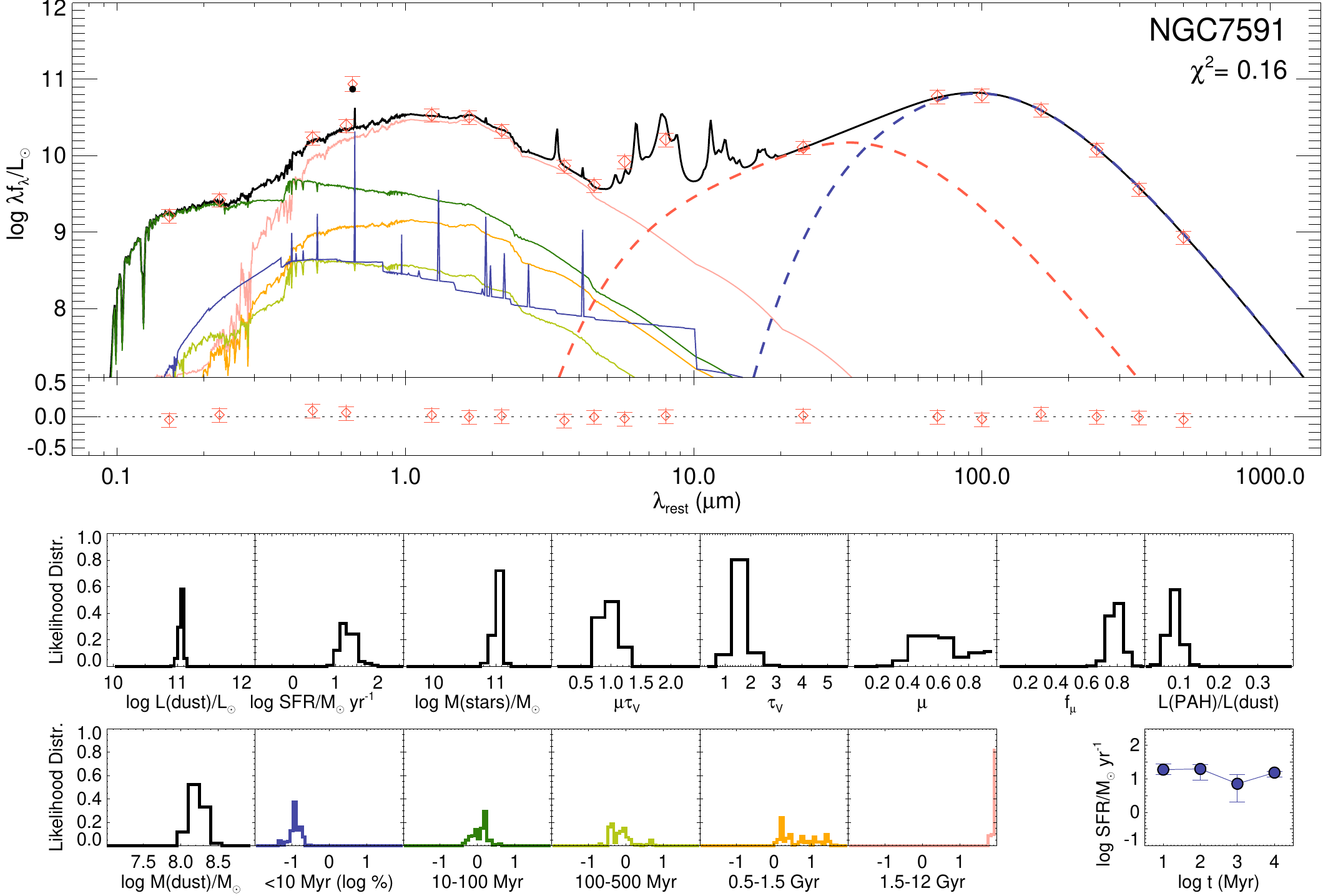}
\caption{Continued.}
\end{figure*}
\clearpage
\addtocounter{figure}{-1}
\begin{figure*}
\centering
\includegraphics[width=0.9\textwidth]{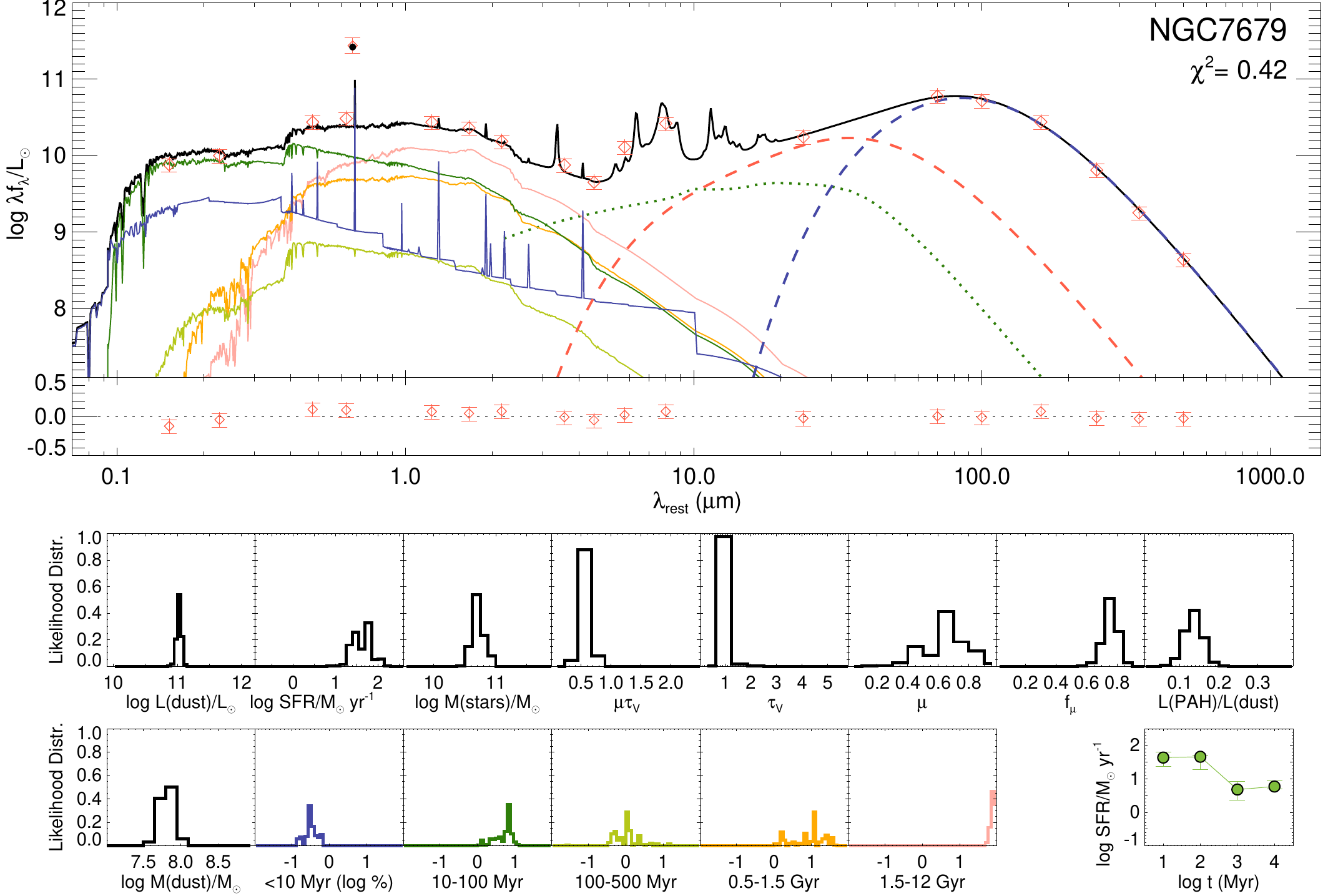}
\includegraphics[width=0.9\textwidth]{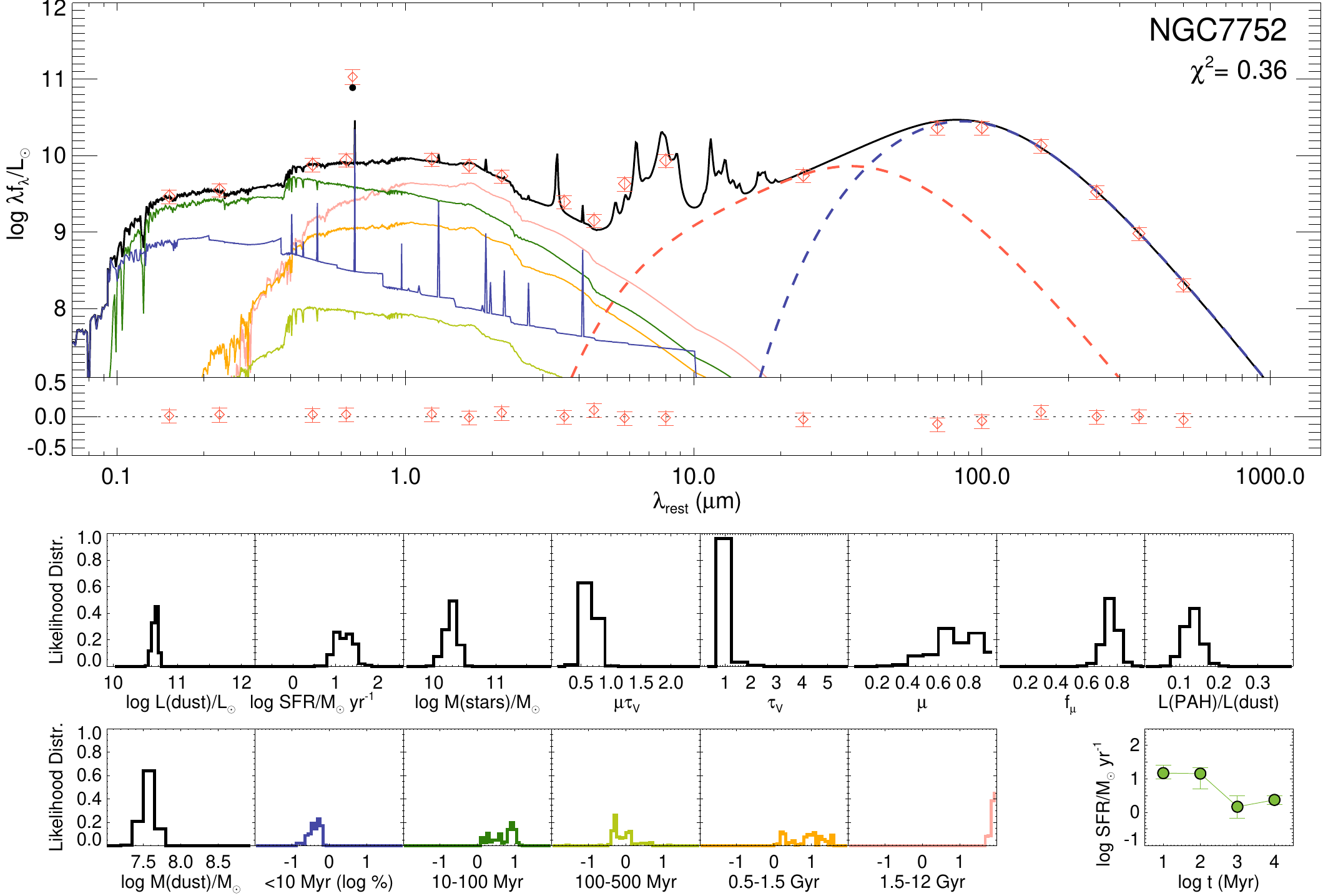}
\caption{Continued.}
\end{figure*}
\clearpage
\addtocounter{figure}{-1}
\begin{figure*}
\centering
\includegraphics[width=0.9\textwidth]{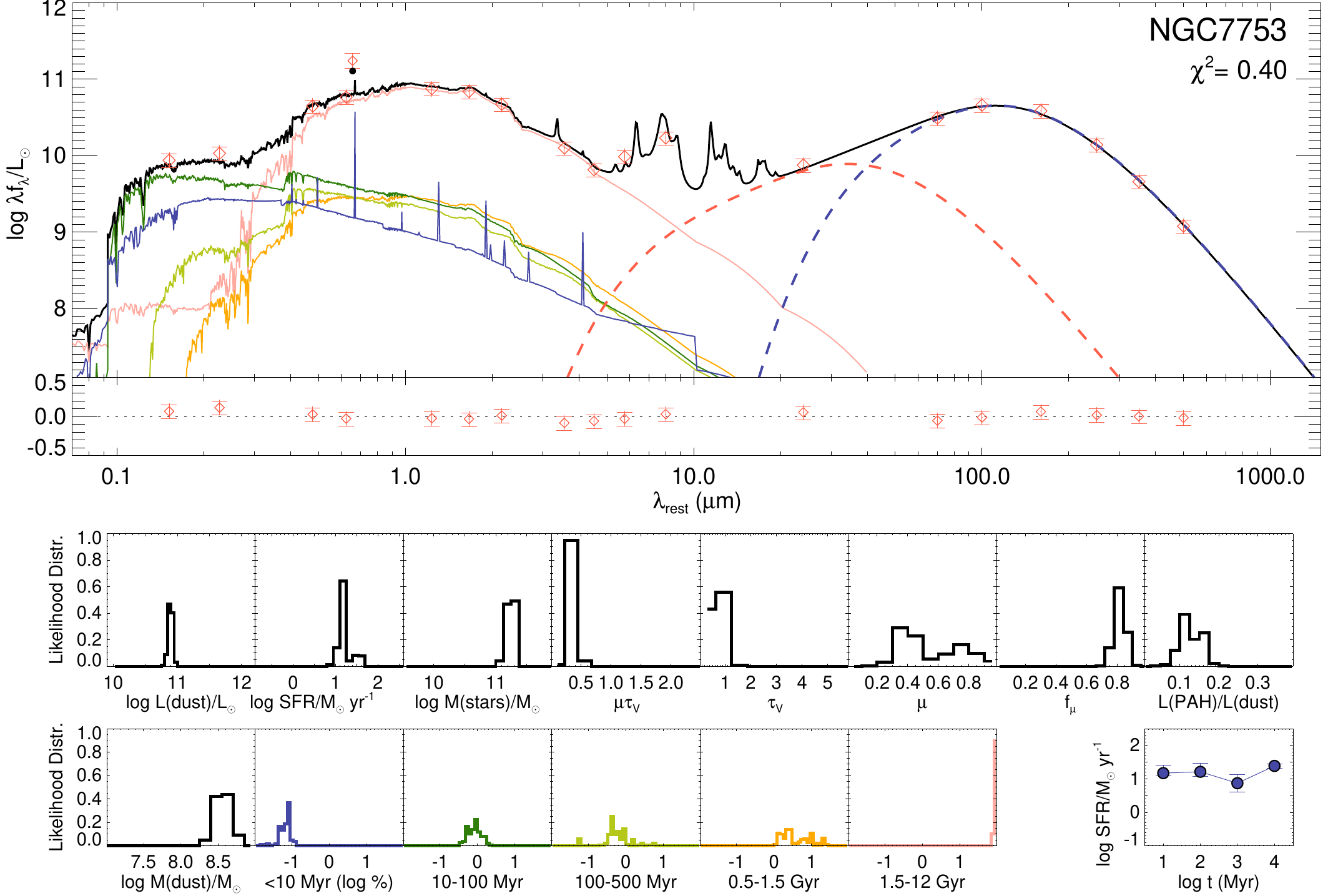}
\includegraphics[width=0.9\textwidth]{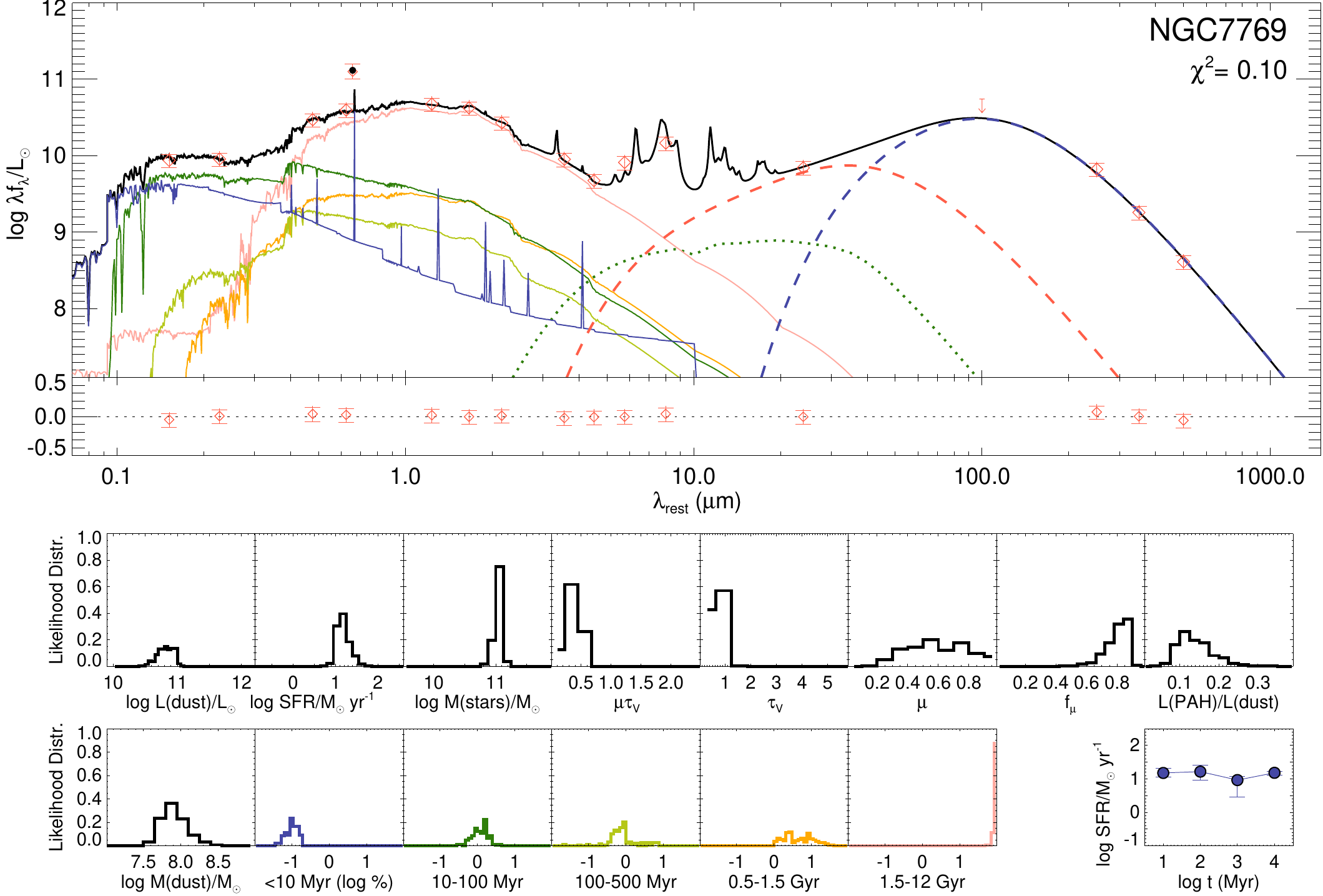}
\caption{Continued.}
\end{figure*}
\clearpage
\addtocounter{figure}{-1}
\begin{figure*}
\centering
\includegraphics[width=0.9\textwidth]{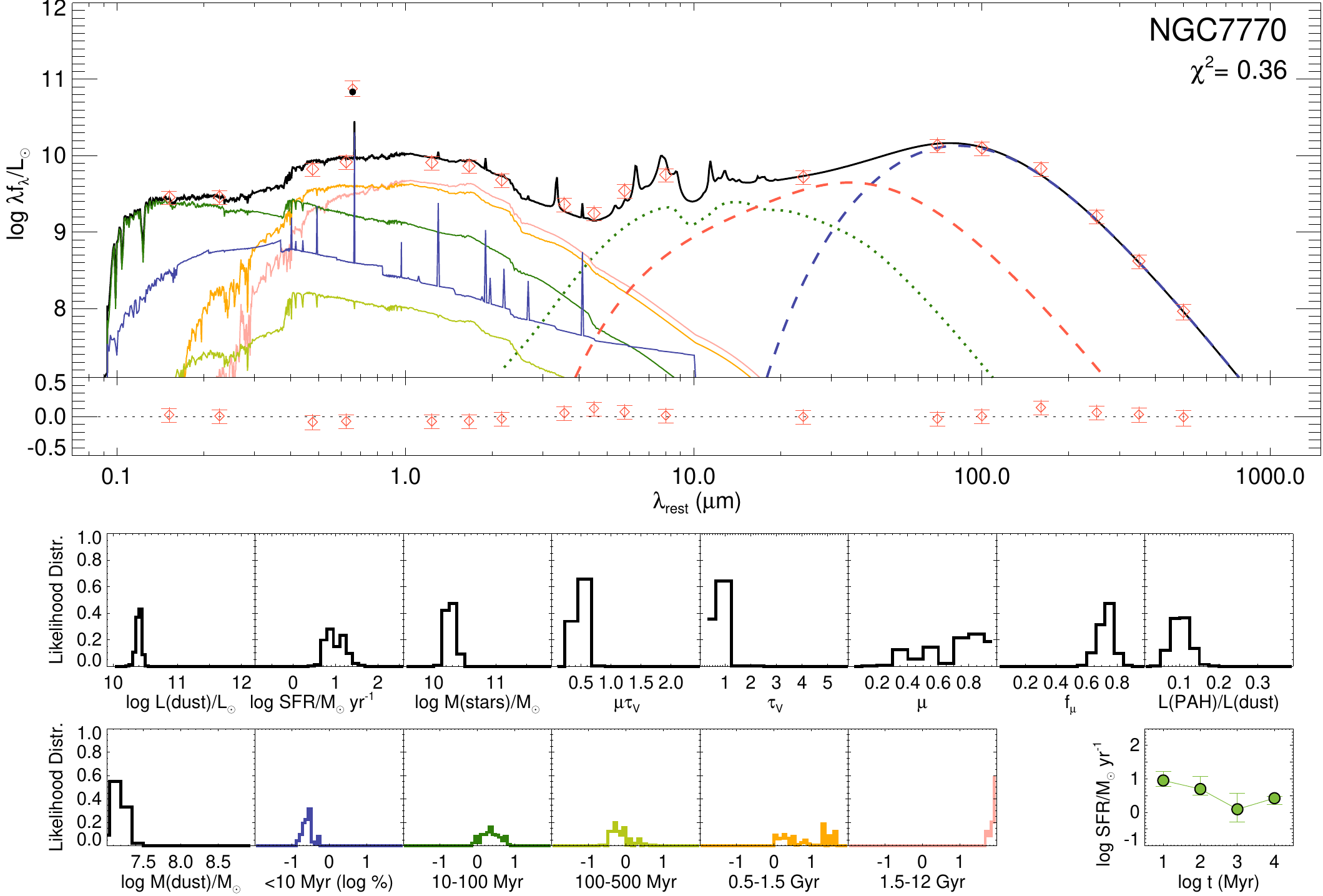}
\includegraphics[width=0.9\textwidth]{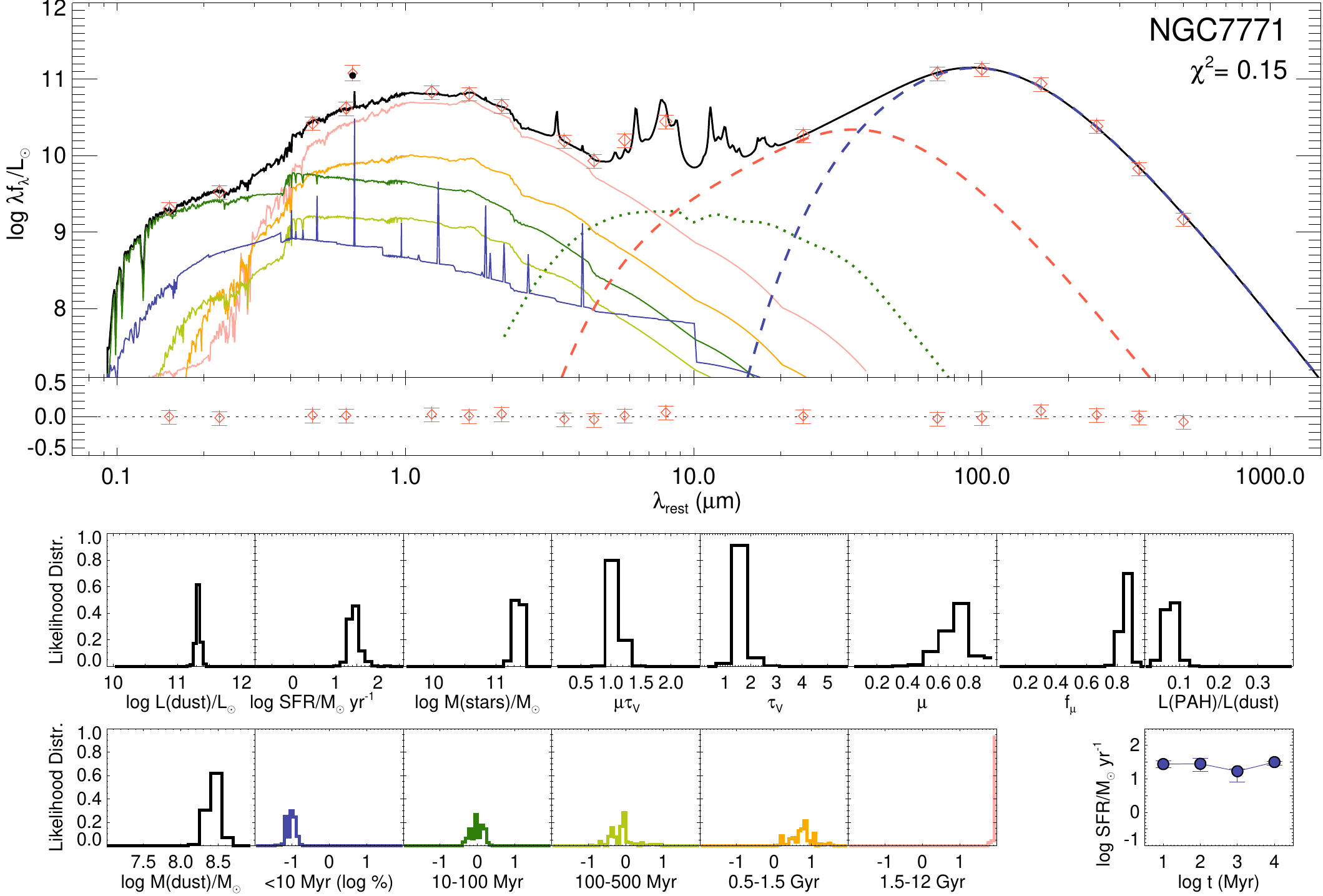}
\caption{Continued.}
\end{figure*}
\clearpage

\end{document}